\documentclass[,nofootinbib,amsmath,amssym,preprint,aps]{revtex4}
\usepackage{hyperref,slashed,color}
\usepackage{graphicx,subfig}
\begin{document}
\preprint{TUHEP-TH-11175}
 \draft
\title{Electroweak Chiral Lagrangian from TC2 Model with nontrivial TC fermion condensation and walking}

\author{Feng-Jun Ge$^1$, Shao-Zhou Jiang$^2$\footnote{{\it Email
address}:~jsz@gxu.edu.cn(S.Z.Jiang).},  Qing Wang$^{1,3}$\footnote{Corresponding author at:
Department of Physics, Tsinghua University, Beijing 100084,
P.R.China\\ {\it Email
address}:~wangq@mail.tsinghua.edu.cn(Q.Wang).}}

\address{$^1$Department of Physics, Tsinghua University, Beijing 100084, P.R.China \\
$^2$College of Physics Science and Technology, Guangxi University, Nanning, Guangxi 530004, P.R.China\\
    $^3$Center for High Energy Physics, Tsinghua University, Beijing 100084, P.R.china}

\date{April 29, 2011}

\begin{abstract}

The electroweak chiral Lagrangian for the topcolor assisted technicolor model proposed by K. Lane, which uses nontrivial patterns of techniquark condensation and walking, was investigated in this study. We found that the features of the model are qualitatively similar to those of Lane's previous natural TC2 prototype model, but there is no limit on the upper bound of the $Z'$ mass. We discuss the phase structure and
possible walking behavior of the model. We obtained the values of all coefficients of the electroweak chiral Lagrangian up to an order of $p^4$. We show that although the walking effect reduces the S parameter to half its original value, it maintains an order of $2$. Moreover, a special hyper-charge arrangement is needed to achieve further reductions in its value.

\end{abstract}
\maketitle


\vspace{1cm}
\section{Introduction}

Modern technicolor (TC) models of dynamical electroweak symmetry breaking require assistance for top-color interactions that are strong in the
TeV energy region to provide the large mass of the top quark, and a walking technicolor (WTC) gauge coupling to aid in the avoidance of large flavor-changing neutral current (FCNC) effects. The first addition consists of a class of topcolor-assisted technicolor (TC2) models made through the careful arrangement of TC, topcolor, extended hypercharge groups, and relevant techniquark and Standard Model (SM) fermion representations.
With the help of extended technicolor (ETC), we expect that technicolor condensates will form and provide the mass for the weak vector bosons. ETC provides the mass for the light quarks and leptons and a bottom-quark-sized mass to the top. The largest contribution to the top-quark mass is from the formation of a top-quark condensate through the dynamics of the topcolor gauge sector. The second addition is based on the phase diagram of strongly coupled TC gauge theories involving fermions in arbitrary representations of the gauge group.  With suitable choices for the TC group and techniquark representations, WTC is a natural option for situations with asymptotic freedom that are nearly conformal. In this case, the TC gauge coupling has an approximate infrared-stable fixed point (the zero of the beta function) $\alpha_*$ which is slightly larger than the critical
value $\alpha_c$ necessary for techniquark condensate formation. In such a theory, for values of $\alpha$ above $\alpha_*$, as the energy scale decreases $\alpha$ increases. However, its rate of increase decreases to zero as $\alpha$ approaches $\alpha_*$. Hence, over an extended energy interval, $\alpha$ is order O(1), and it is slowly varying which leads a large anomalous dimension $\gamma\simeq 1$ for the bilinear local techniquark operator. This results in the enhancement of the SM fermion and those undiscovered pseudo goldstone boson masses, which achieve realistic scales while maintaining sufficient suppression
of FCNC effects.

  The typical gauge group of the TC2 models is
 \begin{eqnarray}
 SU(N)_{\mathrm{TC}}\otimes SU(3)_1\otimes SU(3)_2\otimes SU(2)_L\otimes U(1)_{Y_1}\otimes U(1)_{Y_2}\label{gaugegroup}
 \end{eqnarray}
in which the topcolor and extended hypercharge groups $SU(3)_1\otimes SU(3)_2\otimes U(1)_{Y_1}\otimes U(1)_{Y_2}$ spontaneously break into their diagonal subgroups $SU(3)_C\otimes U(1)_Y$ at an energy of a few TeV. The remaining electroweak groups $SU(2)_L\otimes U(1)_Y$
spontaneously break into their electromagnetic subgroup $U(1)_{\mathrm{em}}$ at electroweak scale because of a combination of a top-quark condensate
and techniquark condensate. In the simplest example of Hill's TC2 model \cite{Hill95}, there are separate color and weak hypercharge gauge groups for the heavy third generation quarks and leptons and for the two lighter generations. The third generation transforms under a strongly coupled $SU(3)_1\otimes U(1)_1$ and maintains its usual charges. However, the light generations
transform conventionally under a weakly coupled $SU(3)_2\otimes U(1)_2$. Near 1 TeV, these four
groups break into a diagonal subgroup of ordinary color and hypercharge,
$SU(3)_C\otimes U(1)_Y$. The desired condensation pattern occurs because the $U(1)_1$ couplings are such that the spontaneously broken $SU(3)_1\otimes U(1)_1$ interactions are supercritical only for the top quark.

After Hill's proposal was made, Chivukula, Dobrescu, and Terning \cite{CDT} claimed that the techniquarks required to break the top and bottom quark chiral symmetries are likely to have custodial-isospin violating couplings to the strong $U(1)_1$. To maintain a $\rho\simeq1$, the $U(1)_1$ interaction must be so weak that it is necessary to fine-tune the $SU(3)_1$ coupling. This results in the implementation of the theory being unnatural. To remedy this isospin violation and improve the suitability of the model, K. Lane proposed a natural prototype TC2 model in Ref.\cite{Lane95}. In that model, the different techniquark isodoublets, $T^t$ and $T^b$, provide ETC mass to the top and bottom quarks. These doublets then could have different $U(1)_1$ charges, which are isospin conserving for the right and left handed parts of each doublet. The $U(1)$ symmetries presented in the model automatically
avoid the problem of $B_d-\bar{B}_d$ mixing raised by Kominis\cite{Kominis}. To achieve the mixing of the magnitude
observed between the heavy and light generations while breaking the strong top-color interactions near 1 TeV, K. Lane also proposed an alternative model based on the nontrivial patterns of techniquark condensation and discussed its phenomenology\cite{Lane96}. In this new model, to break the extended hypercharge groups into $U(1)_Y$, a set of electrically neutral $SU(2)$ singlet techniquarks belonging to the antisymmetric tensor representation of the TC group were added into the model. This, in combination with other techniquarks, further ensures
the technicolor coupling walks. With so many techniquarks, one may wonder whether the $S$ parameter of the model can be small. Although qualitatively the large number of techniquarks will increase the value of $S$, walking effects and certain arrangements of the hypercharges of the techniquarks may compensate for this increase, and result in a small overall $S$ parameter. One aim of this paper is to examine this possibility.

In fact, our interests are not limited to the S parameter, which is one of the low energy constants (LECs) of the bosonic part of the standard electroweak chiral Lagrangian (EWCL)\cite{EWCL}. Rather, our interests include all EWCL LECs. In our previous studies, we compiled a formulation for computing the bosonic part of the EWCL LECs for orders up to $p^4$ for the one-doublet TC model discussed in Ref.\cite{1D}, Hill's schematic TC2 model \cite{Hill95} in Ref.\cite{HongHao08}, K. Lane's natural prototype TC2 model \cite{Lane95} in Ref.\cite{JunYi09} and a hypercharge-universal TC2 model \cite{Sekhar} in Ref.\cite{LangPLB}. Here, the bosonic part of the EWCL is the part that only involves SM electroweak gauge fields and corresponding Goldstone fields. This part describes the electroweak symmetry breaking effects on the electroweak gauge fields, but the parts of the EWCL dealing with matter also include SM fermions which describe the electroweak symmetry breaking effects on the SM fermion fields. In the literature, these two parts are proposed in Refs.\cite{EWCL} and \cite{EWCLfermion} separately because they have independent characteristics. The reason that we choose to compute the bosonic part of the EWCL in isolation is that the matter part is more complex than the bosonic part. Moreover, some of the three-dimensional fermion mass terms and six-dimensional FCNC terms were already discussed in Lane's original paper \cite{Lane96}. In this paper, we only discuss the bosonic part of the EWCL for the first stage of computing the LECs that are generalized from the $S$ parameter, and leave the part dealing with matter for future discussion. The EWCL is an universal platform which enables us to compare different underlying models with experimental data and extract the true physical theory that guides our world. To achieve this comparison, we compute the EWCL coefficients model by model. This study is the fourth paper in a series, starting with Ref.\cite{HongHao08}, in which we compute these strongly coupled physics models. Here, we focus on K. Lane's alternative TC2 model with nontrivial TC fermion condensation and walking\cite{Lane96}, which was mentioned previously. Corresponding to recent advances in the understanding of the phase diagram of the $SU(N)$ gauge theories and the new possibilities for model building\cite{NewWalking}, this work offers a modern way to investigate walking effects in a realistic strongly-coupled theory with complex structures.

In this paper, except for some conventional calculations that are similar to those in our previous papers, we focus on the effects of walking that have not been discussed before. We will compare the different situations of walking, ideal walking, and running; and examine their effects on the $S$ parameter. In the next section, we first review K. Lane's alternative TC2 model with nontrivial condensation and walking\cite{Lane96} and discuss its phase structure. In section III, we apply our formulation developed in Ref.\cite{HongHao08} to Lane's model \cite{Lane96}. We perform these dynamical calculations through several steps: first we integrate in the Goldstone field, U. Then, we integrate out the technigluons and techniquarks by solving the Schwinger-Dyson equation (SDE) for techniquarks. Next, we integrate out the colorons and $Z'$, perform a low energy expansion, and compute the effective action. Finally, we obtain the EWCL coefficients. For simplicity, some details of the derivation and computation in this section are placed in the appendices. Section IV. contains numerical results and discussions. Section V. is a short summary and discussion.

\section{Review of the Model and its phase structure}

Consider K. Lane's TC2 model \cite{Lane96} with nontrivial TC fermion condensation and walking, in which the group is given by (\ref{gaugegroup}).
Because we are only interested in the bosonic part of EWCL, which is independent of the SM fermions, we do not list their representations and $U(1)$ charge arrangements here. The left gauge charges for the techniquarks are shown in Table I. There are three sets of techniquarks. The first set includes $T^1$ and $T^2$. These are the specific techniquarks
of the model and are expected to have twisted condensates that generate $SU(3)_1\otimes SU(3)_2\rightarrow SU(3)_c$ and electroweak breaking, and a sufficient level of generation mixing. The second set includes $T^l$, $T^t$ and $T^b$, which are the standard TC2 techniquarks from Lane's first natural prototype TC2 model \cite{Lane95}. They supply the ETC mass to the SM fermions, including the top and bottom. The third set consists of the high-dimensional representation field $\psi$, which is responsible for generating $U(1)_1\otimes U(1)_2\rightarrow U(1)_Y$ and ensuring theory walking.

\begin{table}[h]
\small{{\bf TABLE I}.~Gauge charge assignments of the techniquarks in
Lane's TC2 model.\\~\vspace*{-0.2cm}~}

\begin{tabular}{|c|cccccc|}\hline
field$\setminus$group & $SU(N)_{\mathrm{TC}} $&$SU(3)_{1}$&$SU(3)_{2}$& $SU(2)_L $ &
$U(1)_{1} $ &
$U(1)_{2} $ \\
\hline\hline {\footnotesize field,coupling} &
$G^{\alpha}_\mu,g_{\mathrm{TC}}$&$A^{A}_{1\mu},h_{1}$&$A^{A}_{2\mu},h_{2}$
& $W_{\mu}^{a},g_{2}$&$B_{1\mu},q_{1}$&~~$B_{2\mu},q_{2}$~~~\\
\hline $T_{L}^{1}$ & N& 3&1& 2&$u_{1}$&$u_{2}$ \\
\hline $U_{R}^{1}$ & N& 3&1& 1&$v_{1}$&$v_{2}+\frac{1}{2}$ \\
\hline $D_{R}^{1}$ & N& 3&1& 1&$v_{1}$&$v_{2}-\frac{1}{2}$ \\
\hline $T_{L}^{2}$ & N& 1&3& 2&$v_{1}$&$v_{2}$ \\
\hline $U_{R}^{2}$ & N& 1&3& 1&$u_{1}$&$u_{2}+\frac{1}{2}$ \\
\hline $D_{R}^{2}$ & N& 1&3& 1&$u_{1}$&$u_{2}-\frac{1}{2}$ \\
\hline $T_{L}^{l}$ & N& 1&1& 2&$x_{1}$&$x_{2}$\\
\hline $U_{R}^{l}$ & N& 1&1& 1&$x_{1}^{\prime}$&$x_{2}^{\prime}+\frac{1}{2}$\\
\hline $D_{R}^{l}$ & N& 1&1& 1&$x_{1}^{\prime}$&$x_{2}^{\prime}-\frac{1}{2}$\\
\hline $T_{L}^{t}$ & N& 1&1& 2&$y_{1}$&$y_{2}$\\
\hline $U_{R}^{t}$ & N& 1&1& 1&$y_{1}^{\prime}$&$y_{2}^{\prime}+\frac{1}{2}$\\
\hline $D_{R}^{t}$ & N& 1&1& 1&$y_{1}^{\prime}$&$y_{2}^{\prime}-\frac{1}{2}$\\
\hline $T_{L}^{b}$ & N& 1&1& 2&$z_{1}$&$z_{2}$\\
\hline $U_{R}^{b}$ & N& 1&1& 1&$z_{1}^{\prime}$&$z_{2}^{\prime}+\frac{1}{2}$\\
\hline $D_{R}^{b}$ & N& 1&1& 1&$z_{1}^{\prime}$&$z_{2}^{\prime}-\frac{1}{2}$\\
\hline $\psi_{L} $ &$\frac{1}{2}N(N-1)$& 1&1& 1&$\xi$&$-\xi$\\
\hline $\psi_{R} $ &$\frac{1}{2}N(N-1)$& 1&1& 1&$\xi^{\prime}$&$-\xi^{\prime}$\\
\hline\hline
 \end{tabular}
 \end{table}

 The details of the ETC interaction are not specified in Lane's original paper\cite{Lane96}; this
prohibits quantitative computations. The effects on the EWCL LECs from these ETC operators can be ignored in our calculation because the relevant operators are small. Unfortunately, although we know from Ref. \cite{JunYi09} that its contribution to the EWCL LECs is small, the effective four-fermion coupling may become strong enough to change the results of the current walking theory\cite{ETC}. When the effective four-fermion coupling exceeds its critical value, the position of the infrared fixed point changes significantly. For the first step of the investigation, we ignore this case by assuming that the four-fermion coupling does not exceed the critical value and leave discussion of more general effects for future studies.

 A number of constraints were given in Lane's original paper\cite{Lane96}  to limit and simplify the charges:
 \begin{itemize}
\item To ensure that the techniquark condensates conserve electric charge, $u_1+u_2=v_1+v_2$, $x_1+x_2=x_1'+x_2'$, $y_1+y_2=y_1'+y_2'$, and $z_1+z_2=z_1'+z_2'$.
\item The $U(1)_1$ charges of the techniquarks respect custodial isospin.
\item For the $U(1)_1$ charges of $T^1$ and $T^2$: while $u_1\neq v_1$, the broken $U(1)_1$ interactions favor the condensation of $T^1$ with $T^2$. If this interaction is stronger than the $SU(3)_1$ attraction of $T^1$ to itself and we neglect the other vacuum-aligning ETC interactions, then
 $\langle\bar{T}^i_LT^j_R\rangle\propto(i\tau^2)_{ij}$ in each charge sector.
\item $u_1\neq v_1$ implies $Y_{1i}\neq Y_{1i}'$ for the fermions.
\item For the $SU(N)_{\mathrm{TC}}$ antisymmetric tensor $\psi$, $\xi'\neq\xi$ guarantees $U(1)_1\otimes U(1)_2\rightarrow U(1)_Y$ when $\langle\overline{\psi_L}\psi_R\rangle$ forms.
\end{itemize}

 The Lagrangian of the model is
\begin{equation}
S[G,A_1,A_2,W,B_1,B_2,\bar{T},T,\bar{\psi},\psi]=\int d^4x[\mathcal {L}_{\mathrm{gauge~kinetic}}+\mathcal {L}_{\mathrm{techniquark}}+\mathcal {L}_{\mathrm{SM~fermion}}]\;,
\end{equation}
with
\begin{eqnarray}
\mathcal {L}_{\mathrm{gauge~kinetic}}=-\frac{1}{4}\bigg[G_{\mu\nu}^{\alpha}G^{\alpha,\mu\nu}+A_{1\mu\nu}^{A}A^{A,1\mu\nu}
+A_{2\mu\nu}^{A}A^{A,2\mu\nu}+W_{\mu\nu}^aW^{a,\mu\nu}+B_{1\mu\nu}B^{1,\mu\nu}+B_{2\mu\nu} B^{2,\mu\nu}\bigg]\nonumber
\end{eqnarray}
and
\begin{eqnarray}
&&\hspace*{-0.5cm}\mathcal
{L}_{\mathrm{techniquark}}=\nonumber\\
&&\hspace*{-0.5cm}+\bar{T}^{1}[i\slashed{\partial}\!-\!g_{\rm TC}t^\alpha
\slashed{G}^\alpha\!\!-\!h_1\frac{\lambda^{A}}{2}\slashed{A}_1^A\!\!-\!g_2\frac{\tau^{a}}{2}\slashed{W}^{a}P_L\!\!
-\!q_1u_1\slashed{B}_1P_L\!\!-\!q_2u_2\slashed{B}_2P_L\!\!-\!q_1v_1\slashed{B}_1P_R\!\!-\!q_2(v_2\!\!+\!\frac{\tau^3}{2})\slashed{B}_{2}P_{R}]T^{1}\nonumber\\
&&\hspace*{-0.5cm}+\bar{T}^{2}[i\slashed{\partial}\!-\!g_{\rm TC}t^\alpha\slashed{G}^\alpha\!\!-\!h_2\frac{\lambda^{A}}{2}\slashed{A}_2^A\!\!
-\!g_2\frac{\tau^a}{2}\slashed{W}^aP_L\!\!-\!q_1v_1\slashed{B}_1P_L\!\!-\!q_2v_2\slashed{B}_2P_L\!\!-\!q_1u_1\slashed{B}_1P_R\!\!
-\!q_2(u_2\!\!+\!\frac{\tau^3}{2})\slashed{B}_2P_R]T^{2}\nonumber\\
&&\hspace*{-0.5cm}+\bar{T}^{l}[i\slashed{\partial}-g_{\rm TC}t^\alpha\slashed{G}^\alpha\!-g_2\frac{\tau^a}{2}\slashed{W}^aP_L\!
-q_1x_1\slashed{B}_1P_L\!-q_2x_2\slashed{B}_2P_L\!-q_1x_1^{\prime}\slashed{B}_1P_R\!-q_2(x_2^{\prime}\!+\!\frac{\tau^3}{2})\slashed{B}_2P_R]T^{l}\nonumber\\
&&\hspace*{-0.5cm}+\bar{T}^{t}[i\slashed{\partial}-g_{\rm TC}t^\alpha\slashed{G}^\alpha\!-g_2\frac{\tau^a}{2}\slashed{W}^aP_L\!
-q_1y_1\slashed{B}_1P_L\!-q_2y_2\slashed{B}_2P_L\!-q_1y_1^{\prime}\slashed{B}_1P_R\!-q_2(y_2^{\prime}\!+\!\frac{\tau^3}{2})\slashed{B}_2P_R]T^{t}\nonumber\\
&&\hspace*{-0.5cm}+\bar{T}^{b}[i\slashed{\partial}-g_{\rm TC}t^\alpha\slashed{G}^\alpha\!-g_2\frac{\tau^a}{2}\slashed{W}^aP_L\!
-q_1z_1\slashed{B}_1P_L\!-q_2z_2\slashed{B}_2P_L\!-q_1z_1^{\prime}\slashed{B}_1P_R\!-q_2(z_2^{\prime}\!+\!\frac{\tau^3}{2})\slashed{B}_2P_R]T^{b}\nonumber\\
&&\hspace*{-0.5cm}+\bar{\psi}[i\slashed{\partial}-g_{\rm TC}\tilde{t}^\alpha\slashed{G}^\alpha\!-q_1\xi\slashed{B}_1P_L\!
+q_2\xi\slashed{B}_2P_L-q_1\xi_1^{\prime}\slashed{B}_1P_R\!+q_2\xi^{\prime}\slashed{B}_2P_R]\psi\;.
\end{eqnarray}
Where $\lambda^A$ is the three-dimensional Gellman matrix for topcolor interaction, $\tau^a$ is the Pauli matrix for the electroweak interaction, $t^{\alpha}$ is the $SU(N)_{\mathrm{TC}}$ fundamental representation matrix, $\tilde{t}^{\alpha}$ is the $SU(N)_{\mathrm{TC}}$ antisymmetric tensor representation matrix. We do not specify $\mathcal {L}_{\mathrm{SM~fermion}}$ which is not relevant to our discussions for the present approximation.

Now we will discuss the phase structure of the model. The two-loop $\beta$ function of the $SU(N)_{\mathrm{TC}}$ coupling, $g_{\mathrm{TC}}$, is\footnote{The reason that we chose the two-loop $\beta$ function instead of the one-loop version is that it can generate the walking effects needed for the model. Otherwise, the model setting must be rearranged. Physically, we
expect that the most significant contribution should come from the TC interaction. The SM particle mass does not reach the TC scale, and the masses of the colorons and $Z'$ slightly exceed this scale, all of their contributions are expected to be smaller than those of the TC interactions. For simplicity in the first stage approximation, we ignore the possible effects from SM particles, colorons, and $Z'$. We also ignore the high-dimension ETC interactions. We will investigate the accuracy of this approximation in a future study of all of these effects.}
\begin{eqnarray}
\beta(\alpha)&=&-\beta_0\frac{g_{\mathrm{TC}}^3}{(4\pi)^2}-\beta_1\frac{g_{\mathrm{TC}}^5}{(4\pi)^4}\hspace*{2cm}\alpha\equiv \frac{g_{\mathrm{TC}}^2}{4\pi}\label{betaF}\;.
\end{eqnarray}
In this case, the two coefficients $\beta_0$ and $\beta_1$\footnote{Here we apply the convention of Ref.\cite{ConformalWindow}.} are
\begin{eqnarray}
2N\beta_0&=&\frac{11}{3}C_2(SU(N)_{\mathrm{TC}})-\frac{4}{3}[T(R_1)+T(R_2)+T(R_3)]\label{beta0}\\
 (2N)^2\beta_1&=&\frac{34}{3}C_2^2(SU(N)_{\mathrm{TC}})-{\displaystyle\sum_{i=1}^3}[\frac{20}{3}C_2(SU(N)_{\mathrm{TC}})T(R_i)+4C_2(R_i)T(R_i)]\;.\label{beta1}
 \end{eqnarray}
The representations of the three sets of techniquarks mentioned above are labeled $R_1$, $R_2$ and $R_3$. Their corresponding parameters are given in Table II.

\begin{table}[h]
\small{{\bf TABLE II}.~The representation parameters of this model. $d(R)$ is the dimension of the representation, and $d(SU(N)_{\mathrm{TC}})$
 is the number of group generators. $C_2(R_i)$ and $C_2(SU(N)_{\mathrm{TC}})$  are the quadratic Casimir operators of the representation
 $R_i$ and the adjoint representation, respectively. $N_f$ is the number of techniquarks in the same representation, $N_fC_2(R)d(R)=T(R)d(G)$\\~\vspace*{-0.2cm}~}

\begin{tabular}{|c|cccccc|}\hline
 $i$ & $d(R_i)$ & $C_2(R_i)$ & $C_2(SU(N)_{\mathrm{TC}})$ & $T(R_i)$ & $d(SU(N)_{\mathrm{TC}})$ & $N_f$\\
 \hline
 1 & $N$ & $N^2-1$ & $2N^2$ & $N_fN$ & $N^2-1$ & 12\\
 2 & $N$ & $N^2-1$ & $2N^2$ & $N_fN$ & $N^2-1$ & 6\\
 3 & $N(N-1)/2$ & $2(N+1)(N-2)$ & $2N^2$ & $N_fN(N-2)$ & $N^2-1$ & 1\\
 \hline
\end{tabular}
 \end{table}

\begin{figure}[t]
\caption{Phases of Lane's alternative TC2 model with nontrivial TC fermion condensation and walking. The blue solid line represents the infrared fixed point $\alpha_*$. The red dashed line denotes the critical coupling of the first and second techniquark sets(fundamental representation of $SU(N)_{\mathrm{TC}})$). The black dashed-dotted line denotes the critical coupling of the third techniquark set(antisymmetric representation of $SU(N)_{\mathrm{TC}})$). The magenta dotted line shows the value $N=32/9$ from $\beta_0=0$.} \label{fig-phase}
\hspace*{-4.5cm}\vspace*{-0.3cm}\begin{minipage}[t]{8cm}
    \includegraphics[scale=0.8]{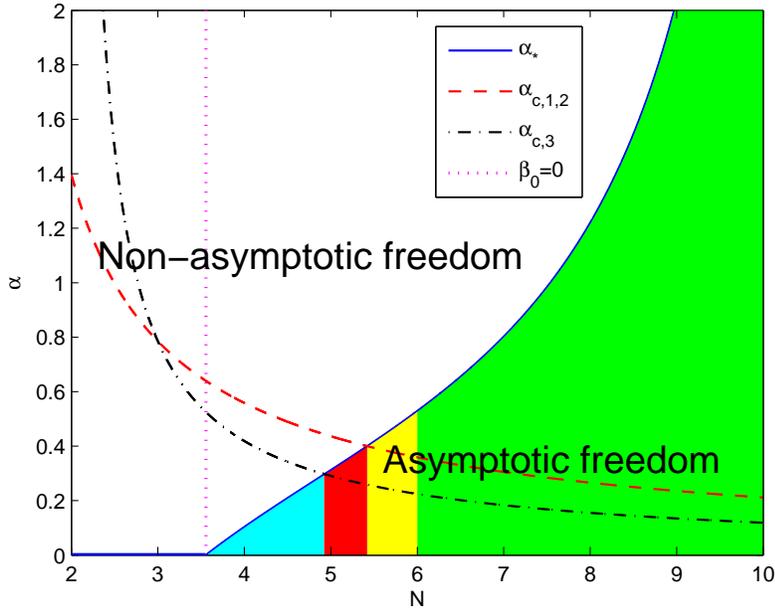}
\end{minipage}
\end{figure}

The reason that we only use the two-loop $\beta$ function is that the three-loop term of the $\beta$ function is scheme dependent. Usually, it is only used for error estimates. The behavior of the TC coupling, $\alpha$, is guided by the renormalization group equation $\mu\frac{\partial\alpha}{\partial\mu}=\beta$. From the equation, we know that $\beta_0>0$ corresponds to the case in which the TC interaction allows asymptotic freedom. However, $\beta_0<0$ corresponds to the loss of asymptotic freedom, or non-asymptotic freedom. From (\ref{beta0})
and Table II, we find that the critical value dividing asymptotic freedom and non-asymptotic freedom is determined by $\beta_0=0$ and leads $N=32/9$. If further ($\beta_0>0$ and $\beta_1<0$), TC interaction creates a Banks-Zaks infrared fixed point $\alpha_*=-\frac{4\pi\beta_0}{\beta_1}$ \cite{BZ}, which corresponds to the zero of the $\beta$ function. In the more general case, an infrared fixed point may not exist , which often happens in the situation in which the number of fermions is small. This is the case for QCD. In this model, because there are already too many technifermions, we have checked that the infrared fixed point always exists. The existence of an infrared fixed point requires that the coupling remains nearly constant over a given range of infrared energy scales, i.e., it walks. This is the modern realization of the walking mechanism. When an infrared fixed point exists, the two-loop $\beta$ function dictates the following energy scale dependence of the TC coupling:
 \begin{eqnarray}
 \frac{1}{\alpha(x)}=\frac{\beta_0}{2\pi}\ln x+\frac{1}{\alpha_*}\ln\frac{\alpha(x)}{\alpha_*-\alpha(x)}\hspace*{3cm}x=\frac{q^2}{\Lambda^2_w}\;.\label{alphaBehavior}
 \end{eqnarray}
 Where the parameter $\Lambda_w$ is roughly the length of the interval of constant coupling in the infrared region. At this scale, the coupling constant completes the walk and begins a fast run in which it exhibits typical asymptotic freedom behavior. In Section IV, we show that in the ideal walking situation, $\Lambda_w$ can be interpreted as the ETC scale. It is often referred to as $\Lambda_{\mathrm{ETC}}$ in the literature\cite{Yamawaki}. Moreover, in the standard running situation, $\Lambda_w$ can be treated as the TC scale (or $\Lambda_{\mathrm{TC}}$). Realistically, in our model, the system is somewhere between the  cases of running and ideal walking, which suggests that
 $\Lambda_{\mathrm{TC}}<\Lambda_w<\Lambda_{\mathrm{ETC}}$. This change from $\Lambda_{\mathrm{ETC}}$ to $\Lambda_w$ also reflects the fact that $\alpha(x)$ in the presence of some walking effects does not depend on the value of $\Lambda_{\mathrm{ETC}}$ too much. However, in the ideal walking theory they are very much correlated. Furthermore, the existence of both asymptotic freedom and an infrared fixed point will divide
the theory into two different phases. One phase is the asymptotic freedom phase in which $\alpha\leq\alpha_*$. In this case, the coupling $\alpha$ increases from zero to $\alpha_*$ monotonically while the energy scale decreases from the ultraviolet region to the infrared region. The other phase is the non-asymptotic freedom phase, where $\alpha\geq\alpha_*$. In this case, the coupling $\alpha$ decreases from infinity to $\alpha_*$ monotonically while the energy scale decreases from the ultraviolet region to the infrared region. Furthermore, the ladder approximation
Schwinger-Dyson equation (SDE) for techniquark self-energy predicts a critical coupling:
\begin{eqnarray}
\alpha_c=\frac{2\pi N}{3C_2(R)}
\end{eqnarray}
for techniquarks that belong to the techni-gauge group representation, $R$. While the infrared fixed point $\alpha_*$  exceeds its critical coupling $\alpha_c$, spontaneous chiral symmetry breaking occurs, and the SDE automatically develops nonzero techniquark self-energies and condensates. However, when $\alpha_*$ is less than $\alpha_c$, there is no spontaneous chiral symmetry breaking, and the techniquark self-energy vanishes. Later, we will see that to ensure the correctness of our $\beta$ function, the nonzero values of the techniquark self-energy and condensate must be small enough compare to $\Lambda_w$. This dictates that $\alpha_*$ can only be larger than $\alpha_c$ by a small amount. In practice, $\alpha_*$ may not be so close in value to $\alpha_c$, this will cause inaccuracy in our computations. We will estimate this error in later calculations. For the cases discussed above for different values of TC coupling and different choices of $N$, our model may exhibit different behaviors and then form different phases. We present\footnote{Because $N_f$ is fixed in the model, we depict the phase diagram in terms of $N$ and $\alpha$, instead of $N$ and $N_f$, which is more commonly done in the literature. Comparing our Fig.\ref{fig-phase} to the phase diagram depicted by Fig.1 in Ref.\cite{ConformalWindow}, our phase diagram corresponds to a horizontal line with a fixed $N_f$ in their diagram. Their phase diagram only provides information about $N_f$ and $N$. Our phase diagram does not provide information about $N_f$ , but does provide more information about the running coupling constant.} a phase diagram of our model in Fig.\ref{fig-phase}.

From Fig.\ref{fig-phase}, we can see that the blue line (infrared fixed point) divides the phase space into two parts: the region above the blue line represents the non-asymptotic freedom phase and that below the blue line represents the asymptotic freedom phase.

In the asymptotic freedom phase, $\alpha$ runs from  $\alpha_*$ (blue line) to zero, as the energy scale increases. The blue line crosses the red dashed line (critical coupling of the first
and second techniquark sets) and the black dashed-dotted line (critical coupling of the third
 techniquark set) at two points, which divide the blue line into three segments. The trapezoids (and triangle) under these segments form the three sub-regions of the asymptotic freedom phase. From left to right, the blue region is the conformal region, where $\alpha$ is always below its critical value and no techniquark condensation forms. Therefore, there is no spontaneous chiral symmetry breaking. The second red region is the intermediate mixture region, where $\alpha$ is always below the critical value $\alpha_{c,1}=\alpha_{c,2}$, but will cross $\alpha_{c,3}$ as the energy scale decreases. This means the third set of techniquarks forms condensates, but the first and second sets do not. The yellow and green regions are the ones that we mainly focus on in this paper. In these regions, $\alpha$ will cross all its critical values as the energy scale decreases. Thus, all techniquarks have nonzero self-energies and condensates. Therefore, this is the model required for spontaneous chiral symmetry breaking.

\begin{figure}[t]
\caption{Energy scale dependence of the TC coupling, $\alpha$, determined using (\ref{alphaBehavior}). } \label{fig-alpha}
\hspace*{-4.5cm}\vspace*{-0.3cm}\begin{minipage}[t]{8cm}
    \includegraphics[scale=0.8]{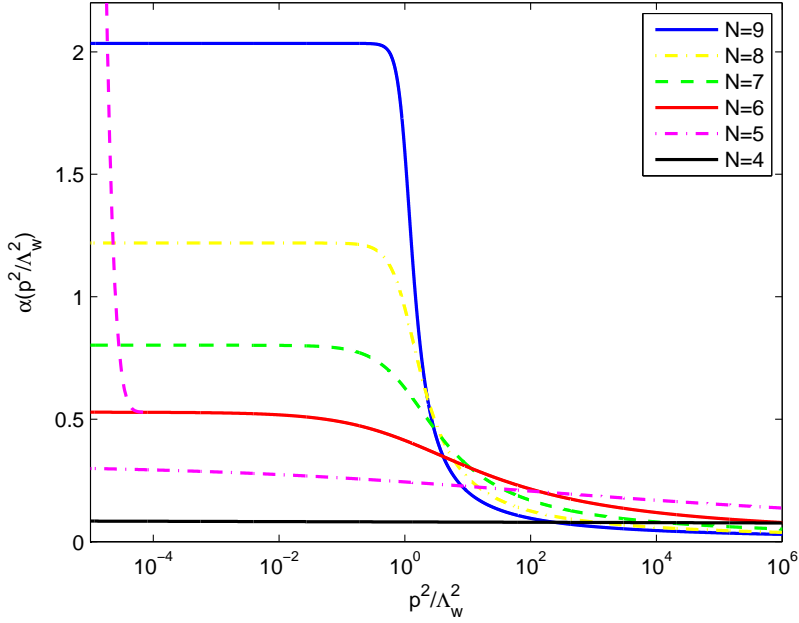}
\end{minipage}
\end{figure}

In the yellow region, the unique TC coupling in the infrared energy region approaches that of the infrared fixed point, critical values $\alpha_c$ of the first and second techniquark sets (within a magnitude of 0.2 ), and that of the third techniquark set (within a magnitude of 0.4 ), as the energy scale decreases. This causes a near conformal behavior in which the value of the techniquark self-energy is very small (corresponding to a tiny mass). For at least two reasons, this region is the most important to the investigation of the walking effect. First, the lower the techniquark self-energy, the more accurate and reliable our estimate of the $\beta$ function over the energy region will be. This is because we have used the $\overline{\mathrm{MS}}$ scheme, which assumes massless techniquarks, to obtain the coefficients of the $\beta$ function in (\ref{beta0}) and  (\ref{beta1}). Second, if a techniquark has a significant mass, it will decouple and not contribute to the $\beta$ function in the low energy region. Therefore, in the extreme infrared region, because of spontaneous chiral symmetry breaking, we cannot treat techniquarks as massless. Therefore, we need to ignore techniquark contributions if they have mass. The coupling without these techniquark contributions will run (rather than walk) to a very large value and will not reach its original infrared fixed point. We show this special running behavior in the infrared energy region for $N=6$ using a dashed magenta line near the vertical axis in Fig.\ref{fig-alpha}. A techniquark self-energy on the order of  $F_{\mathrm{TC}}$ leads to an infrared interval of the same
order size, which is small in comparison to the typical scale for $\Lambda_w$. The
 smaller the $F_{\mathrm{TC}}$ is,  the more accurately (\ref{alphaBehavior}) describes the coupling walking behavior. Therefore, we expect that replacing the running behavior in this region with an infrared fixed point will only cause errors of order $F_{\mathrm{TC}}/\Lambda_w$ in the solution of the SDE for the techniquark self-energy. In this model, because our techniquarks belong to different representations of the TC group, which leads to different critical couplings, there is not a unique point where the $\alpha_*$ is equal to all the critical coupling values. Usually this is a necessary component of modern walking theory.

Furthermore, the minimum integer $N$ closest to the conformal region is $N=6$, but the value $N=4$ was chosen in Lane's original paper\cite{Lane96} and does not satisfy the walking requirements of this study. Although we do not have an unique $\alpha_*$ that is equal to all the critical coupling values and $N=6$ is perhaps too far from the conformal region, our numerical results given in section IV show that walking effects are present. Therefore, we do achieve the situation where the infrared fixed point is not enough but sufficiently close to the critical coupling. In fact, even if we found a unique infrared fixed point $\alpha_*$ meets all the critical couplings and an integer $N$ very near the conformal region, the walking results would not be significantly more reliable. This is because of the large number of assumptions made in our calculations. These assumptions include: ignoring higher-order loops (error of $1/16\pi^2$), SM particles of mass $m$ (error of $m^2/F^2_{\mathrm{TC}}$),  and gauge fields such as coloron and $Z'$ (error of $F^2_{\mathrm{TC}}/M^2_{\mathrm{coloron}}$ and $F^2_{\mathrm{TC}}/M^2_{Z'}$ in the $\beta$ function). The precision in the critical value is now only at the two-loop level. As we mentioned before, the ETC effects may also play a role. One known effect from the ETC interaction\cite{ETC} is that while the coupling of the ETC-induced effective four-fermion interaction exceeds its critical value, the area of the conformal window will be substantially reduced. In this sense, we must include all the above-mentioned corrections before we can quantitatively improve the precision of the present calculation of the possible walking effects of the model. In the asymptotic freedom phase, we show the scale dependence of the TC coupling according to formula (\ref{alphaBehavior}) for different values of $N$ in Fig.\ref{fig-alpha}.
From Fig.\ref{fig-alpha}, it can be seen that in the asymptotic freedom phase, the smaller the value of $N$, the flatter the curve. In other words, the smaller the slope of the curve or corresponding value of the $\beta$ , the larger the impact on the walking effect. From Fig.\ref{fig-phase}, we know that when $N\leq 5$, there is no overall spontaneous chiral symmetry breaking. Therefore, the minimum value of $N$ at which spontaneous chiral symmetry breaking occurs and results in the largest walking effect is $N=6$. Throughout this paper, we will use $N=6$ in our quantitative computations.
\section{Derivation of the EWCL from Lane's Model}

Our goal is to obtain
\begin{eqnarray}
\exp\bigg(iS_{\mathrm{EW}}[W_\mu^a,B_\mu]\bigg)
&=&\int\mathcal{D}\bar{\psi}\mathcal{D}\psi\mathcal{D}\bar{T}^{1}\mathcal{D}T^{1}\mathcal{D}\bar{T}^{2}\mathcal{D}T^{2}\mathcal{D}\bar{T}^{l}\mathcal{D}T^{l}\mathcal{D}\bar{T}^{t}\mathcal{D}T^{t}\mathcal{D}\bar{T}^{b}\mathcal{D}T^{b}
\mathcal{D}G_{\mu}^\alpha
\mathcal{D}B_\mu^A\mathcal{D}Z_\mu^\prime\nonumber\\
&&\times\exp\bigg(iS[G_{\mu}^\alpha,A_{1\mu}^A,A_{2\mu}^A,
W_\mu^a,B_{1\mu},B_{2\mu},\bar{T},T,\bar{\psi},\psi]\bigg)\bigg|_{A^A_\mu=0}\label{Pathintegral}\\
&=&\mathcal{N}[W_\mu^a,B_\mu]\int\mathcal{D}\mu(U)\exp\bigg(iS_{\mathrm{eff}}[U,W_\mu^a,B_\mu]\bigg)\label{Pathintegral1}\;,
\label{strategy-TC2}
\end{eqnarray}
where $S_{\mathrm{eff}}[U,W_\mu^a,B_\mu]\equiv\int d^4x{\displaystyle\sum_i}\mathcal{L}_i$ is the action of the EWCL. $B_\mu$ is the gauge field of $U(1)_Y$ and $Z'_\mu$ is the gauge field of $U(1)'\equiv U(1)_{Y_1}\otimes
U(1)_{Y_2}/U(1)_Y$. They are related to $B_{1\mu}$ and $B_{2\mu}$ through the mixing angle $\theta$ by
\begin{eqnarray}
\begin{pmatrix}B_{1\mu} &
B_{2\mu}\end{pmatrix}=\begin{pmatrix}Z_\mu^\prime &
B_\mu\end{pmatrix}
\begin{pmatrix}\cos\theta & -\sin\theta\\ \sin\theta &
\cos\theta\end{pmatrix}\hspace*{2cm}g_1\equiv q_1\sin\theta=q_2\cos\theta\;.\label{B1B2-BZpri}
\end{eqnarray}
In (\ref{Pathintegral}) $A^A_\mu$ is the gluon field of $SU(3)_c$ and $B^A_\mu$ is the gauge field of $SU(3)_1\otimes
SU(3)_2/SU(3)_c$. They are related to $A^A_{1\mu}$ and $A^A_{2\mu}$ through the mixing angle $\theta'$ by
\begin{eqnarray}
\begin{pmatrix}A_{1\mu}^A &
A_{2\mu}^A\end{pmatrix}=\begin{pmatrix}B_\mu^A &
A_\mu^A\end{pmatrix}
\begin{pmatrix}\cos\theta^{\prime} & -\sin\theta^{\prime}\\ \sin\theta^{\prime} &
\cos\theta^{\prime}\end{pmatrix}\hspace*{2cm}g_3\equiv
h_1\sin\theta^{\prime}=h_2\cos\theta^{\prime}\;.\label{A1A2-AB}
\end{eqnarray}
In the next section, we will use Schwinger-Dyson analysis that the $SU(N)_{\rm TC}$ interaction induces techniquark condensates
$\langle\overline{\psi_L}\psi_{R}\rangle\neq0$ and $\langle\overline{T_L}^iT^j_R\rangle\neq 0$ for $i,j=1,2$. They trigger the
extended hypercharge symmetry breaking, $U(1)_{Y_1}\otimes U(1)_{Y_2}\rightarrow U(1)_Y$, and the topcolor symmetry breaking, $SU(3)_1\otimes SU(3)_2\rightarrow SU(3)_c$, at a TeV energy scale. These processes leave a singlet heavy state $Z_\mu^\prime$ in broken $U(1)'$ and colorons $B^A_\mu$ in the broken
$SU(3)_1\otimes SU(3)_2/SU(3)_c$, respectively. Because this work is only concerned with the EWCL, we ignored the gluon field by taking $A^A_\mu=0$.

In (\ref{Pathintegral1}), $U$ is the standard electroweak Goldstone boson, which can be expressed in terms of a dimensionless unitary unimodular
$2\times2$ matrix field,  $\mathcal{D}\mu$ denotes the normalized functional integration measure on $U$. The normalization factor $\mathcal{N}[W_\mu^a,B_\mu]$ is determined through the requirement that when the TC interaction is switched off, $S_{\mathrm{eff}}[U,W_\mu^a,B_\mu]$ must vanish. This fixes it at:
\begin{eqnarray}
\mathcal{N}[W_\mu^a,B_\mu]
&=&\int\mathcal{D}\bar{\psi}\mathcal{D}\psi\mathcal{D}\bar{T}^{1}\mathcal{D}T^{1}\mathcal{D}\bar{T}^{2}\mathcal{D}T^{2}\mathcal{D}\bar{T}^{l}\mathcal{D}T^{l}\mathcal{D}\bar{T}^{t}\mathcal{D}T^{t}\mathcal{D}\bar{T}^{b}\mathcal{D}T^{b}
\mathcal{D}G_{\mu}^\alpha
\mathcal{D}B_\mu^A\mathcal{D}Z_\mu^\prime\nonumber\\
&&\times\exp\bigg(iS[G_{\mu}^\alpha,A_{1\mu}^A,A_{2\mu}^A,
W_\mu^a,B_{1\mu},B_{2\mu},\bar{T},T,\bar{\psi},\psi]\bigg)\bigg|_{A^A_\mu=0, \mbox{\tiny ignore TC interation}}\;.~~~
\end{eqnarray}
In Ref.\cite{EWCL}, the EWCL was constructed with building blocks which are $SU(2)_L$ covariant and $U(1)_Y$ invariant as $T\equiv U\tau^3U^\dag$, $V_\mu\equiv(D_\mu U)U^\dag$, $g_1B_{\mu\nu}$, $g_2W_{\mu\nu}\equiv g_2\frac{\tau^a}{2}W_{\mu\nu}^a$. Where $B_{\mu\nu}$ and $W_{\mu\nu}$ are the field strengths of the $U(1)_Y$ and $SU(2)_L$ gauge fields, respectively. Alternatively, in Ref.\cite{HongHao08}, we reformulated the EWCL equivalently using $SU(2)_L$ invariant and $U(1)_Y$ covariant building blocks as
$\tau^3$, $X_\mu\equiv U^\dag(D_\mu U)$, $g_1B_{\mu\nu}$, $\overline{W}_{\mu\nu}\equiv U^\dag g_2W_{\mu\nu}U$. In which, $\tau^3$ and $g_1B_{\mu\nu}$ are both $SU(2)_L$ and $U(1)_Y$ invariant, but $X_\mu$ and $\overline{W}_{\mu\nu}$ are bilinearly $U(1)_Y$ covariant. The second formulation was used throughout this paper. In Table III, we detail the  relationship between the two formalisms.

\begin{table}[h]
\small{{\bf TABLE III}.~ Symmetry breaking sector of the EWCL $S_{\mathrm{eff}}[U,W_\mu^a,B_\mu]=\int d^4x{\displaystyle\sum_i}\mathcal{L}_i$\\~\vspace*{-0.2cm}~}
\label{table-ewcl}
\begin{tabular}{*{3}{|l}|}
\hline
& Formulation in Ref.\cite{EWCL} & Formulation in Ref.\cite{HongHao08}\\
\hline \hline
${\cal L}^{(2)}$ & $\frac{1}{4}f^2{\rm tr}[(D_\mu U^\dag) (D^\mu
U)]=-\frac{1}{4}f^2{\rm tr}(V_\mu V^\mu)$ &
$-\frac{1}{4}f^2{\rm tr}(X_\mu X^\mu)$\\
\hline
${\cal L}^{(2)\prime}$ & $\frac{1}{4}\beta_1f^2[{\rm tr}(TV_\mu)]^2$
& $\frac{1}{4}\beta_1f^2[{\rm tr}(\tau^3
X_\mu)]^2$\\
\hline
${\cal L}_1$ & $\frac{1}{2}\alpha_1g_2g_1 B_{\mu\nu}{\rm
tr}(TW^{\mu\nu})$ & $\frac{1}{2}\alpha_1g_1 B_{\mu\nu}{\rm
tr}(\tau^3\overline{W}^{\mu\nu})$\\
\hline
${\cal L}_2$ & $\frac{1}{2}i\alpha_2g_1 B_{\mu\nu}{\rm
tr}(T[V^\mu,V^\nu])$ & $i\alpha_2g_1 B_{\mu\nu}{\rm
tr}(\tau^3X^\mu X^\nu)$\\
\hline
${\cal L}_3$ & $i\alpha_3g_2{\rm tr}(W_{\mu\nu}[V^\mu,V^\nu])$ &
$2i\alpha_3{\rm
tr}(\overline{W}_{\mu\nu}X^\mu X^\nu)$\\
\hline
${\cal L}_4$ & $\alpha_4[{\rm tr}(V_\mu V_\nu)]^2$ &
$\alpha_4[{\rm tr}(X_\mu X_\nu)]^2$\\
\hline
${\cal L}_5$ & $\alpha_5[{\rm tr}(V_\mu V^\mu)]^2$ &
$\alpha_5[{\rm tr}(X_\mu X^\mu)]^2$\\
\hline
${\cal L}_6$ & $\alpha_6{\rm tr}(V_\mu V_\nu){\rm tr}(TV^\mu){\rm
tr}(TV^\nu)$ & $\alpha_6{\rm tr}(X_\mu X_\nu){\rm tr}(\tau^3
X^\mu){\rm
tr}(\tau^3 X^\nu)$\\
\hline
${\cal L}_7$ & $\alpha_7{\rm tr}(V_\mu V^\mu){\rm tr}(TV_\nu){\rm
tr}(TV^\nu)$ & $\alpha_7{\rm tr}(X_\mu X^\mu){\rm
tr}(\tau^3X_\nu){\rm
tr}(\tau^3X^\nu)$\\
\hline
${\cal L}_8$ & $\frac{1}{4}\alpha_8g_2^2[{\rm tr}(TW_{\mu\nu})]^2$ &
$\frac{1}{4}\alpha_8[{\rm
tr}(\tau^3\overline{W}_{\mu\nu})]^2$\\
\hline
${\cal L}_9$ & $\frac{1}{2}i\alpha_9g_2{\rm tr}(TW_{\mu\nu}){\rm
tr}(T[V^\mu,V^\nu])$ & $i\alpha_9{\rm
tr}(\tau^3\overline{W}_{\mu\nu}){\rm tr}(\tau^3X^\mu X^\nu)$\\
\hline
${\cal L}_{10}$ & $\frac{1}{2}\alpha_{10}[{\rm tr}(TV_\mu){\rm
tr}(TV_\nu)]^2$ & $\frac{1}{2}\alpha_{10}[{\rm tr}(\tau^3X_\mu){\rm
tr}(\tau^3X_\nu)]^2$\\
\hline
${\cal L}_{11}$ & $\alpha_{11}g_2\epsilon^{\mu\nu\rho\lambda}{\rm
tr}(TV_\mu){\rm tr}(V_\nu W_{\rho\lambda})$ &
$\alpha_{11}\epsilon^{\mu\nu\rho\lambda}{\rm
tr}(\tau^3X_\mu){\rm tr}(X_\nu \overline{W}_{\rho\lambda})$\\
\hline
${\cal L}_{12}$ & $\alpha_{12}g_2{\rm tr}(TV_\mu){\rm tr}(V_\nu
W^{\mu\nu})$ & $\alpha_{12}{\rm tr}(\tau^3X_\mu){\rm tr}(X_\nu
\overline{W}^{\mu\nu})$\\
\hline
${\cal L}_{13}$ &
$\alpha_{13}g_2g_1\epsilon^{\mu\nu\rho\sigma}B_{\mu\nu}{\rm
tr}(TW_{\rho\sigma})$ &
$\alpha_{13}\epsilon^{\mu\nu\rho\sigma}g_1B_{\mu\nu}{\rm
tr}(\tau^3\overline{W}_{\rho\sigma})$\\
\hline
${\cal L}_{14}$ & $\alpha_{14}g_2^2\epsilon^{\mu\nu\rho\sigma}{\rm
tr}(TW_{\mu\nu}){\rm tr}(TW_{\rho\sigma})$ &
$\alpha_{14}\epsilon^{\mu\nu\rho\sigma}{\rm
tr}(\tau^3\overline{W}_{\mu\nu}){\rm
tr}(\tau^3\overline{W}_{\rho\sigma})$\\
\hline
\end{tabular}
\end{table}

From (\ref{Pathintegral}) and (\ref{Pathintegral1}), it can be seen that to obtain the EWCL, we must integrate in the electroweak Goldstone boson field, $U$. We also need to integrate out the series of fields which include the three sets of techniquarks, $\psi$, $T^1$, $T^2$,  $T^l$, $T^t$, $T^b$ and the technigluon $G_\mu^\alpha$, and the colorons $B^A_\mu$ and $Z_\mu'$. In the following subsections, we divide this work into five steps.
\subsection{Integrating in the electroweak Goldstone boson field $U$}

We introduce a local $2\times2$ operator
\begin{equation}
  O(x)\equiv \mathrm{tr}[T^1_L\bar{T}^1_R+T^2_L\bar{T}^2_R+T^l_L\bar{T}^l_R+T^t_L\bar{T}^t_R+T^b_L\bar{T}^b_R](x)
\end{equation}
In this case, $\mathrm{tr}$ are the traces with respect to the Lorentz, $SU(N)_{\mathrm{TC}}$, $SU(3)_1$ and $SU(3)_2$ indices. The transformation of $O(x)$ under $SU(2)_L\times U(1)_Y$ is
\begin{equation}
O(x)\rightarrow V_L(x)O(x)V_R^\dag(x)\hspace*{2cm} V_L(x)=e^{i\frac{\tau^a}{2}\theta^a(x)}\qquad V_R(x)=e^{-i\frac{\tau^3}{2}\theta^0(x)}\;.
\end{equation}
Then we decompose $O(x)$ as
\begin{equation}
  O(x)=\xi^\dag_L(x)\sigma(x)\xi_R(x)
\end{equation}
Where $\sigma(x)$ which is represented using a Hermitian matrix, describes the modular degree of freedom; and $\xi_L(x)$ and $\xi_R(x)$, which are represented using unitary matrices, describe the phase degrees of freedom of $SU(2)_L$ and $U(1)_Y$ respectively. Their transformations under $SU(2)_L\otimes U(1)_Y$ are
\begin{eqnarray}
  \sigma(x)\rightarrow h(x)\sigma(x)h^\dag(x)\hspace*{1cm}
  \xi_L(x)\rightarrow h(x)\xi_L(x)V^\dag_L(x)\hspace*{1cm}
  \xi_R(x)\rightarrow h(x)\xi_R(x)V^\dag_R(x)~~~~~~
\end{eqnarray}
where
\begin{equation}
  h(x)=e^{i\theta_h(x)\frac{\tau^3}{2}}
\end{equation}
belongs to an induced hidden local $U(1)$ symmetry group. Next, we define a new field
\begin{equation}
  U(x)\equiv \xi_L^\dag(x)\xi_R(x)\;,
\end{equation}
which is the nonlinear realization of the Goldstone boson field in the EWCL. Subtracting the $\sigma(x)$ field, we find that the present decomposition results in a constraint $\xi_L(x)O(x)\xi_R^\dag(x)-\xi_R(x)O^\dag(x)\xi^\dag_L(x)=0$ and its functional expression is
\begin{equation}
  \int \mathcal{D}_\mu(U)\mathcal{F}[O]\delta(\xi_LO\xi^\dag_R-\xi_RO^\dag\xi^\dag_L)=\mathrm{const}\;,
\end{equation}
where $\mathcal{D}_\mu(U)$ is an effective invariant integration measure; and $\mathcal{F}[O]$ only depends on $O$ and is invariant under $SU(2)_L\otimes U(1)_Y$ transformations. This causes the value of the integrated quantity to be a constant. Inserting the above identity into (\ref{Pathintegral}), we have
\begin{eqnarray}
e^{iS_{\mathrm{EW}}[W_\mu^a,B_\mu]}
&=&\int\mathcal{D}\bar{\psi}\mathcal{D}\psi\mathcal{D}\bar{T}^{1}\mathcal{D}T^{1}\mathcal{D}\bar{T}^{2}\mathcal{D}T^{2}\mathcal{D}\bar{T}^{l}\mathcal{D}T^{l}\mathcal{D}\bar{T}^{t}\mathcal{D}T^{t}\mathcal{D}\bar{T}^{b}\mathcal{D}T^{b}
\mathcal{D}G_{\mu}^\alpha
\mathcal{D}B_\mu^A\mathcal{D}Z_\mu^\prime\nonumber\\
&&\times\int \mathcal{D}_\mu(U)\mathcal{F}[O]\delta(\xi_LO\xi^\dag_R-\xi_RO^\dag\xi^\dag_L)
e^{iS[G_{\mu}^\alpha,A_{1\mu}^A,A_{2\mu}^A,W_\mu^a,B_{1\mu},B_{2\mu},\bar{T},T,\bar{\psi},\psi]}\bigg|_{A^A_\mu=0}\!.~~~
\end{eqnarray}
Using a special $SU(2)_L\otimes U(1)_Y$ rotation for $V_L(x)=\xi_L(x)$ and $V_R(x)=\xi_R(x)$ and labeling the fields after rotation with the subscript, $_\xi$, the above path integral becomes:
\begin{eqnarray}
e^{iS_{\mathrm{EW}}[W_\mu^a,B_\mu]}
&=&\int\mathcal{D}\bar{\psi}\mathcal{D}\psi\mathcal{D}\bar{T}^1_\xi\mathcal{D}T^1_\xi\mathcal{D}\bar{T}^2_\xi\mathcal{D}T^2_\xi
\mathcal{D}\bar{T}^l_\xi\mathcal{D}T^l_\xi\mathcal{D}\bar{T}^t_\xi\mathcal{D}T^t_\xi\mathcal{D}\bar{T}^b_\xi\mathcal{D}T^b_\xi
\mathcal{D}G_{\mu}^\alpha\mathcal{D}B_\mu^A\mathcal{D}Z_\mu^\prime\nonumber\\
&&\times\int \mathcal{D}_\mu(U)\mathcal{F}[O_\xi]\delta(O_\xi-O^\dag_\xi)
e^{iS[G_{\mu}^\alpha,A_{1\mu}^A,A_{2\mu}^A,W_{\xi,\mu}^a,B_{1\xi,\mu},B_{2\xi,\mu},\bar{T}_\xi,T_\xi,\bar{\psi},\psi]}\bigg|_{A^A_\mu=0}\!.~~~
\label{IntegralAfterRotation}
\end{eqnarray}
where we have used the result that the functional integration measure, $\mathcal{F}[O]$ and the action on the exponential of the integrand are invariant under $SU(2)_L\otimes U(1)_Y$ transformations. From Table I, it can be seen that:
\begin{eqnarray}
  T^1_{\xi L}=e^{-i(u_1+u_2)\theta_0}P_L\xi_L T^1_L\hspace*{2cm} T^1_{\xi R}=e^{-i(v_1+v_2)\theta_0}P_R\xi_R T^1_R\nonumber\\
  T^2_{\xi L}=e^{-i(v_1+v_2)\theta_0}P_L\xi_L T^2_L\hspace*{2cm} T^2_{\xi R}=e^{-i(u_1+u_2)\theta_0}P_R\xi_R T^2_R\nonumber\\
  T^l_{\xi L}=e^{-i(x_1+x_2)\theta_0}P_L\xi_L T^l_L\hspace*{2cm} T^l_{\xi R}=e^{-i(x'_1+x'_2)\theta_0}P_R\xi_R T^l_R\\
  T^t_{\xi L}=e^{-i(y_1+y_2)\theta_0}P_L\xi_L T^t_L\hspace*{2cm} T^t_{\xi R}=e^{-i(y'_1+y'_2)\theta_0}P_R\xi_R T^t_R\nonumber\\
  T^b_{\xi L}=e^{-i(z_1+z_2)\theta_0}P_L\xi_L T^b_L\hspace*{2cm} T^b_{\xi R}=e^{-i(z'_1+z'_2)\theta_0}P_R\xi_R T^b_R\;,\nonumber
\end{eqnarray}
Furthermore,
\begin{eqnarray}
&&\hspace*{-1cm}g_2\frac{\tau^a}{2}W^a_{\xi,\mu}=\xi_L[g_2\frac{\tau^a}{2}W^a_\mu-i\partial_\mu]\xi_L^\dag\\
&&\hspace*{-1cm}g_1\frac{\tau^3}{2}B_{\xi,\mu}=\xi_R[g_1\frac{\tau^3}{2}B_\mu-i\partial_\mu]\xi_R^\dag\hspace*{1cm}
\begin{pmatrix}B_{1\xi,\mu} &
B_{2\xi,\mu}\end{pmatrix}=\begin{pmatrix}Z_\mu^\prime &
B_{\xi,\mu}\end{pmatrix}
\begin{pmatrix}\cos\theta & -\sin\theta\\ \sin\theta &
\cos\theta\end{pmatrix}\;.~~~~
\end{eqnarray}
Note the fields without the subscript $_\xi$ in (\ref{IntegralAfterRotation}) are the fields that are invariant under $SU(2)_L\otimes U(1)_Y$ rotation.
\subsection{Integrating out the technigluons}

As a second step,we integrate out the technigluon in (\ref{IntegralAfterRotation}) using:
\begin{eqnarray}
\int\mathcal{D}G_{\mu}^{\alpha}e^{iS[G_{\mu}^\alpha,A_{1\mu}^A,A_{2\mu}^A,
W_{\xi,\mu}^a,B_{1\xi,\mu},B_{2\xi,\mu},\bar{T}_\xi,T_\xi,\bar{\psi},\psi]}
=e^{iS_{\mathrm{TC}}[\bar{T}_\xi,T_\xi,\bar{\psi},\psi]
+iS_{\mathrm{TC1}}[A_{1\mu}^A,A_{2\mu}^A,W_{\xi,\mu}^a,B_{1\xi,\mu},B_{2\xi,\mu},\bar{T}_\xi,T_\xi,\bar{\psi},\psi]}\;,~~~\label{Gintout}
\end{eqnarray}
where we choose
\begin{eqnarray}
&&\hspace*{-0.5cm}e^{iS_{\mathrm{TC}}[\bar{T}_\xi,T_\xi,\bar{\psi},\psi]}=\int\mathcal{D}G_{\mu}^\alpha~e^{i\int
d^{4}x(-\frac{1}{4}G_{\mu\nu}^{\alpha}G^{\alpha,\mu\nu}-g_{TC}G_{\mu}^{\alpha}J^{\mu\alpha})}\label{STCdef}\\
&&\hspace*{-0.5cm}
S_{\mathrm{TC1}}[A_{1\mu}^A,A_{2\mu}^A,W_{\xi,\mu}^a,B_{1\xi,\mu},B_{2\xi,\mu},\bar{T}_\xi,T_\xi,\bar{\psi},\psi]=
S[G_{\mu}^\alpha,A_{1\mu}^A,A_{2\mu}^A,W_{\xi,\mu}^a,B_{1\xi,\mu},B_{2\xi,\mu},\bar{T}_\xi,T_\xi,\bar{\psi},\psi]\bigg|_{G_{\mu}^\alpha=0}~~
\end{eqnarray}
and
\begin{eqnarray}
J^{\mu\alpha}&=&\bar{\psi}\tilde{t}^{\alpha}\gamma^{\mu}\psi+\tilde{J}^{\mu\alpha}\label{Jdef}\\
\tilde{J}^{\mu\alpha}&=&\bar{T}^1_{\xi}t^{\alpha}\gamma^{\mu}T^1_\xi
+\bar{T}^2_{\xi}t^{\alpha}\gamma^{\mu}T^2_\xi+\bar{T}^l_{\xi}t^{\alpha}\gamma^{\mu}T^l_\xi
+\bar{T}^t_{\xi}t^{\alpha}\gamma^{\mu}T^t_\xi+\bar{T}^b_{\xi}t^{\alpha}\gamma^{\mu}T^b_\xi\;.\label{tildeJdef}
\end{eqnarray}
Integrating out the technigluon fields in (\ref{STCdef}), we get
\begin{eqnarray}
iS_{\mathrm{TC}}[\bar{T}_\xi,T_\xi,\bar{\psi},\psi]=\sum_{n=2}^\infty\int d^4x_1\ldots
d^4x_n\frac{(-ig_{\mathrm{TC}})^n}{n!}G_{\mu_1\ldots\mu_n}^{\alpha_1\ldots\alpha_n}(x_1,\ldots,x_n)
J_{\alpha_1}^{\mu_1}(x_1)\ldots
J_{\alpha_n}^{\mu_n}(x_n)\;,~~~\label{action-TC1}
\end{eqnarray}
where $G_{\mu_1\ldots\mu_n}^{\alpha_1\ldots\alpha_n}(x_1,\ldots,x_n)$ is a n-point Green's function for the technigluons.

\subsection{Integrating out the techniquarks}

Combining (\ref{IntegralAfterRotation}) and (\ref{Gintout}), our starting $S_{\mathrm{EW}}[W_\mu^a,B_\mu]$, after integrating in the electroweak Goldstone boson field $U$ and integrating out the technigluons, becomes
\begin{eqnarray}
e^{iS_{\mathrm{EW}}[W_\mu^a,B_\mu]}
&=&\int\mathcal{D}\bar{\psi}\mathcal{D}\psi\mathcal{D}\bar{T}^1_\xi\mathcal{D}T^1_\xi\mathcal{D}\bar{T}^2_\xi\mathcal{D}T^2_\xi
\mathcal{D}\bar{T}^l_\xi\mathcal{D}T^l_\xi\mathcal{D}\bar{T}^t_\xi\mathcal{D}T^t_\xi\mathcal{D}\bar{T}^b_\xi\mathcal{D}T^b_\xi
\mathcal{D}B_\mu^A\mathcal{D}Z_\mu^\prime\label{IntegralAfterRotation1}\\
&&\hspace*{-1cm}\times\int \mathcal{D}_\mu(U)\mathcal{F}[O_\xi]\delta(O_\xi-O^\dag_\xi)
e^{iS_{\mathrm{TC}}[\bar{T}_\xi,T_\xi,\bar{\psi},\psi]
+iS_{\mathrm{TC1}}[A_{1\mu}^A,A_{2\mu}^A,W_{\xi,\mu}^a,B_{1\xi,\mu},B_{2\xi,\mu},\bar{T}_\xi,T_\xi,\bar{\psi},\psi]}\bigg|_{A^A_\mu=0}\!.
\nonumber
\end{eqnarray}
After some detailed derivations and approximations which can be found in Appendix \ref{InteOutTCq}, we get:
\begin{eqnarray}
e^{iS_{\mathrm{EW}}[W_\mu^a,B_\mu]}
&=&\int \mathcal{D}_\mu(U)\mathcal{F}[O_\xi]\delta(O_\xi-O^\dag_\xi)\int\mathcal{D}B_\mu^A\mathcal{D}Z_\mu^\prime
~\exp\bigg[i\int d^4x[-\frac{1}{4}(A_{1\mu\nu}^{A}A^{A,1\mu\nu}
\nonumber\\
&&+A_{2\mu\nu}^{A}A^{A,2\mu\nu}+W_{\mu\nu}^aW^{a,\mu\nu}+B_{1,\mu\nu}B^{1,\mu\nu}+B_{2,\mu\nu}B^{2,\mu\nu})]\nonumber\\
&&+\mathrm{Trln}[i\slashed{\partial}+g_1(\cot\theta\!+\tan\theta)\xi\slashed{Z}^\prime\gamma^5-\tilde{\Sigma}(\partial^2)]
+\mathrm{Tr"ln}[i\slashed{\partial}+\slashed{V}_{2\xi}\!+\slashed{A}_{2\xi}\gamma^5\!-\hat{\Sigma}(\overline{\nabla}^2)]\nonumber\\
&&+\mathrm{Tr'ln}[i\slashed{\partial}+\!\slashed{V}_{1\xi}\!+\slashed{A}_{1\xi}\gamma^5\!
-\bar{\Sigma}(\hat{\nabla}^2)\!-i\gamma_5\tau^2\bar{\Sigma}_5(\hat{\nabla}^2)]\bigg]_{A^A_\mu=0}\;,\label{SEW0}
\end{eqnarray}
The various quantities appearing in (\ref{SEW0}) are defined at the end of  Appendix \ref{InteOutTCq}. Furthermore, in Appendix \ref{DerSDE}, we have shown that the techniquark self energies $\tilde{\Sigma}$,  $\hat{\Sigma}$,  $\bar{\Sigma}$ and  $\bar{\Sigma}_5$ satisfy the following SDEs,
\begin{eqnarray}
\tilde{\Sigma}(p_E^2)&=&\frac{3(N+1)(N-2)}{4\pi^3N}\int{d^4q_E}\frac{\alpha[(p_E-q_E)^2]}{(p_E-q_E)^2}
\frac{\tilde{\Sigma}(q_E^2)}{q_E^2+\tilde{\Sigma}^2(q_E^2)}\label{SDEtildeSigma}\\
\hat{\Sigma}(p_E^2)&=&\frac{3(N^2-1)}{8\pi^3N}\int{d^4q_E}\frac{\alpha[(p_E-q_E)^2]}{(p_E-q_E)^2}
\frac{\hat{\Sigma}(q_E^2)}{q_E^2+\hat{\Sigma}^2(q_E^2)}\label{SDEhatSigma}\\
\bar{\Sigma}(p_E^2)&=&\frac{3(N^2-1)}{8\pi^3N}\int{d^4q_E}\frac{\alpha[(p_E-q_E)^2]}{(p_E-q_E)^2}
\frac{\bar{\Sigma}(q_E^2)}{q_E^2+\bar{\Sigma}^2(q_E^2)+\bar{\Sigma}_5^2(q_E^2)}\label{SDEbarSigma}\\
\bar{\Sigma}_5(p_E^2)&=&\frac{3(N^2-1)}{8\pi^3N}\int{d^4q_E}\frac{\alpha[(p_E-q_E)^2]}{(p_E-q_E)^2}
\frac{\bar{\Sigma}_5(q_E^2)}{q_E^2+\bar{\Sigma}^2(q_E^2)+\bar{\Sigma}_5^2(q_E^2)}\;,\label{SDEbarSigma5}
\end{eqnarray}
where the technigluon propagator is parameterized though the TC running coupling constant $\alpha$ as
\begin{eqnarray}
G_{\mu\nu}^{\alpha\beta}(x,y)
=\!\int\!\frac{d^4p}{(2\pi)^4}e^{-ip(x-y)}\frac{-i\delta^{\alpha\beta}}{p^2[1\!+\!\Pi(-p^2)]}\bigg(g_{\mu\nu}\!-\frac{p_\mu
p_\nu}{p^2}\bigg)\hspace*{1cm}\alpha(p_E^2)\equiv\frac{g^2_{\mathrm{TC}}}{4\pi[1\!+\!\Pi(p_E^2)]}\;.~~~ \label{alphaDef}
\end{eqnarray}
\subsection{Integrating out the colorons and the low energy expansion}

Before integrating out the coloron field, we first discuss its mass which is determined by the kinetic and mass terms.  From the exponential of the integrand in (\ref{SEW0}), it can be seen that there is already a standard coloron kinetic term from $-\frac{1}{4}(A_{1\mu\nu}^{A}A^{A,1\mu\nu}
+A_{2\mu\nu}^{A}A^{A,2\mu\nu})$. The first set of techniquarks contributes to the quantum loop corrections to the coloron kinetic and mass terms through the term $\mathrm{Tr'ln}[i\slashed{\partial}+\!\slashed{V}_{1\xi}\!+\slashed{A}_{1\xi}\gamma^5\!
-\bar{\Sigma}(\hat{\nabla}^2)\!-i\gamma_5\tau^2\bar{\Sigma}_5(\hat{\nabla}^2)]$ in (\ref{SEW0}). Through detailed computations, we find that these corrections are
\begin{eqnarray}
&&\hspace{-0.5cm}\mathrm{Tr'ln}[i\slashed{\partial}+\!\slashed{V}_{1\xi}\!+\slashed{A}_{1\xi}\gamma^5\!
-\bar{\Sigma}(\hat{\nabla}^2)\!-i\gamma_5\tau^2\bar{\Sigma}_5(\hat{\nabla}^2)]\bigg|_{\mbox{\tiny coloron kinetic and mass terms}}\nonumber\\
&&\hspace{-0.5cm}=\frac{i}{4}\int d^4x\bigg[
Cg_3^2(\tan\theta'\!+\!\tan\theta')^2B^A_\mu B_A^\mu
  -(\partial^\mu B^A_\nu-\partial^\nu B^A_\mu)^2
  [\mathcal{K}g_3^2(\cot^2\theta'\!+\!\tan^2\theta')\nonumber\\
  &&+\hat{\mathcal{K}}_{13}^{\Sigma\neq0}g_3^2(\tan\theta'\!-\!\cot\theta')^2
  +\frac{1}{2}\hat{E}g_3^2(\tan\theta'+\cot\theta')^2]\bigg]\;,\label{coloronAction}
\end{eqnarray}
In this case, the coefficients are given at the beginning of Appendix \ref{LEexp}. Combining the standard coloron kinetic term in (\ref{SEW0}) and the techniquark quantum loop correction given by (\ref{coloronAction}), we find the formula for the coloron mass to be:
\begin{eqnarray}
M_{\mathrm{coloron}}^2=\frac{C}{\hat{E}+2(\mathcal{K}+\hat{\mathcal{K}}_{13}^{\Sigma\neq0})
+(2/g_3^2-8\hat{\mathcal{K}}_{13}^{\Sigma\neq0})/(\cot\theta'+\tan\theta')^2}
\;.\label{coloronmass}
\end{eqnarray}
In Appendix \ref{LEexp}, we integrate out the coloron fields and perform the low energy expansion. Finally we obtain,
\begin{eqnarray}
e^{iS_{\mathrm{EW}}[W_\mu^a,B_\mu]}
&=&e^{i\int d^4x[-\frac{1}{4}W_{\mu\nu}^aW^{a,\mu\nu}-\frac{1}{4}B_{\mu\nu}B^{\mu\nu}]}\int \mathcal{D}_\mu(U)\mathcal{F}[O_\xi]\delta(O_\xi-O^\dag_\xi)\int\mathcal{D}Z_\mu^{\prime}~e^{iS_0+iS_{Z'}}\;.~~~~~\label{SEW2}
\end{eqnarray}
Where detailed expressions of $S_0$ and $S_{Z'}$ are given in (\ref{S0def}) and (\ref{SZ'}) respectively in Appendix \ref{LEexp}.
\subsection{Integrating out $Z'$}

We denote the resulting action after the integration over $Z'$ as
\begin{eqnarray}
\int\mathcal{D}Z'_\mu~e^{iS_{Z'}}=e^{i\bar{S}_{Z'}}\;.\label{barSZ'def}
\end{eqnarray}
We can use the loop expansion to calculate the above integral:
\begin{eqnarray}
\bar{S}_{Z'}=S_{Z'}\bigg|_{Z'=Z'_c}+\mbox{loop corrections}
\end{eqnarray}
where the classical field $Z'_c$ satisfies:
\begin{eqnarray}
\frac{\partial }{\partial
Z'_{c,\mu}(x)}\bigg[S_{Z'}+\mbox{loop corrections}\bigg]=0\;.\label{Z'Eq}
\end{eqnarray}
Using this method, we integrate out the $Z'$ field in Appendix \ref{IntOutZ'} and
simplify the result $\bar{S}_{Z'}$ given in (\ref{DeltaSZ'}) into the form of EWCL. Furthermore, combining (\ref{barSZ'def}) and (\ref{SEW2}) together, we find
 \begin{eqnarray}
e^{iS_{\mathrm{EW}}[W_\mu^a,B_\mu]}
&=&e^{i\int d^4x[-\frac{1}{4}W_{\mu\nu}^aW^{a,\mu\nu}-\frac{1}{4}B_{\mu\nu}B^{\mu\nu}]}\int \mathcal{D}_\mu(U)\mathcal{F}[O_\xi]\delta(O_\xi-O^\dag_\xi)~e^{iS_0+i\bar{S}_{Z'}}\;.~~~~~\label{SEW3}
\end{eqnarray}
Comparing this with (\ref{strategy-TC2}) and Table.III, we obtain all the EWCL LECs. Our final analytical results for the EWCL LECs (up to an order of $p^4$) are
\begin{eqnarray}
&&\hspace*{-0.5cm}f^2=5\hat{F}_0^2\hspace*{1cm}\beta_1=\frac{10a_3^2\hat{F}_0^2}{\bar{M}_{Z'}^2}
\hspace*{1cm}\alpha_1=\frac{5}{2}(1-2\beta_1)(\hat{\mathcal{K}}_{2}^{\Sigma\neq0}-\hat{\mathcal{K}}_{13}^{\Sigma\neq0})
  +\frac{\beta_1f^2}{2M_{Z'}^2}-\frac{\gamma\beta_1}{2a_3}\nonumber\\
 &&\hspace*{-0.5cm}\alpha_2=(\beta_1-\frac{1}{2})(\frac{5}{2}\hat{\mathcal{K}}_{13}^{\Sigma\neq0}-\frac{5}{8}\hat{\mathcal{K}}_{14}^{\Sigma\neq0})
   +\frac{\beta_1f^2}{2M_{Z'}^2}-\frac{\gamma\beta_1}{2a_3}
  \hspace*{1cm}\alpha_3=(\beta_1-\frac{1}{2})(\frac{5}{2}\hat{\mathcal{K}}_{13}^{\Sigma\neq0}-\frac{5}{8}\hat{\mathcal{K}}_{14}^{\Sigma\neq0})\nonumber~~~~~~\\
  &&\hspace*{-0.5cm}\alpha_4=(2\beta_1+\frac{1}{4})(\frac{5}{2}\hat{\mathcal{K}}_{13}^{\Sigma\neq0}-\frac{5}{8}\hat{\mathcal{K}}_{14}^{\Sigma\neq0})
  +(\frac{5}{16}\hat{\mathcal{K}}_{4}^{\Sigma\neq0}-\frac{5}{32}\hat{\mathcal{K}}_{14}^{\Sigma\neq0})+\frac{\beta_1f^2}{2M_{Z'}^2}\nonumber\\
  &&\hspace*{-0.5cm}\alpha_5=-\frac{5}{2}(4\beta_1+\frac{1}{4})\hat{\mathcal{K}}_{13}^{\Sigma\neq0}
   +\frac{5}{4}(3\beta_1+\frac{1}{4})\hat{\mathcal{K}}_{14}^{\Sigma\neq0}
    +\frac{5}{32}(\hat{\mathcal{K}}_{3}^{\Sigma\neq0}-\hat{\mathcal{K}}_{4}^{\Sigma\neq0})-\frac{\beta_1f^2}{2M_{Z'}^2}\nonumber\\
  &&\hspace*{-0.5cm}\alpha_6=-\frac{\beta_1f^2}{2M_{Z'}^2}
  -\frac{\beta_1^2}{4a_3^2}[-(2a_0^2+\hat{a}_0^2)\hat{\mathcal{K}}_{3}^{\Sigma\neq0}
  -(2a_0^2+\hat{a}_0^2+5a_3^2)\hat{\mathcal{K}}_{4}^{\Sigma\neq0}
  -10a_3^2\hat{\mathcal{K}}_{13}^{\Sigma\neq0}+5a_3^2\hat{\mathcal{K}}_{14}^{\Sigma\neq0}\nonumber\\
  &&\hspace{0.5cm}+2a_0^2\hat{D}_4]  -\frac{\beta_1}{2}(\frac{5}{2}\hat{\mathcal{K}}_{4}^{\Sigma\neq0}+15\hat{\mathcal{K}}_{13}^{\Sigma\neq0}
  -5\hat{\mathcal{K}}_{14}^{\Sigma\neq0})\nonumber
  \end{eqnarray}
  \begin{eqnarray}
  &&\hspace*{-0.5cm}\alpha_7=\frac{\beta_1f^2}{2M_{Z'}^2}-\frac{\beta_1^2}{4a_3^2}[
  (\frac{5}{2}a_3^2+a_0^2+\frac{1}{2}\hat{a}_0^2)\hat{\mathcal{K}}_{3}^{\Sigma\neq0}+(a_0^2+\frac{1}{2}\hat{a}_0^2-\frac{5}{2}a^2_3)\hat{\mathcal{K}}_{4}^{\Sigma\neq0}
  -10a_3^2\hat{\mathcal{K}}_{13}^{\Sigma\neq0}+5a_3^2\hat{\mathcal{K}}_{14}^{\Sigma\neq0}+a_0^2\hat{D}_3]\nonumber\\
   &&\hspace{0.5cm}-\frac{\beta_1}{2}(\frac{5}{4}\hat{\mathcal{K}}_{3}^{\Sigma\neq0}-\frac{5}{4}\hat{\mathcal{K}}_{4}^{\Sigma\neq0}
  -15\hat{\mathcal{K}}_{13}^{\Sigma\neq0}+5\hat{\mathcal{K}}_{14}^{\Sigma\neq0})\nonumber\\
    &&\hspace*{-0.5cm}\alpha_8=-\frac{\beta_1f^2}{2M_{Z'}^2}+10\beta_1(\hat{\mathcal{K}}_{2}^{\Sigma\neq0}\!
  -\hat{\mathcal{K}}_{13}^{\Sigma\neq0})\hspace*{1cm}\alpha_9=-\frac{\beta_1f^2}{2M_{Z'}^2}\!
  +\beta_1(5\hat{\mathcal{K}}_{2}^{\Sigma\neq0}\!-10\hat{\mathcal{K}}_{13}^{\Sigma\neq0}\!
  +\frac{5}{4}\hat{\mathcal{K}}_{14}^{\Sigma\neq0})\nonumber~~~~\\
  &&\hspace*{-0.5cm}\alpha_{10}=\frac{5\beta_1^2}{4}(\hat{\mathcal{K}}_{3}^{\Sigma\neq0}+\hat{\mathcal{K}}_{4}^{\Sigma\neq0})
 +\frac{\beta_1^4}{8a_3^4}g_{4Z}-\frac{\beta_1^3}{2a_3^3}[(2a_3^3+6a_0^2a_3+3\hat{a}_0^2a_3)(\hat{\mathcal{K}}_{3}^{\Sigma\neq0}
 +\hat{\mathcal{K}}_{4}^{\Sigma\neq0})+2a_0^2a_3\hat{D}_2]\nonumber\\
  &&\hspace*{-0.5cm}\alpha_{11}=\alpha_{12}=\alpha_{13}=\alpha_{14}=0\;.\label{alphaResult}
\end{eqnarray}
\section{Numerical results and discussion}

We first analyze the general features of the EWCL LECs obtained in the previous section, which are similar to those in Lane's first natural prototype TC2 model\cite{JunYi09}:
\begin{itemize}
\item  The contributions of the $p^4$-order coefficients are divided into two parts: the contribution from the three sets of techniquarks and the $Z'$ contribution
\item  All correction terms from the $Z'$ particle to the EWCL LECs are proportional to powers of $\beta_1$ which vanish if the mixing disappear ($\theta=0$). This can be seen from (\ref{alphaResult}) and (\ref{a0a3}) which show that:   $\beta_1=\frac{10g_1^2\hat{F}_0^2\tan^2\theta}{16\bar{M}_{Z'}^2}$. By using the relation $\alpha_\mathrm{em}T=2\beta_1$, we can express all LECs in terms of the $T$ parameter. Later in the paper, we show the $T$ dependence of the LECs.
\item From (\ref{alphaResult}) (for $f^2$ and $\beta_1$), combined with (\ref{a0a3}), (\ref{barMZ'}) , the relation $\alpha_\mathrm{em}T=2\beta_1$ and the relationships of the hyper-charges from Ref.\cite{Lane96}, we have
\begin{eqnarray}
\alpha_\mathrm{em}T&=&\bigg[1+\frac{2}{5}[\frac{81\tilde{F}_0^2}{4\hat{F}_0^2}+716+4(1-\frac{F_0^{\prime2}}{\hat{F}_0^2})]
(u_1-v_1)^2(1+\cot^2\theta)^2 \bigg]^{-1}\;.~~~\nonumber
\end{eqnarray}
If we include the numerical result that $F^{\prime2}_0<\hat{F_0^2}$, the above result implies that $T$ must be positive and has an upper bound. The upper bound is:
\begin{eqnarray}
\alpha_\mathrm{em}T_{\mathrm{Max}}=\frac{1}{1+\frac{2}{5}[\frac{81\tilde{F}_0^2}{4\hat{F}_0^2}+716+4(1-\frac{F_0^{\prime2}}{\hat{F}_0^2})]
(u_1-v_1)^2}\;.~~\label{Tmax}
\end{eqnarray}
    \item Because numerical calculation shows that $\hat{\mathcal{K}}_{2}^{\Sigma\neq0}\!-\hat{\mathcal{K}}_{13}^{\Sigma\neq0}<0$ and $\beta_1$ is positive, $\alpha_8$ is negative based on (\ref{alphaResult}). Then $U=-16\pi\alpha_8$ which is a coefficient given in Ref.\cite{EWCL} , is always positive in the present model.
\end{itemize}
Combining (\ref{barMZ'}), (\ref{cZ'}) and (\ref{Z'mass}), we find,
\begin{eqnarray}
&&\hspace{-0.5cm}2\frac{\tilde{F}_0^2}{M_{Z'}^2}g_1^2(\cot\theta+\tan\theta)^2\xi^{2}+4\frac{\hat{F}_0^2}{M_{Z'}^2}(2a_0^2+\hat{a}_0^2+5a_3^2)
-8\frac{F^{\prime2}_0}{M_{Z'}^2}a_0^2\label{CalKconstraint}\\
&&\hspace{-0.5cm}=1+[4(\cot\theta+\tan\theta)^2\xi^2+2\tan^2\theta+8\hat{v}+3\tan^2\theta+\hat{y}]\mathcal{K}g_1^2
+4(\cot\theta+\tan\theta)^2\xi^2\tilde{\mathcal{K}}_2^{\Sigma\neq0}g_1^2\nonumber\\
&&\hspace*{0cm}+8(2a_0^2+\hat{a}_0^2+5a_3^2)\hat{\mathcal{K}}_2^{\Sigma\neq0}+[40a_3^2+2(\hat{t}+\hat{s})g_1^2]\hat{\mathcal{K}}_{13}^{\Sigma\neq0}
-15\hat{D}_0a_0^2\;.
\nonumber
\end{eqnarray}
We treat the above equation as a constraint on $\mathcal{K}$. This is done as following: A suitable choice is made for the hypercharges (this will be discussed later), electroweak gauge coupling, $T$ and $M_{Z'}$.  We already know most of the parameters in (\ref{CalKconstraint}), except $\tilde{F}_0$, $\hat{F}_0$, $F^{\prime2}_0$, $\tilde{\mathcal{K}}_2^{\Sigma\neq0}$, $\hat{\mathcal{K}}_2^{\Sigma\neq0}$,  $\hat{\mathcal{K}}_{13}^{\Sigma\neq0}$ and $\hat{D}_0$. By solving the SDEs, (\ref{SDEtildeSigma}), (\ref{SDEhatSigma}), (\ref{SDEbarSigma}), (\ref{SDEbarSigma5}), we can obtain the techniquark self-energies, $\tilde{\Sigma}$, $\hat{\Sigma}$, $\bar{\Sigma}$, $\bar{\Sigma}_5$. Furthermore, substituting the resulting techniquark self-energies into the formulae given in Appendix \ref{Kexpression} and (\ref{D0def}), we can obtain  $\tilde{F}_0$, $\hat{F}_0$, $F^{\prime2}_0$, $\tilde{\mathcal{K}}_2^{\Sigma\neq0}$, $\hat{\mathcal{K}}_2^{\Sigma\neq0}$,  $\hat{\mathcal{K}}_{13}^{\Sigma\neq0}$ and $\hat{D}_0$ from (\ref{CalKconstraint}). Now, aside from $\mathcal{K}$ all the parameters in (\ref{CalKconstraint}) are known. Then we can use (\ref{CalKconstraint}) to fix the value of $\mathcal{K}$. Once $\mathcal{K}$ is fixed, with the help of (\ref{kappDef}), we can determine the ratio of the infrared cutoff $\kappa$ and ultraviolet cutoff $\Lambda$. Numerical calculations show that this is unlike the results in Refs.\cite{JunYi09,LangPLB},  where the condition $\Lambda>\kappa$ occurs through the definitions used for the calculations and offers stringent constraints on the allowed region for $T$ and the upper bound for $M_{Z'}$. In our model, $\Lambda>\kappa$  is naturally satisfied for real values of $M_{Z'}$. For example, we find that $\ln\kappa/\Lambda$ is about $-7.6$ and $-9.0$ for $M_{Z'}$ values of 0.5TeV and  1TeV, respectively.

\begin{figure}[t]
\caption{Coloron mass for Lane's model.} \label{MassV}
\hspace*{-4.5cm}\vspace*{-0.7cm}\begin{minipage}[t]{8cm}
    \includegraphics[scale=0.8]{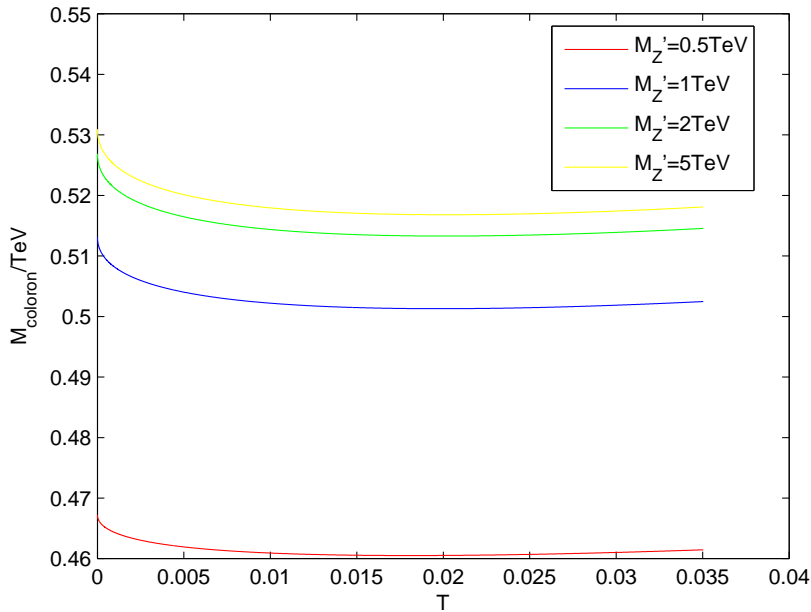}
\end{minipage}
\end{figure}

With the above qualitative features, we now can generate numerical results. First, we take $N=6$ which yields an infrared fixed point of $\alpha_w=88\pi/523$. Then, we take $f=250$GeV. This completely fixes the two-loop value at  $\Lambda_w=5.5$TeV through the running behavior of (\ref{alphaBehavior}), SDE (\ref{SDEhatSigma}), $f^2=5\hat{F}_0^2$ and (\ref{F0}) which sets up the relationship between $\hat{F}_0^2$ techniquark self-energy. This value of $\Lambda_w$ is smaller than the expected conventional ETC scale. Therefore, we cannot interpret it as $\Lambda_{\mathrm{ETC}}$. Later, we will see that this is because the walking effect is not large enough, and more ideal walking can lead to a larger $\Lambda_w$. The current result with $\Lambda_w\ll\Lambda_{\mathrm{ETC}}$ shows that our running coupling constant cannot always walk from extreme infrared energy regions to the ETC scale, $\Lambda_{\mathrm{ETC}}$. Instead, it can only walk a shorter distance to the scale, $\Lambda_w$. Beyond $\Lambda_w$, it will run and fall quickly exhibiting conventional asymptotic freedom behavior. Another theoretical parameter is the coloron mass given by (\ref{coloronmass}), which theoretically depends on the values $\theta'$, introduced in (\ref{A1A2-AB}) and $\Theta$, introduced in (\ref{ThetaDef}). We find the largest coloron mass occurs for
$\Theta=\pi/2$, i.e., the self-energies for the first set of techniquarks are completely contributed by the twisted part of the set, $\bar{\Sigma}_5=\hat{\Sigma}\sin\Theta$ and $\bar{\Sigma}=0$. Using this value of $\Theta=\pi/2$, in Fig.\ref{MassV}, we plot the coloron mass in terms of the $T$ parameter.
We used four values of  $M_{Z'}=0.5,1,2,5$TeV (corresponding to $\ln\kappa/\Lambda \sim$ -7.6,-9.0,-9.4 and -9.5). We found that that the coloron mass is not sensitive to $\theta'$.
From Fig.\ref{MassV}, it can be seen that the coloron mass is roughly half the 1 TeV expected in Lane's original paper\cite{Lane96}. The reason is that we included a techniquark loop correction in the coloron kinetic term, which appeared in (\ref{coloronmass}) with the coefficients $\hat{E}$, $\mathcal{K}$ and $\hat{\mathcal{K}}_{13}^{\Sigma\neq0}$. If we denote the coloron mass without this correction as $M_{\mathrm{bare~coloron}}$ which was the notation used in Lane's original work\cite{Lane96}, then our numerical calculation shows that: $M_{\mathrm{bare~coloron}}/M_{\mathrm{coloron}}\sim
\frac{2}{3}(\tan\theta'+\cot\theta')$. This leads to a larger value for $M_{\mathrm{bare~coloron}}$. In fact, if we carefully examine the denominator of (\ref{coloronmass}), the structure of this kinetic term correction can be divided into three parts: the tree order term $2/g_3^2(\cot\theta'+\tan\theta')^2$, the techniquark self-energy dependent part $\hat{E}+2\hat{\mathcal{K}}_{13}^{\Sigma\neq0}-8\hat{\mathcal{K}}_{13}^{\Sigma\neq0}/(\cot\theta'+\tan\theta')^2$, and the techniquark self-energy independent part $2\mathcal{K}$. The numerical calculation shows that the main contribution comes from the techniquark self-energy dependent part, which is an order of magnitude larger than the contributions from the other two parts. Because the coloron mass is small\footnote{The small coloron mass forces us to switch the order of integration over the coloron and Z', i.e., instead of integrating out the coloron before the Z' boson, we need to integrate out Z' and then the coloron. We have performed the computation using this new procedure and found the same result as that of the present paper, i.e. switching the order of integration yields no correction. We found that the possible correction from switching this order of integration depends on the classical field $B_{A,c}^\mu$ caused by the coloron integration. These classical coloron fields are determined by stationary equations. In both cases, the stationary equations offer the null solution, $B_{A,c}^\mu=0$ , which was used in our results.}, we will use $\Theta=\pi/2$ to give the largest coloron mass for all the following computations.

\begin{figure}[t]
\caption{$S$ parameter for Lane's model.} \label{S}
\hspace*{-7cm}\vspace*{-0.7cm}\begin{minipage}[t]{8cm}
    \includegraphics[scale=0.8]{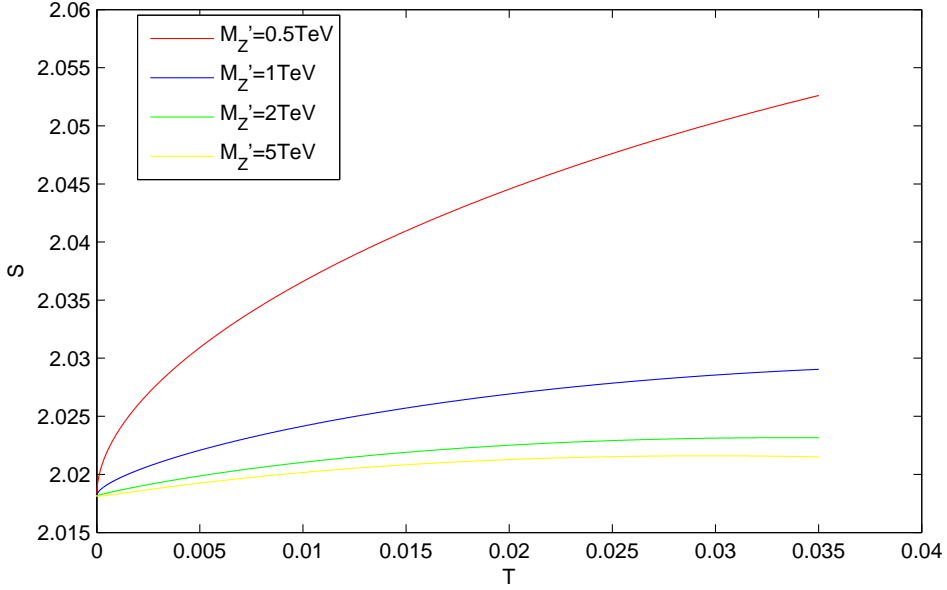}
\end{minipage}
\end{figure}

\begin{figure}[t]
\caption{$U$ parameter for Lane's model.} \label{U}
\hspace*{-7cm}\vspace*{-0.7cm}\begin{minipage}[t]{8cm}
    \includegraphics[scale=0.8]{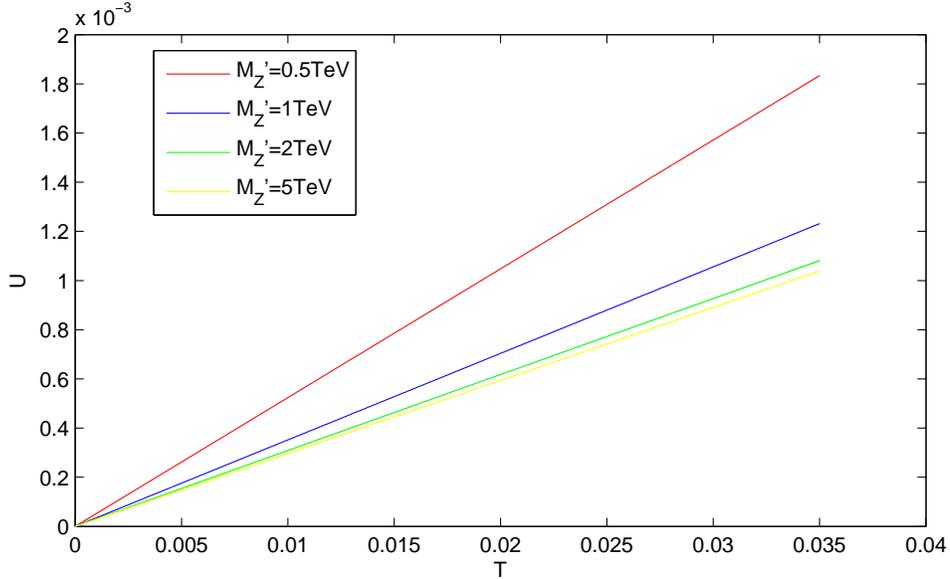}
\end{minipage}
\end{figure}

\begin{figure}[t]
\caption{$S$ parameter for various choices of the hyper-charges: $x_1=-50,y_1=36,y_2=-12$.} \label{Ssmall}
\hspace*{-7cm}\vspace*{-0.7cm}\begin{minipage}[t]{8cm}
    \includegraphics[scale=0.8]{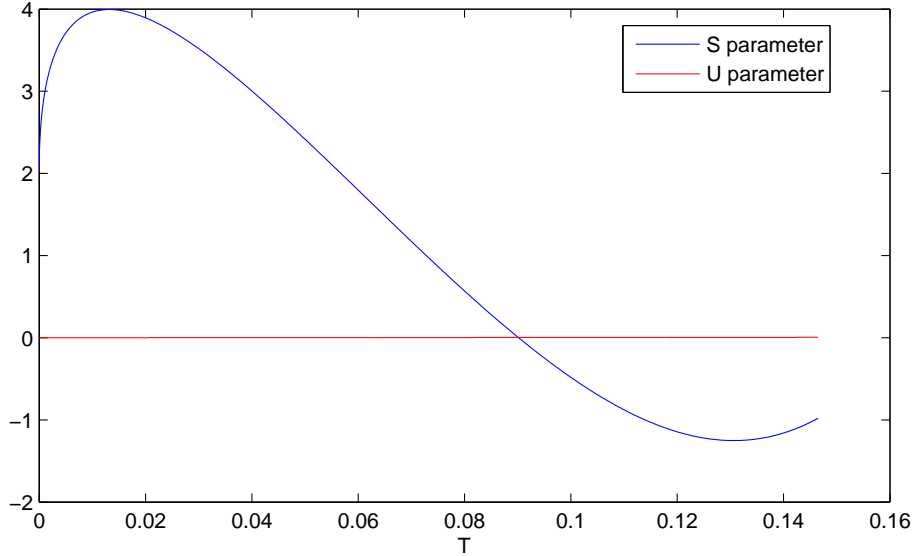}
\end{minipage}
\end{figure}

\begin{figure}[t]
\caption{$\alpha_2$ parameter for Lane's model.} \label{alpha2}
\hspace*{-4.5cm}\vspace*{-0.7cm}\begin{minipage}[t]{8cm}
    \includegraphics[scale=0.8]{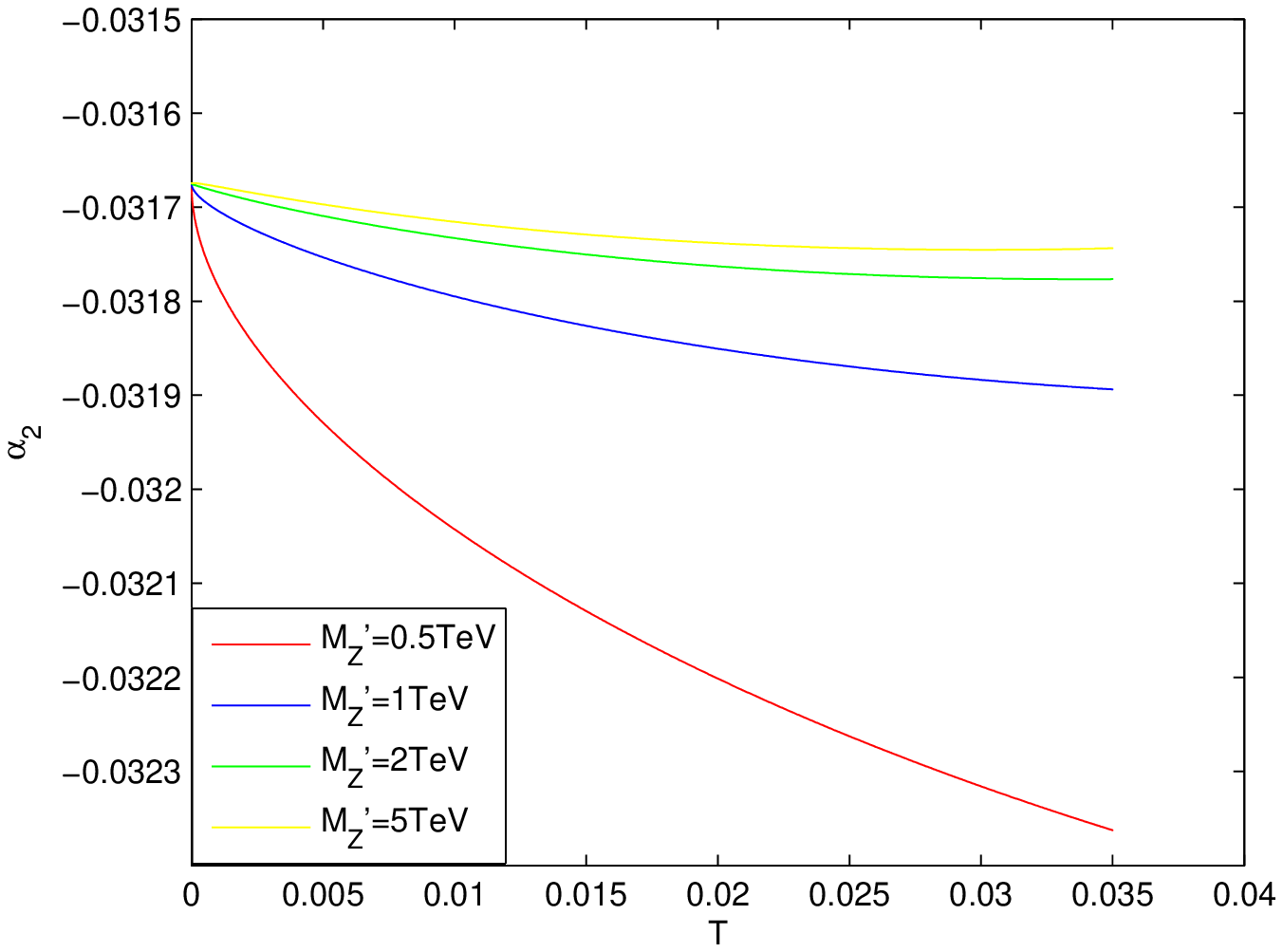}
\end{minipage}
\end{figure}

\begin{figure}[t]
\caption{$\alpha_3$ and $\alpha_{10}$ parameters for Lane's model.} \label{alpha310}
\hspace*{-4.5cm}\vspace*{-0.7cm}\begin{minipage}[t]{8cm}
    \includegraphics[scale=0.8]{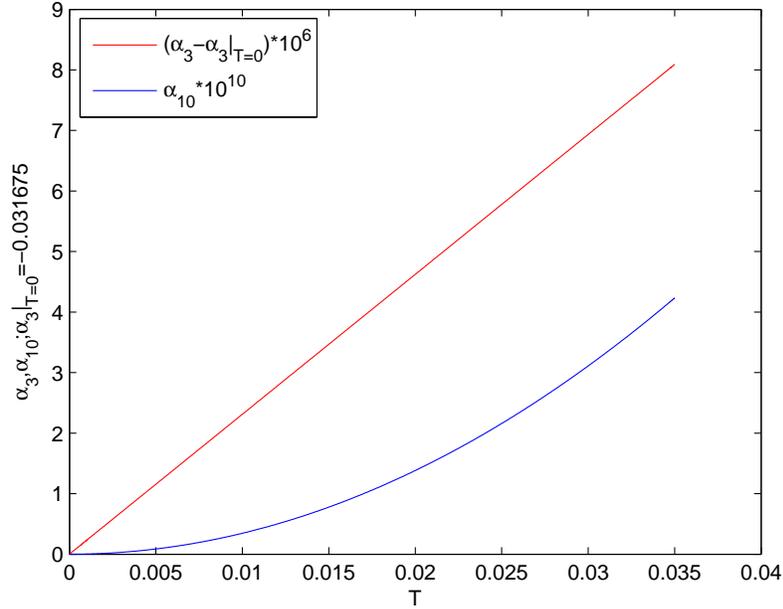}
\end{minipage}
\end{figure}

\begin{figure}[t]
\caption{$\alpha_4$ parameter for Lane's model.} \label{alpha4}
\hspace*{-4.5cm}\vspace*{-0.7cm}\begin{minipage}[t]{8cm}
    \includegraphics[scale=0.8]{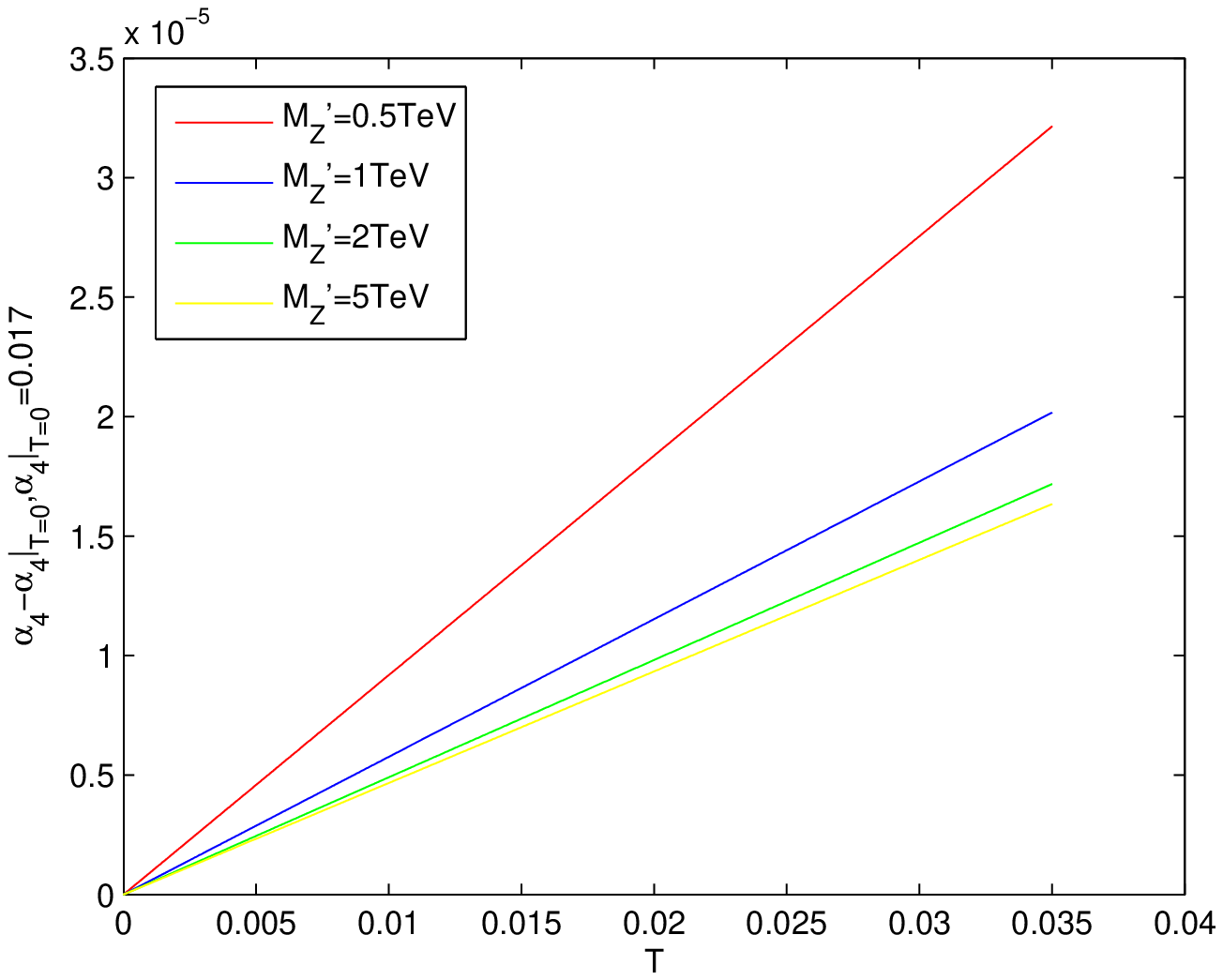}
\end{minipage}
\end{figure}

\begin{figure}[t]
\caption{$\alpha_5$ parameter for Lane's model.} \label{alpha5}
\hspace*{-4.5cm}\vspace*{-0.7cm}\begin{minipage}[t]{8cm}
    \includegraphics[scale=0.8]{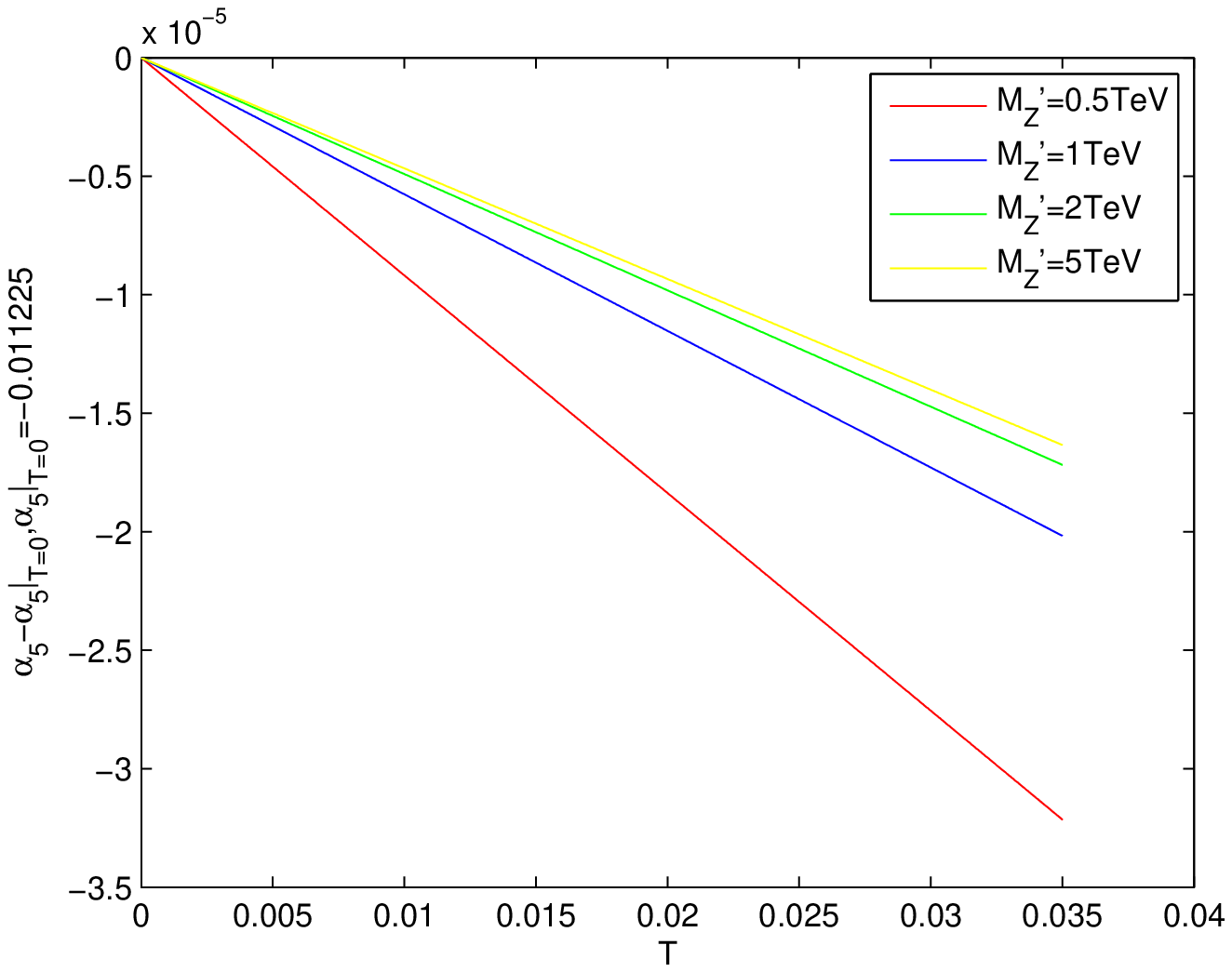}
\end{minipage}
\end{figure}

\begin{figure}[t]
\caption{$\alpha_6$ parameter for Lane's model.} \label{alpha6}
\hspace*{-4.5cm}\vspace*{-0.7cm}\begin{minipage}[t]{8cm}
    \includegraphics[scale=0.8]{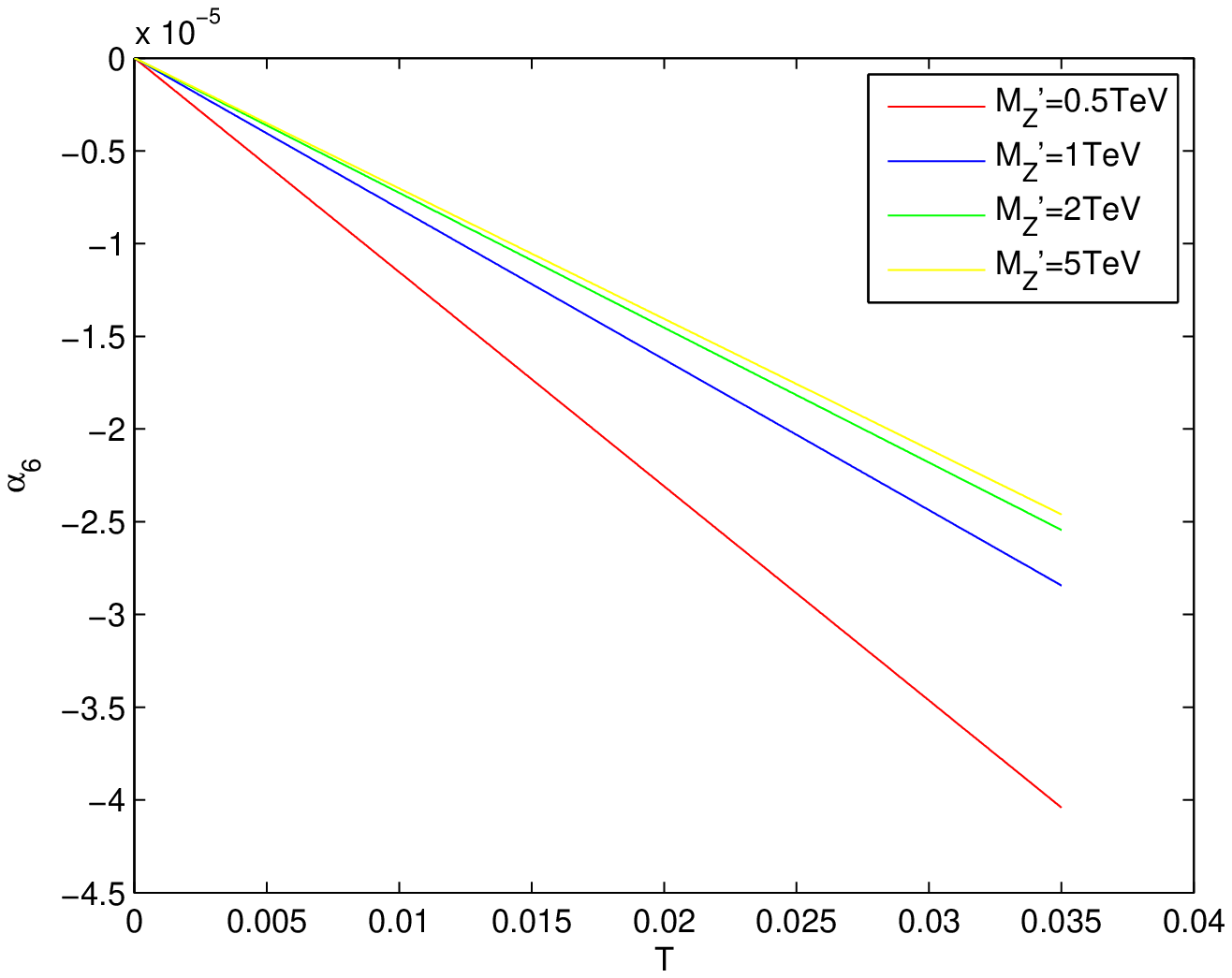}
\end{minipage}
\end{figure}

\begin{figure}[t]
\caption{$\alpha_7$ parameter for Lane's model.} \label{alpha7}
\hspace*{-4.5cm}\vspace*{-0.7cm}\begin{minipage}[t]{8cm}
    \includegraphics[scale=0.8]{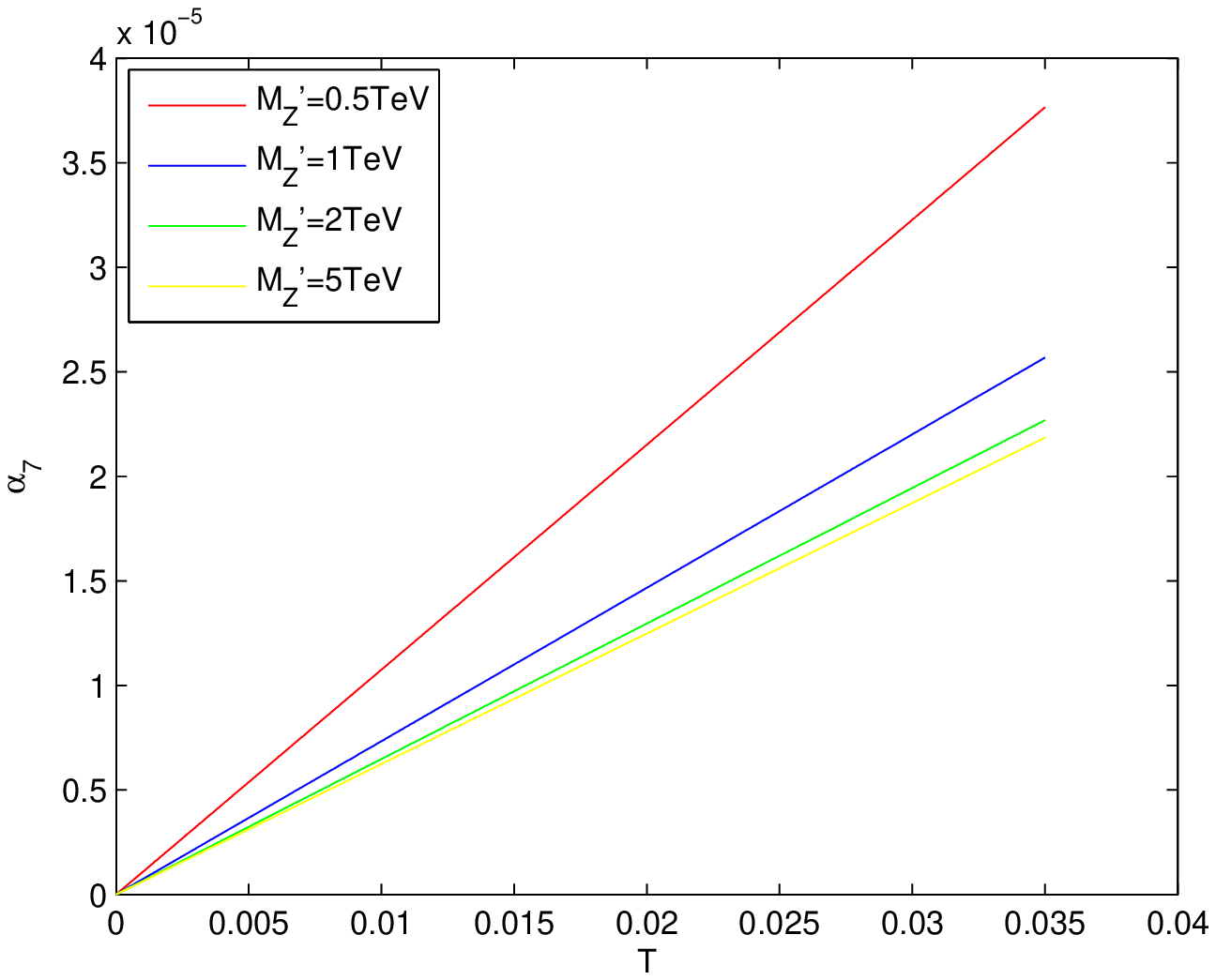}
\end{minipage}
\end{figure}

\begin{figure}[t]
\caption{$\alpha_9$ parameter for Lane's model.} \label{alpha9}
\hspace*{-4.5cm}\vspace*{-0.7cm}\begin{minipage}[t]{8cm}
    \includegraphics[scale=0.8]{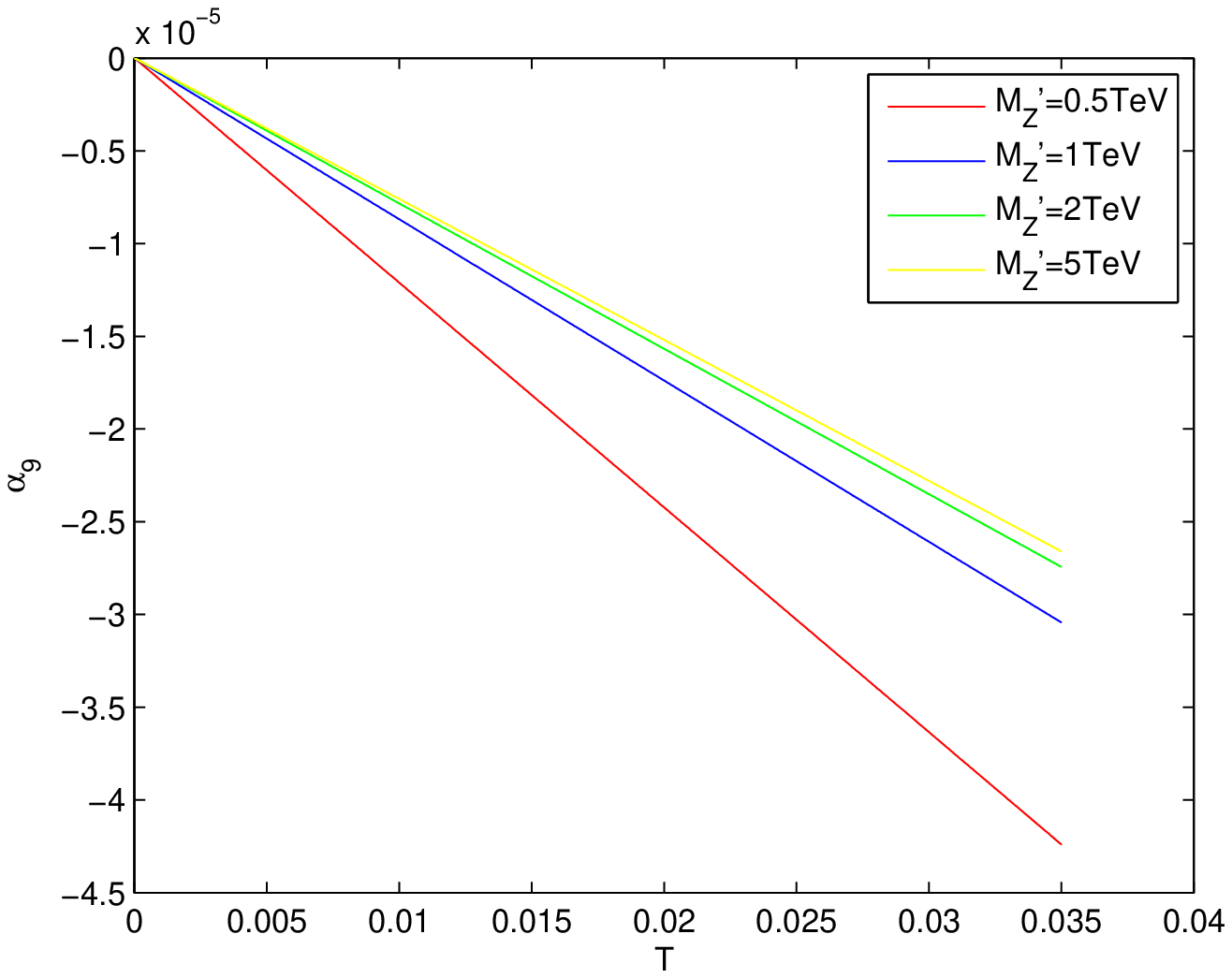}
\end{minipage}
\end{figure}

To provide numerical values for all the EWCL LECs, we need to choose the various hyper-charges for the model. Note that the arrangement of the hyper-charges given in Lane's original paper\cite{Lane96} is not suitable here because that result used $N=4$. We showed in Section II that for the modern interpretation of our two-loop based phase structure model, we use $N=6$, and recalculate the hyper-charges. According to a series of relations among different hyper-charges given by K. Lane in Ref.\cite{Lane96}, we need to use three hyper-charges $x_1$, $y_1$ and $y_1+y_2$. We use a treatment similar to the one used by K. Lane in Ref.\cite{Lane96}. Namely, we use $x_1=y_1$, $y_1+y_2=0$. Furthermore, this requires that $u=(u_1-v_1)/2\sim 1$. These fully fix the typical values of all the hyper-charges. By "typical" we mean that the value of the hyper-charges must satisfy all 23 constraint equations given in Ref. \cite{Lane96} and two more constraints: $x_1=y_1$, $y_1+y_2=0$. The last two constraints were not explicitly mentioned in Ref.\cite{Lane96}, but the detailed example used them. These typical hyper-charges are: $a=-39$, $a'=-46$, $b=14$, $b'=8.2$, $c=-39$, $c'=-46$, $d=-12$, $d'=-14$, $\xi=4.6$, $\xi'=-4.6$, $x_1=25$, $x'_1=19$, $x_2=-26$,  $x'_2=-19$, $y_1=25$,  $y'_1=23$, $y_2=-25$, $y'_2=-23$, $z_1=-7.7$, $z'_1=19$, $z_2=7.7$, $z'_2=-19$, $u_1=-4.1$, $v_1=-6.1$, $u_2=4.2$, $v_2=6.2$.Using this set of typical hyper-charges, combined with the other necessary inputs for the model, which were discussed in the previous paragraph, (\ref{Tmax}) yields an upper bound,  $T_{\mathrm{max}}=0.035$. We show $S=-16\pi\alpha_1$ in Fig.\ref{S}, and $U=-16\pi\alpha_8$ in Fig.\ref{U}.
From Fig.\ref{S}, it can be seen that the value of $S$ is generally larger than 2, which is not in agreement with experimental data. This value of the $S$ parameter already includes the walking effects in the model, which we will discuss later. To examine the possibility of reducing the value of the $S$ parameter through the choice of hyper-charges, we found that when the input hyper-charges $x_1,y_1$ are not constrained by the requirement $x_1=y_1$ and are much larger than 1, $S$ may achieve small values. Fig.\ref{Ssmall} shows the case with: $x_1=-50,y_1=36,y_2=-12$ which leads $a=-19$, $a'=-22$, $b=7$, $b'=4$, $c=-19$, $c'=-22$, $d=-6$, $d'=-7$, $\xi=2.3$,  $\xi'=-2.3$, $x_1=-50$, $x'_1=-53$, $x_2=2.7$, $x'_2=5.7$,$y_1=36$,$y'_1=35$,$y_2=-12$, $y'_2=-11$, $z_1=20$, $z'_1=33$, $z_2=3.6$, $z'_2=-9.4$, $u_1=0.41$, $v_1=-0.59$, $u_2=-0.41$, $v_2=0.59$. The $S$ parameter can achieve negative values with larger values of $T$. There may be other sets of hyper-charges which can also yield small or even negative values of $S$, but typically these hyper-charges have large values.
Excluding the $S$ and $U$ parameters, the leftmost eight non-zero parameters $\alpha_2,\alpha_3,\alpha_4,\alpha_5,\alpha_6,\alpha_7,\alpha_9,\alpha_{10}$
are shown in Fig.\ref{alpha2} to Fig.\ref{alpha9}. $\alpha_3$ and $\alpha_{10}$ are independent of $M_{Z'}$ and are shown in the same figure.

We found that $\alpha_2,\alpha_3,\alpha_4,\alpha_5,$ are on the order of $10^{-2}$, $\alpha_6,\alpha_7,\alpha_9$ are on the order of $10^{-5}$ and
$\alpha_{10}$ is on the order of $10^{-10}$.

\begin{figure}[t]
\caption{$\alpha_1$ and $\alpha_2$ of the TC2 model \cite{Hill95}-Hill, \cite{Lane95}-Lane(I), \cite{Sekhar}-Chiv and \cite{Lane96}-Lane(II). The numbers on each curve are the masses of the $Z'$ boson in TeV.} \label{alpha12comparison}
\hspace*{-0.7cm}\vspace*{-0.7cm}\begin{minipage}[t]{16cm}
    \includegraphics[scale=0.51]{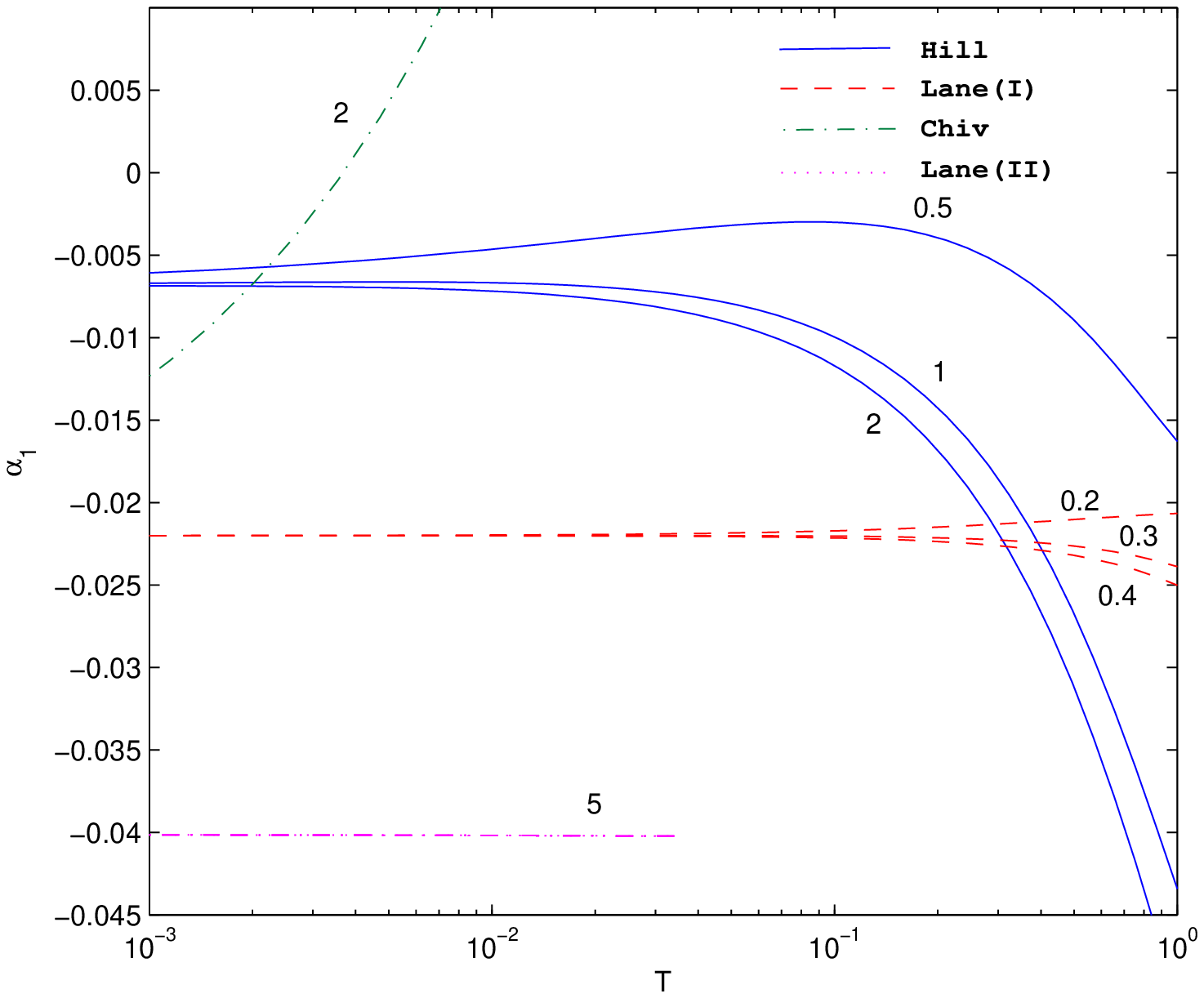}\includegraphics[scale=0.58]{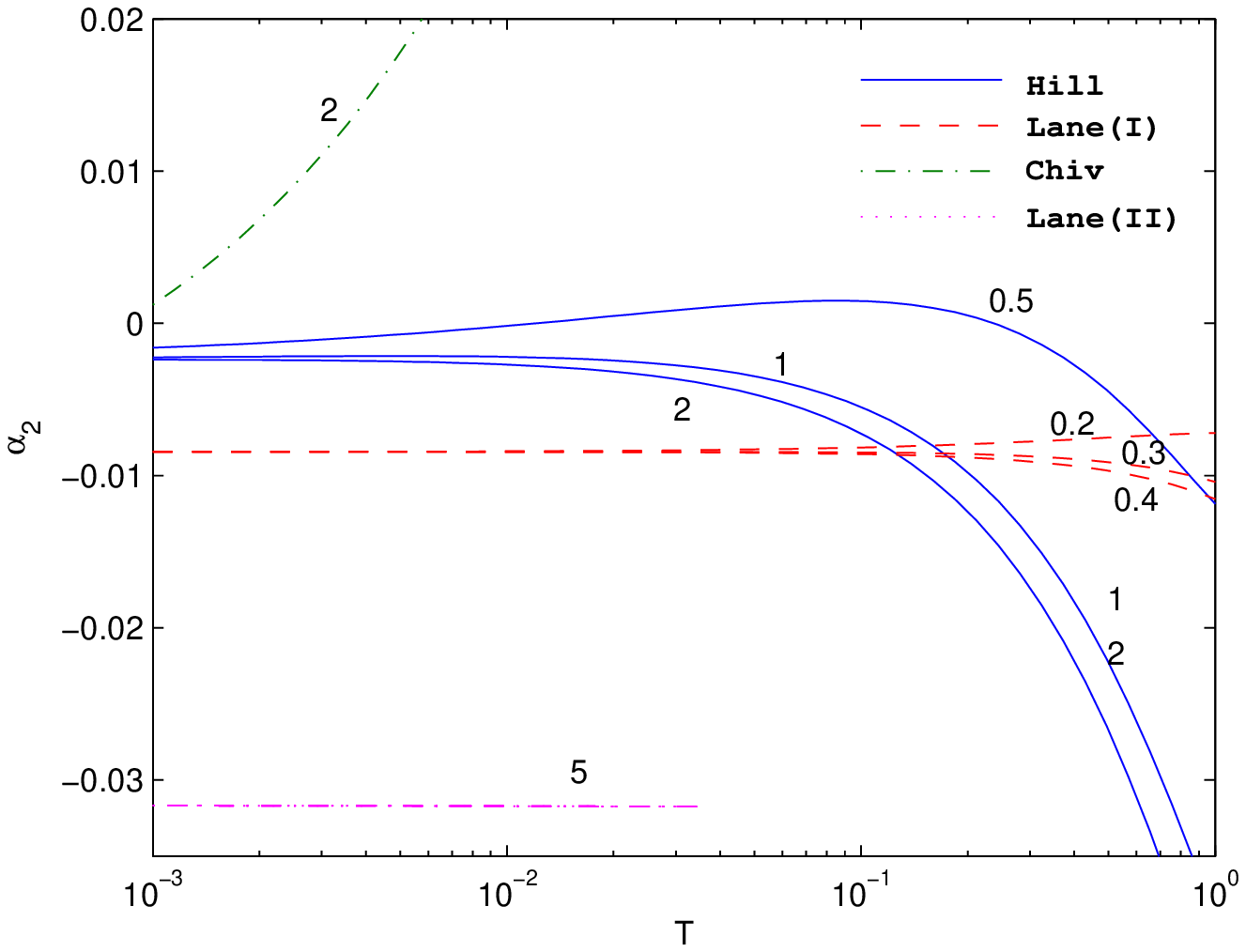}
\end{minipage}
\end{figure}

\begin{figure}[t]
\caption{$\alpha_3$ and $\alpha_4$ of the TC2 model \cite{Hill95}-Hill, \cite{Lane95}-Lane(I), \cite{Sekhar}-Chiv and \cite{Lane96}-Lane(II). The numbers on each curve are the masses of the $Z'$ boson in TeV.} \label{alpha34comparison}
\hspace*{-1cm}\vspace*{-0.7cm}\begin{minipage}[t]{16cm}
    \includegraphics[scale=0.58]{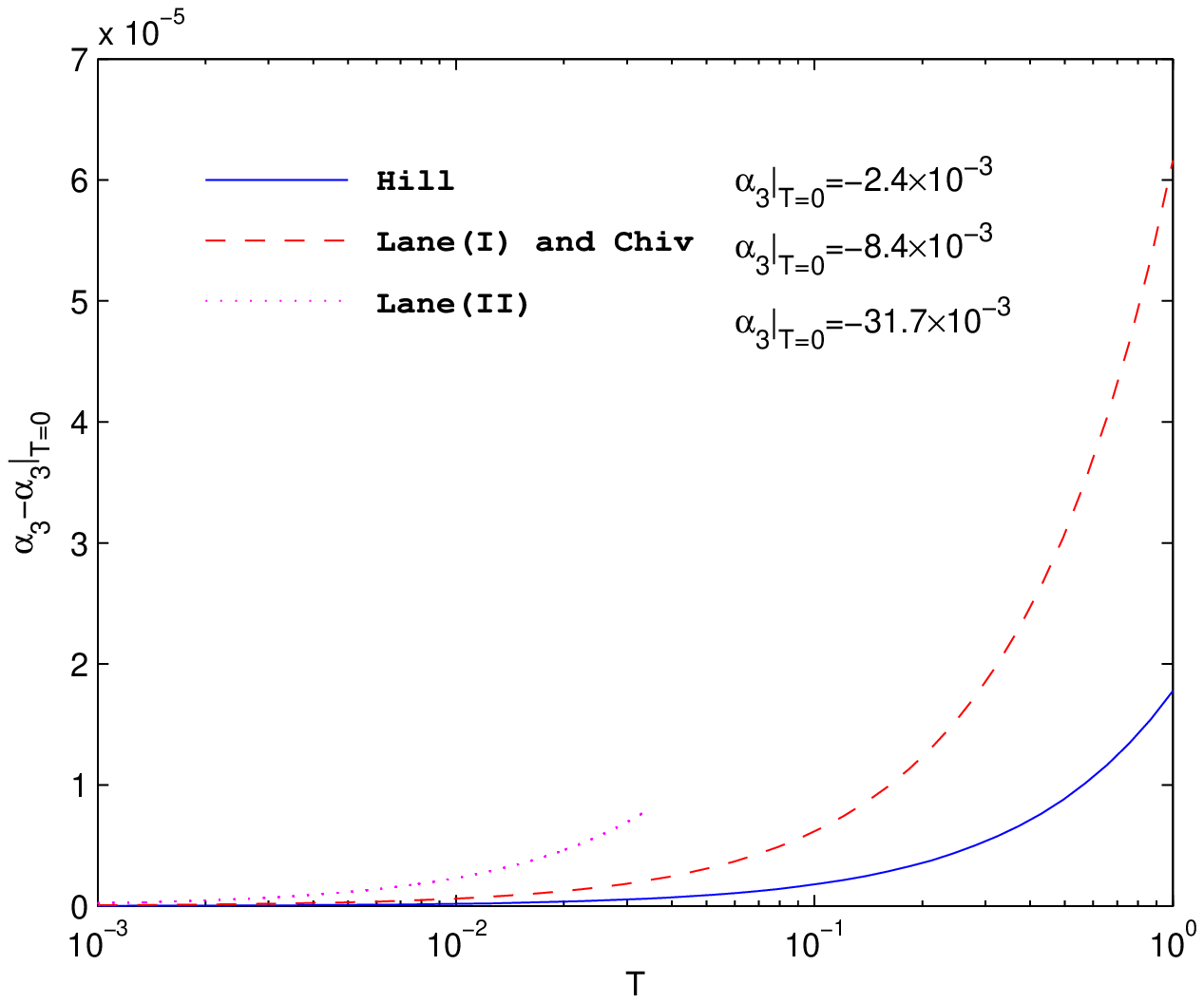}\includegraphics[scale=0.58]{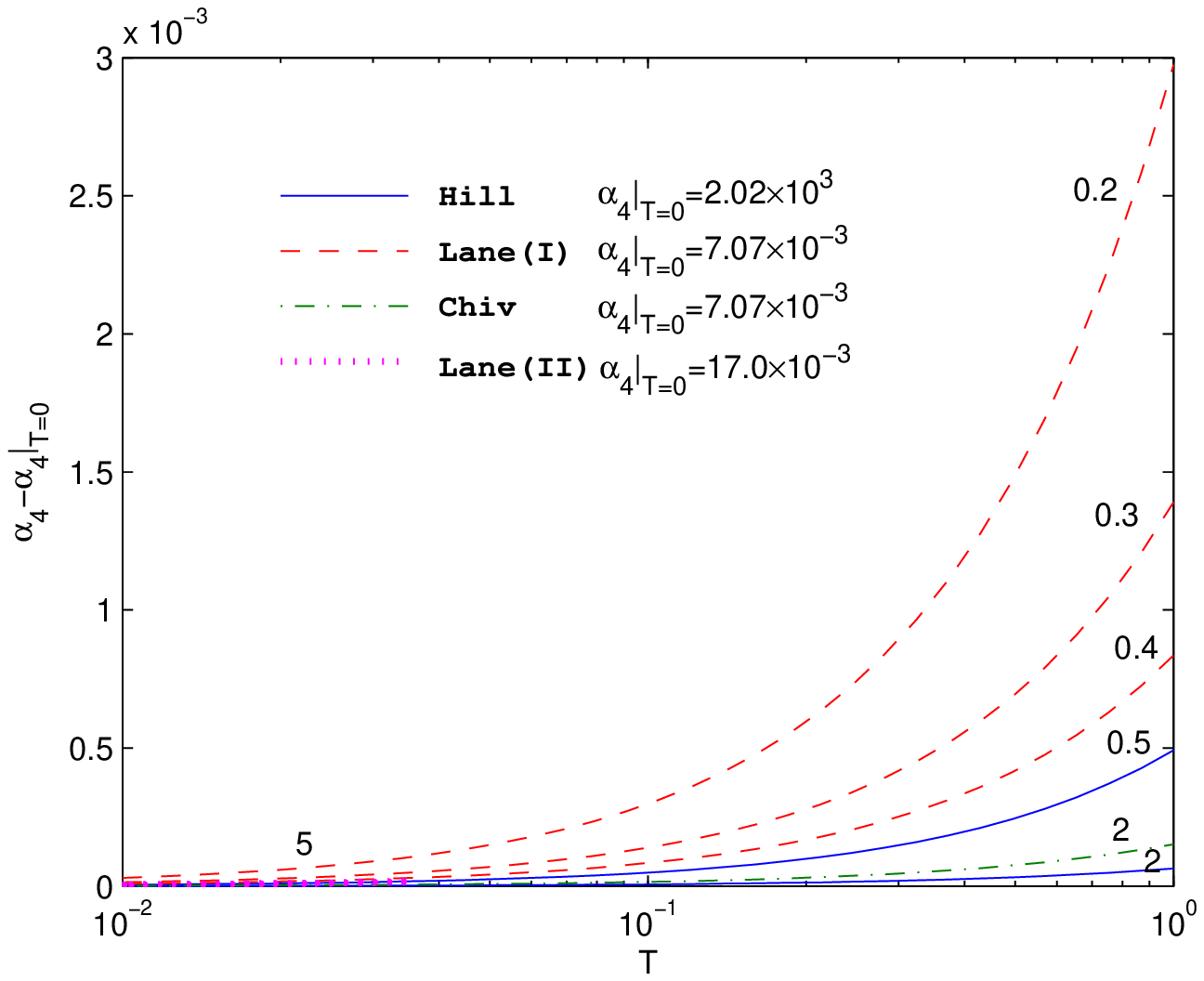}
\end{minipage}
\end{figure}

\begin{figure}[t]
\caption{$\alpha_5$ and $\alpha_6$ of the TC2 model \cite{Hill95}-Hill, \cite{Lane95}-Lane(I), \cite{Sekhar}-Chiv and \cite{Lane96}-Lane(II). The numbers on each curve are the masses of the $Z'$ boson in TeV.} \label{alpha56comparison}
\hspace*{-1cm}\vspace*{-0.7cm}\begin{minipage}[t]{16cm}
    \includegraphics[scale=0.58]{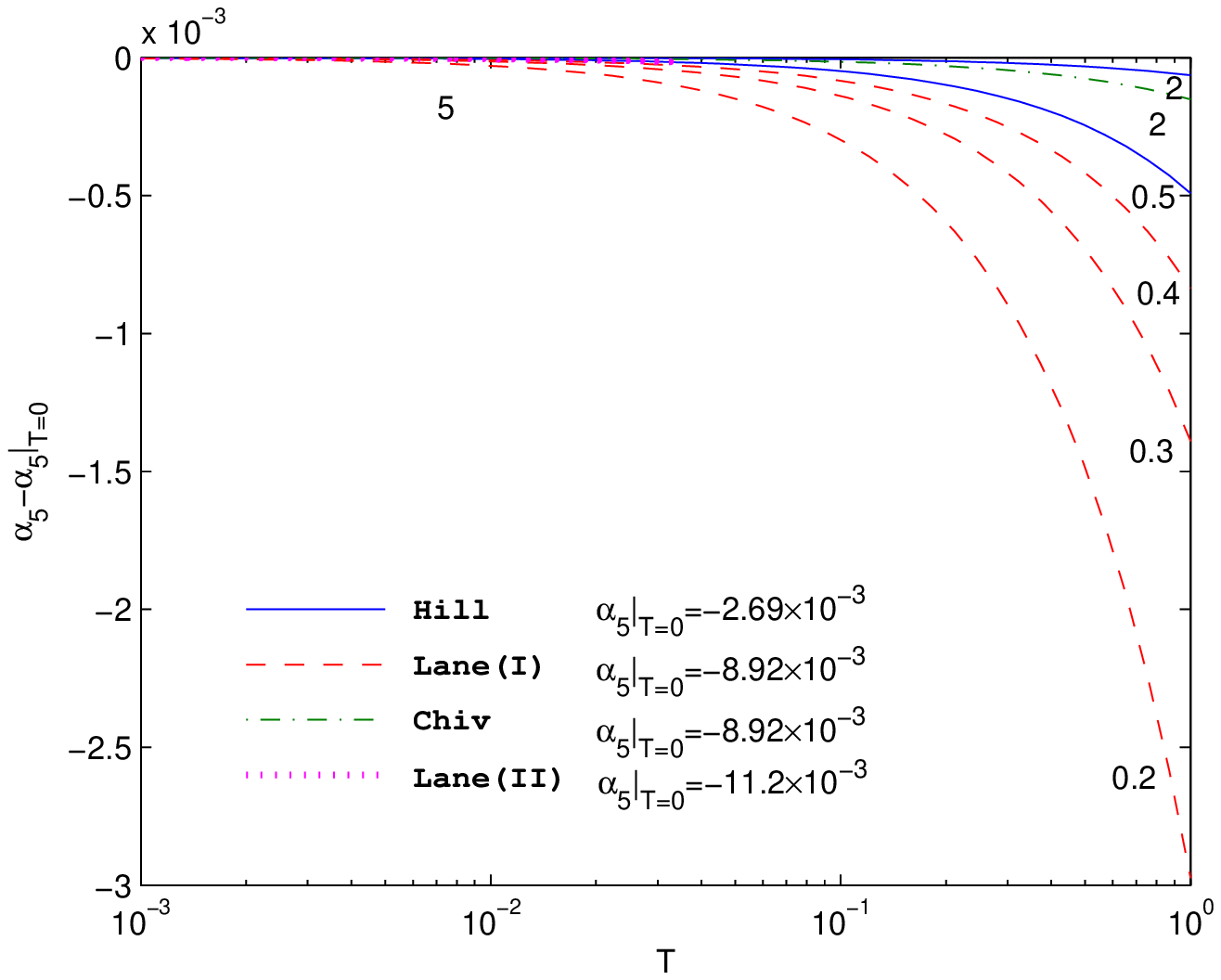}\includegraphics[scale=0.58]{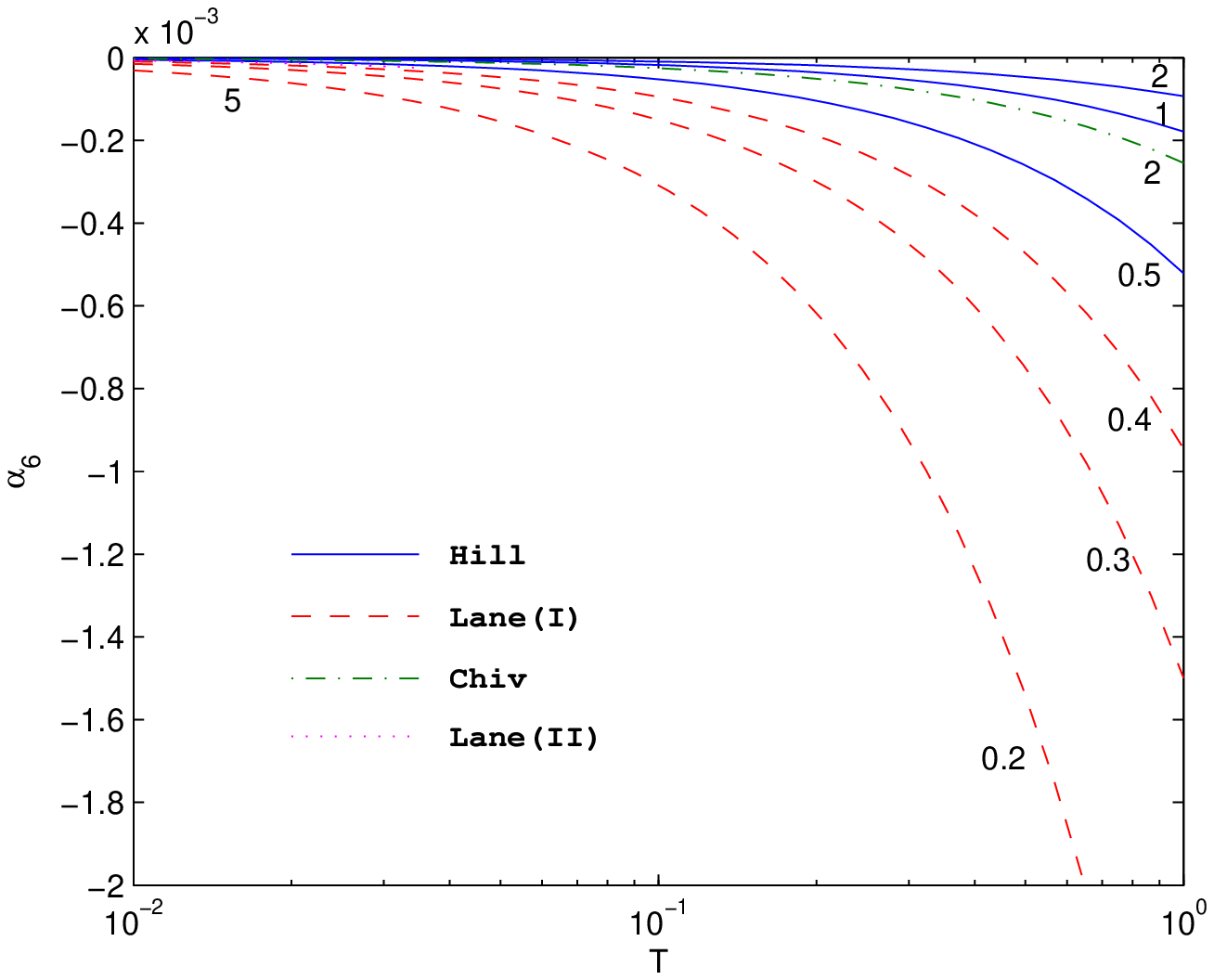}
\end{minipage}
\end{figure}

\begin{figure}[t]
\caption{$\alpha_7$ and $\alpha_8$ of the TC2 model \cite{Hill95}-Hill, \cite{Lane95}-Lane(I), \cite{Sekhar}-Chiv and \cite{Lane96}-Lane(II). The numbers on each curve are the masses of the $Z'$ boson in TeV.} \label{alpha78comparison}
\hspace*{-1cm}\vspace*{-0.7cm}\begin{minipage}[t]{16cm}
    \includegraphics[scale=0.58]{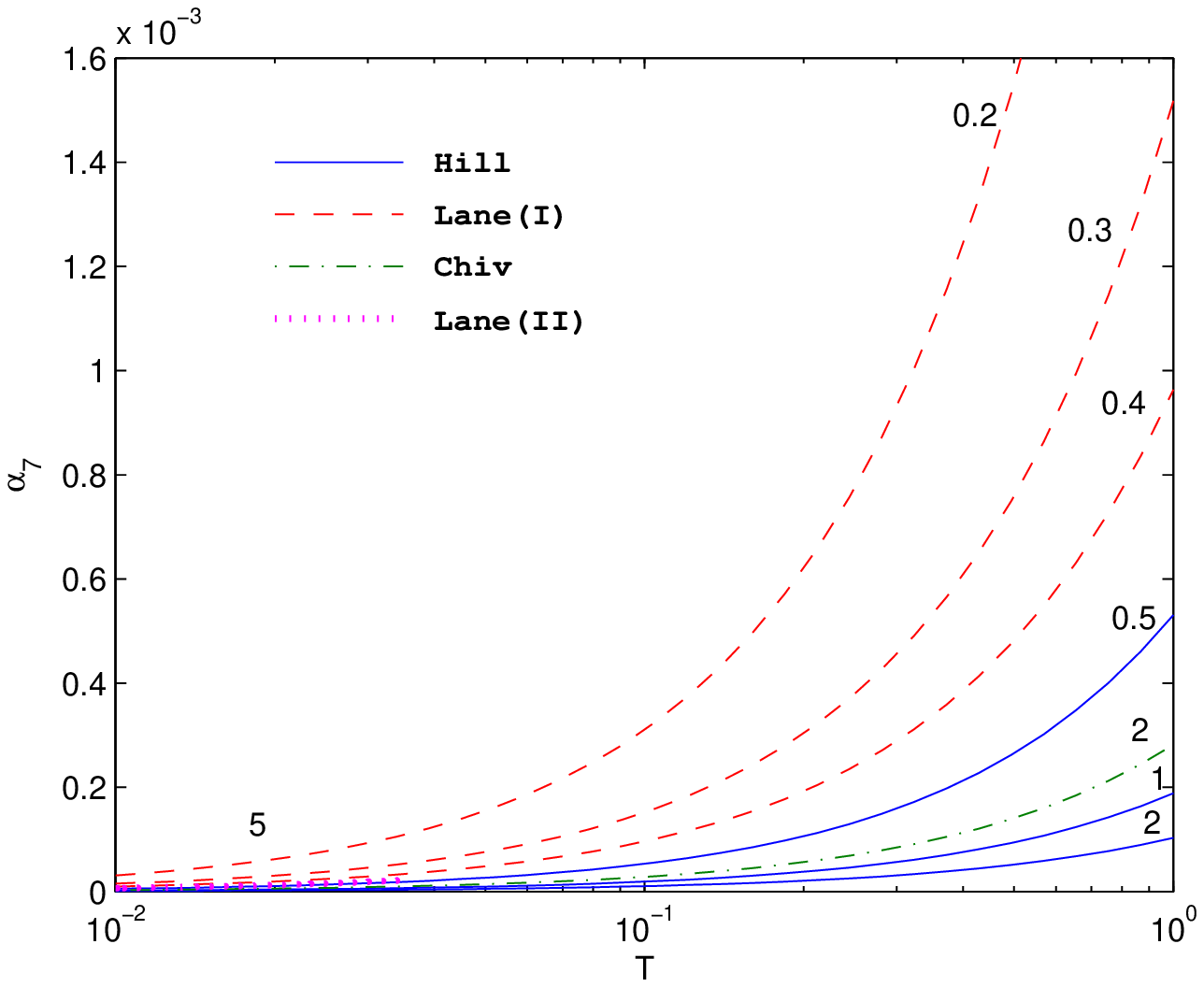}\includegraphics[scale=0.58]{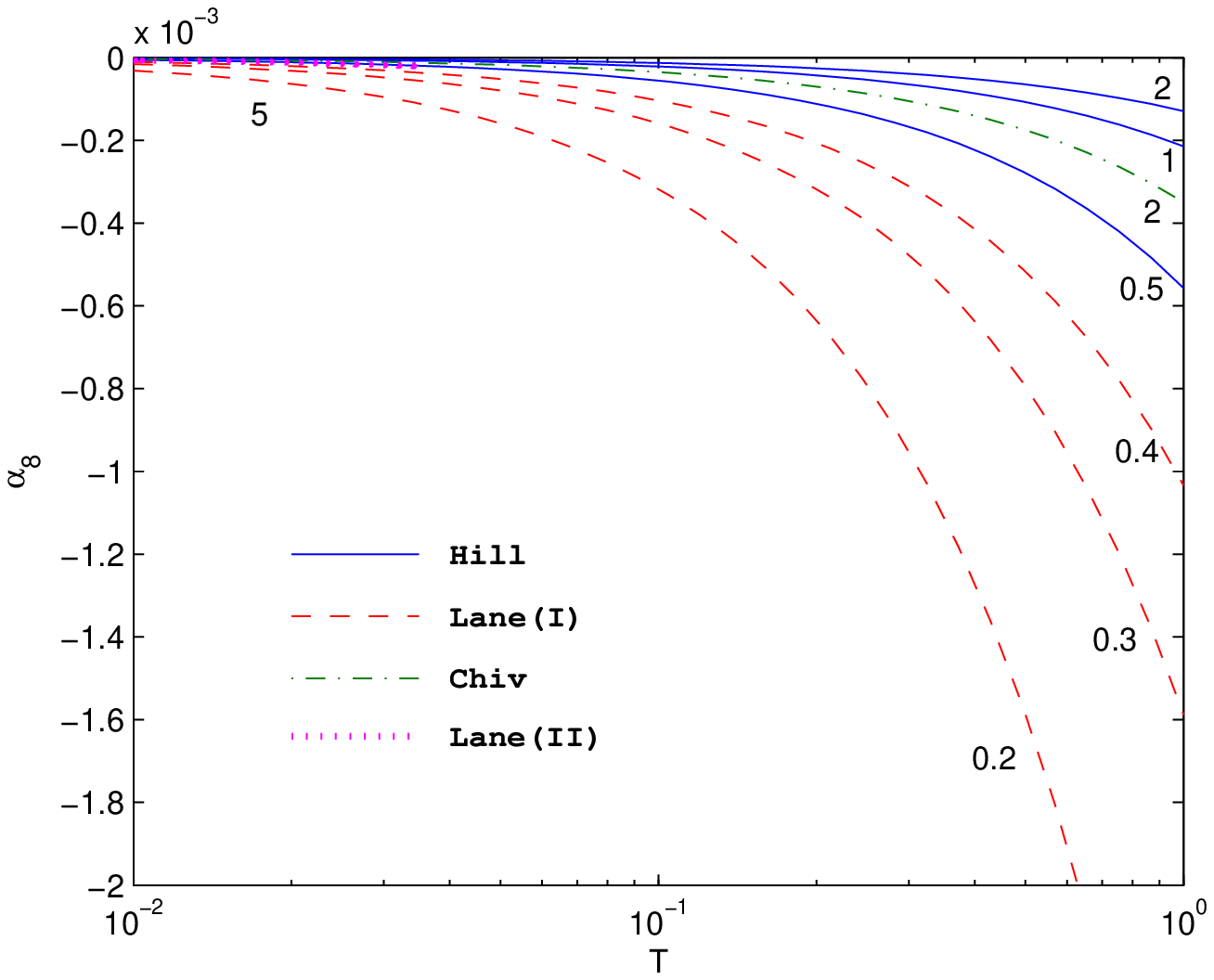}
\end{minipage}
\end{figure}

\begin{figure}[t]
\caption{$\alpha_9$ and $\alpha_{10}$ of the TC2 model \cite{Hill95}-Hill, \cite{Lane95}-Lane(I), \cite{Sekhar}-Chiv and \cite{Lane96}-Lane(II). The numbers on each curve are the masses of the $Z'$ boson in TeV.} \label{alpha910comparison}
\hspace*{-1cm}\vspace*{-0.7cm}\begin{minipage}[t]{16cm}
    \includegraphics[scale=0.58]{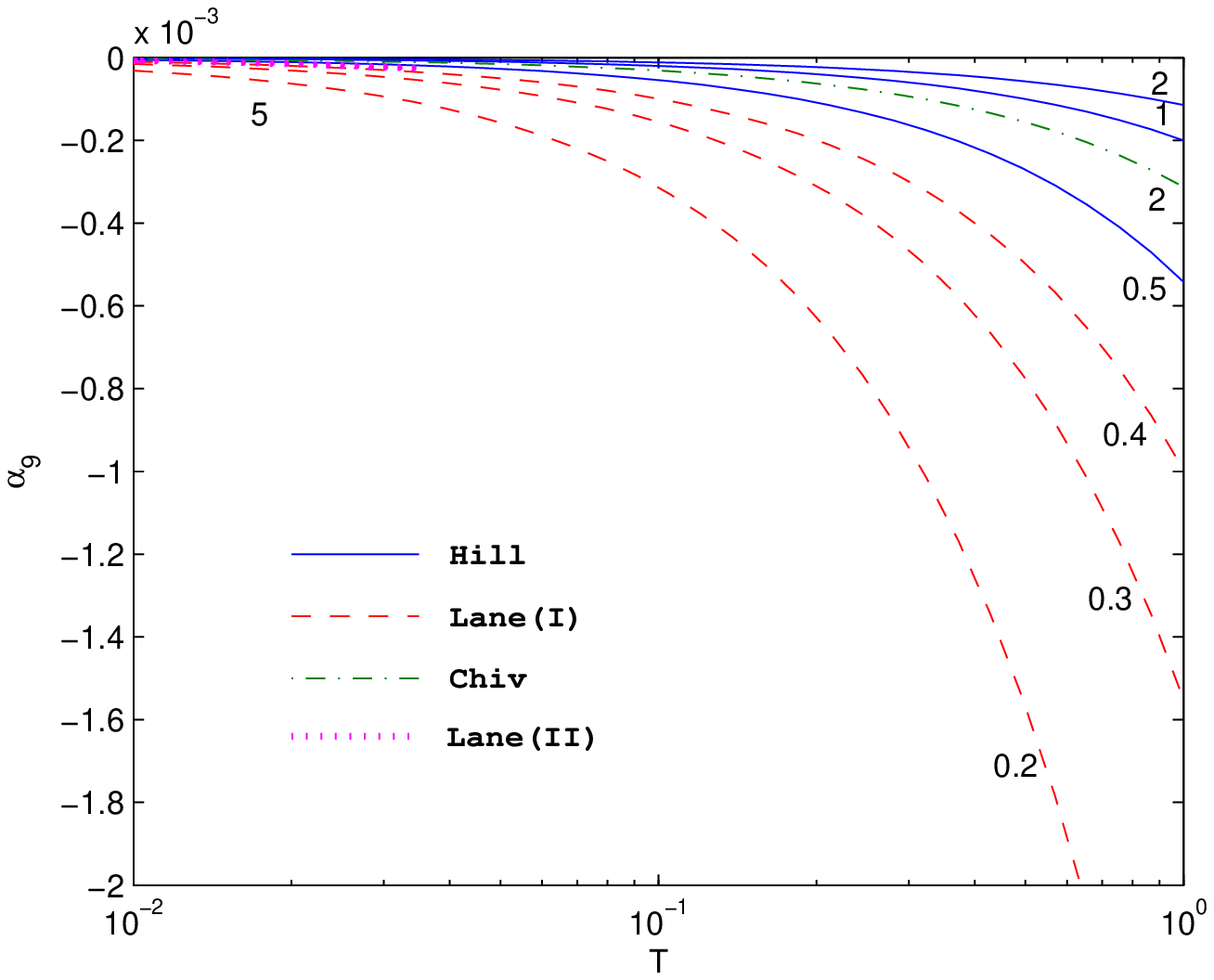}\includegraphics[scale=0.58]{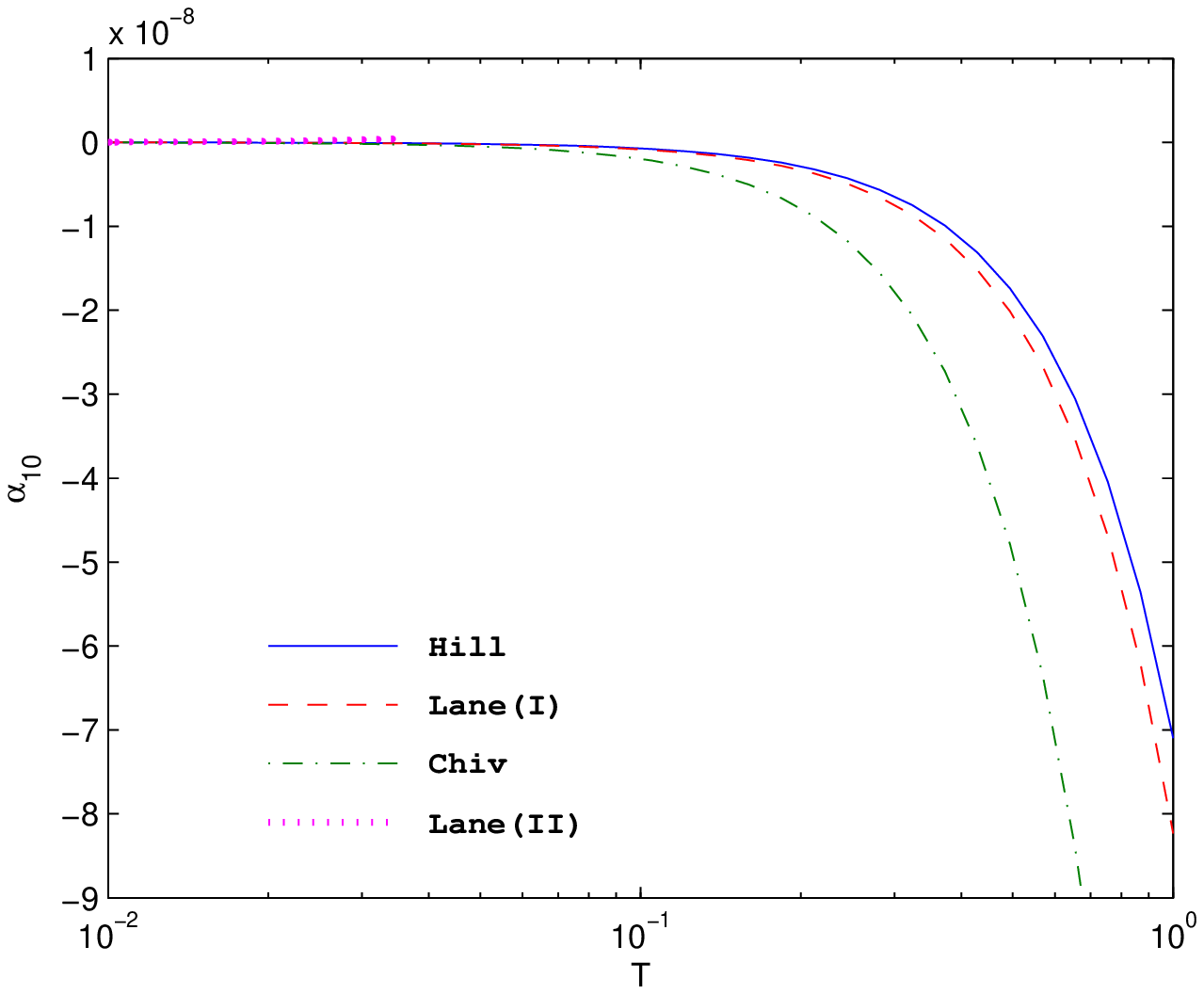}
\end{minipage}
\end{figure}

Previously, we discussed the three other TC2 models\cite{Hill95,Lane95,Sekhar}. In Table IV., we list the different features and the orders of magnitude for all the LECs of these TC2 models. In Fig.\ref{alpha12comparison}, Fig.\ref{alpha34comparison}, Fig.\ref{alpha56comparison},Fig.\ref{alpha78comparison} and Fig.\ref{alpha910comparison}, we show the ten nonzero LECs from these four TC2 models for comparison. This comparison may be useful to other researchers as they consider the needs of future models.

\begin{table}[h]
\small{{\bf TABLE IV}.~Features and LECs of the TC2 models \cite{Hill95}, \cite{Lane95}, \cite{Sekhar} and \cite{Lane96}\\~\vspace*{-0.2cm}~}

\hspace*{0cm}\begin{tabular}{|c|cccc|}\hline
  Property or LEC& \hspace*{-0.5cm}Schematic TC2$^{\mbox{\tiny\cite{Hill95}}}$ & \hspace*{-0.7cm}Natural TC2$^{\mbox{\tiny\cite{Lane95}}}$ & {\footnotesize Hypercharge} Universal$^{\mbox{\tiny\cite{Sekhar}}}$ & Present$^{\mbox{\tiny\cite{Lane96}}}$ \\
 \hline
 Upper bound of $M_{Z'}$ & $\surd$ & $\surd$ & \hspace*{-1.5cm}$\surd$ & $\times$\\
   Negative $S$ & $M_{Z'}\!<\!0.44$TeV or $T\!>\!0.17$ & $\times$ & \hspace*{-1.5cm}$T\geq 10^{-1}$ & \hspace*{-2cm}choose hypercharges\\
  Typical $S\!=\!-16\pi\alpha_1$    & $\sim 0.3$ & $\sim 0.8$ & \hspace*{-1.5cm}$\sim 1$ & $\sim 2$\\
   $\alpha_2$ & $-10^{-3}$ & $-10^{-3}$ & \hspace*{-1.5cm}$-10^{-3}$ & $-10^{-2}$ \\
   $\alpha_3$ & $-10^{-3}$ & \hspace*{-0.5cm}$3\times$ result of \cite{Hill95} & \hspace*{-1.5cm}$-10^{-3}$ & $-10^{-2}$ \\
  $\alpha_4$ & $10^{-3}$ & \hspace*{-0.5cm}$3\times$ result of \cite{Hill95} & \hspace*{-1.5cm}$10^{-3}$ & $10^{-2}$ \\
    $\alpha_5$ & $-10^{-3}$ & \hspace*{-0.5cm}$3\times$ result of \cite{Hill95} & \hspace*{-1.5cm}$-10^{-3}$ & $-10^{-2}$ \\
       $\alpha_6$ & $\sim-10^{-4}$ & $\sim-10^{-3}$ & \hspace*{-1.5cm}$\sim-10^{-4}$ & \hspace*{-0.5cm}$\sim-10^{-5}$ \\
         $\alpha_7$ & $\sim 10^{-4}$ & $\sim 10^{-3}$ & \hspace*{-1.5cm}$\sim 10^{-4}$ & $\sim 10^{-5}$ \\
             $\alpha_8=-\frac{U}{16\pi}$ & $\sim-10^{-4}$ & \hspace*{-0.5cm}$3\times$ result of \cite{Hill95} & \hspace*{-1.5cm}$\sim-10^{-4}$ & \hspace*{-0.5cm}$\sim-10^{-5}$ \\
  $\alpha_9$ & $\sim-10^{-4}$ & \hspace*{-0.5cm}$3\times$ result of \cite{Hill95} & \hspace*{-1.5cm}$\sim-10^{-4}$ & \hspace*{-0.5cm}$\sim-10^{-5}$ \\
    $\alpha_{10}$ & $\sim-10^{-8}$ & $\sim-10^{-8}$ & \hspace*{-1.5cm}$\sim-10^{-7}$ & $\sim 10^{-10}$ \\
  \hline
\end{tabular}
 \end{table}


\begin{figure}[t]
\caption{Dependence of the $\Lambda_w$ (TeV) on the degree of walking.} \label{Lambdaw}
\hspace*{-4cm}\vspace*{-0.7cm}\begin{minipage}[t]{8cm}
    \includegraphics[scale=0.8]{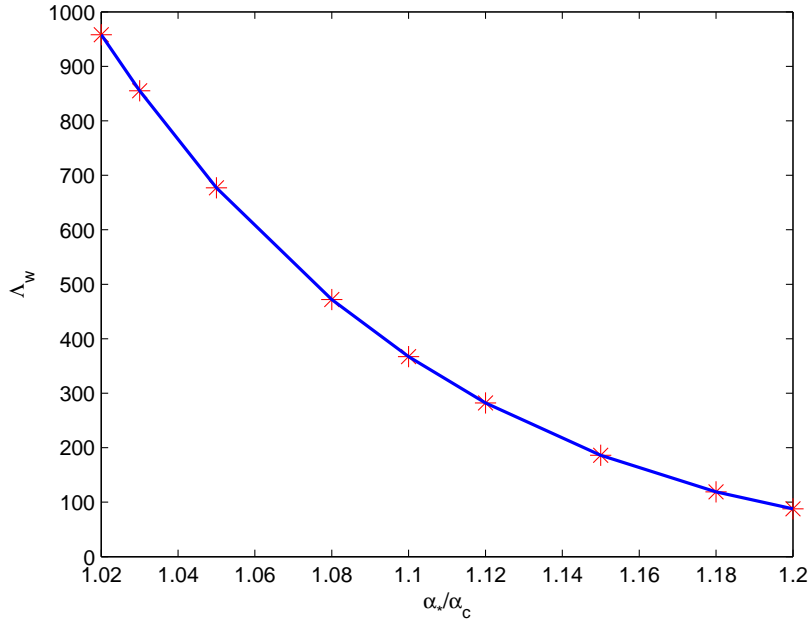}
\end{minipage}
\end{figure}

\begin{figure}[t]
\caption{Three different couplings.~ $\alpha_w$  is the coupling used in our calculation. $\alpha_r$ is the running coupling, which is given in (\ref{alphar}). Here, we show $\alpha_r/5$ to facilitate comparison between the couplings. $1.02\alpha_c$ is the ideal walking coupling, where $\alpha_*=1.02\alpha_c$.} \label{alphaAll}
\hspace*{-5cm}\vspace*{-0.7cm}\begin{minipage}[t]{8cm}
    \includegraphics[scale=0.8]{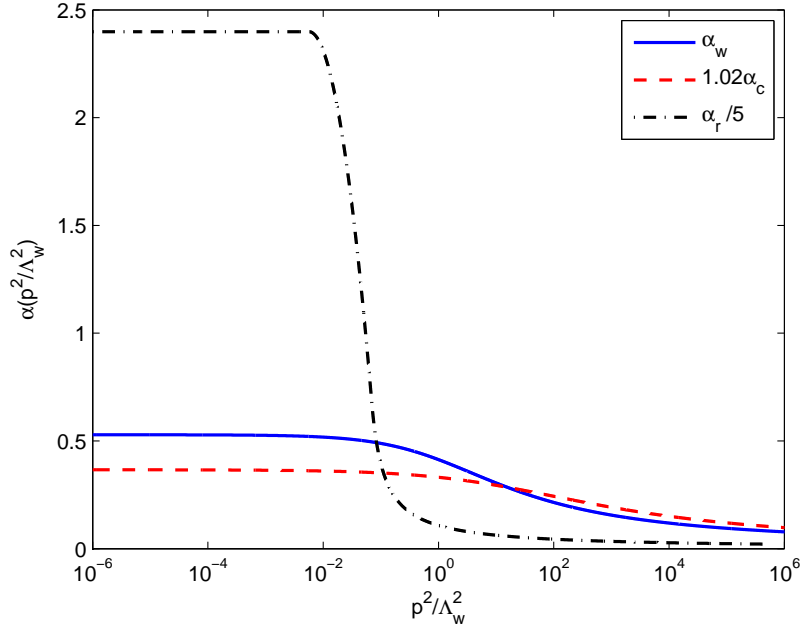}
\end{minipage}
\end{figure}

\begin{figure}[t]
\caption{Techniquark self-energies for three different couplings:~ $\hat{\Sigma}_w$ and $\tilde{\Sigma}_w$ the self-energies for the second and third sets of techniquarks for the coupling that we used in our calculation. $\hat{\Sigma}_r$ and $\tilde{\Sigma}_r$ are the self-energies for the second and third sets of techniquarks for the running coupling, which is given in (\ref{alphar}).  Here, we show $\hat{\Sigma}_r/5$ and $\tilde{\Sigma}_r/5$ to facilitate comparison between the self-energies. $\hat{\Sigma}_{1.02\alpha_c}$ and $\tilde{\Sigma}_{1.02\alpha_c}$ are the self-energies for the second and third sets of techniquarks for the ideal walking coupling, where $\alpha_*=1.02\alpha_c$.} \label{Sigma}
\hspace*{-5cm}\vspace*{-0.7cm}\begin{minipage}[t]{8cm}
    \includegraphics[scale=0.8]{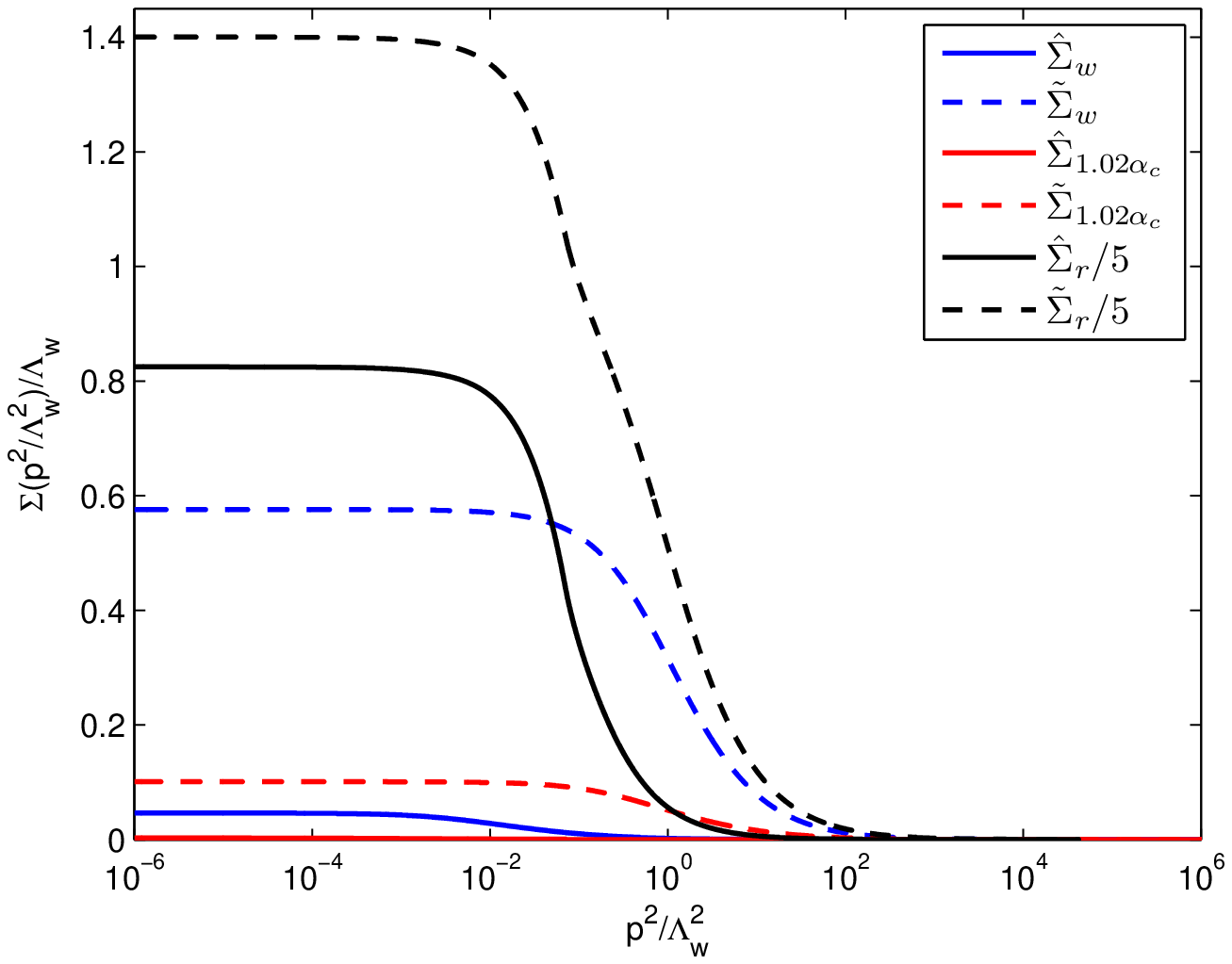}
\end{minipage}
\end{figure}

Finally, we estimate the magnitude of the walking effect in the present model. Because the primary contribution to the walking effect is from the running coupling constant, which appears in the kernel of the SDE, we can measure the walking effect by comparing two other running behaviors:
\begin{itemize}
\item Running $\alpha$:~Rather than using a two-loop running coupling constant (\ref{alphaBehavior}) which exhibits an approximation of walking behavior in $N=6$ and spontaneous chiral symmetry breaking, we used the one-loop running coupling constant used in our previous work\cite{HongHao08,JunYi09,LangPLB} as
    \begin{eqnarray}
    \alpha(x)=\frac{4\pi}{\beta_0}\times\left\{\begin{array}{lll}7&\hspace*{1cm}&\ln x\leq-2\\7-\frac{4}{5}(2+\ln x)^2&&-2\leq\ln x\leq 0.5\\\frac{1}{\ln x}&&\ln x\geq 0.5\end{array}\right.\hspace*{2cm}x=\frac{p^2}{\Lambda^2_{\mathrm{TC}}}\;.~~~~\label{alphar}
    \end{eqnarray}
    Eq.(\ref{alphar}) was originally introduced in Ref.\cite{Runalpha}. The general principle of the technique is to use a plateau in the low energy region to normalize the possibly infinite value in the infrared region that is predicted using the perturbative result and smoothly connect this infrared plateau with the ultraviolet asymptotic freedom running behavior. Note that if we ignore the two-loop term in the $\beta$ function in this model and normalize the infrared coupling constant such that it has a finite value, we can qualitatively obtain the above form of the running coupling constant. Furthermore, this approximation at the one-loop level suggests that $\Lambda_w$ must be treated as $\Lambda_{\mathrm{TC}}$ in this running situation. The change from one-loop running to two-loop walking reflects the evolution of our understanding of the gauge-coupling running behavior in non-abelian gauge theory. In addition, the decision to use the latter model in this study is important because it confirms the existence of the infrared fixed point\cite{Lattice} which qualitatively supports the modern two-loop-based explanation of walking.
\item Ideal walking $\alpha$:~Rather than using a two-loop running coupling constant (\ref{alphaBehavior}) and a value of $\alpha_*=88\pi/523$ that is not close in value to the critical coupling $\alpha_c=4\pi/35$ for the first and second set of techniquarks, we use the same running coupling constant but change the value of $\alpha_*$ in (\ref{alphaBehavior}) by
artificially requiring that
 $\alpha_*=1.02\alpha_c=1.02*4\pi/35$.  Although this is not a realistic
case for the model, it is closer to the conformal situation, and therefore, ideal walking.
\end{itemize}
The reason we must consider the above cases is because our analytical estimation using the $\beta$  may cause some error. Therefore, we can use these two extremes to investigate the effect of changes in the situation on our results. We show three different behaviors of $\alpha$ in Fig.\ref{alphaAll}. It can be seen that $\alpha_r$ is much bigger than $\alpha_w$ only in the extreme infrared region, and that the running behavior corresponding to $1.02\alpha_c$ is smaller than that corresponding to $\alpha_w$ over most of the energy region. From a comparison of Fig.\ref{alphaAll} with Fig.\ref{fig-alpha}, it can be seen that the running effect increases the height of the infrared plateau and narrows its length. To contrast other differences resulting from these different couplings, in Fig.\ref{Sigma}, we show the techniquark self-energies,  $\tilde{\Sigma}$ and $\hat{\Sigma}$, which are determined by the SDEs (\ref{SDEtildeSigma}) and (\ref{SDEhatSigma}). We found that the closer the system came to walking, the lower and wider the techniquark self-energy plateau was. By contrast, during running, the plateau was higher and narrower. For fixed $f=250$GeV, we found that the running situation produces a value of $\Lambda_{\mathrm{TC}}=0.21$TeV ($\Lambda_{\mathrm{ETC}}$ in the running case cannot be determined solely by the running behavior and requires some other physical parameters to be known). This result is consistent with the estimate of $\Lambda_{\mathrm{TC}}\simeq 2f\sqrt{3/N}$ given in Ref.\cite{LambdaTC}. Our walking and ideal walking situations yield:
\begin{eqnarray}
\Lambda_w=\left\{\begin{array}{lll}5.5\mathrm{TeV}&\hspace*{1cm}&\mbox{walking}\\
958\mathrm{TeV}&\hspace*{1cm}&\mbox{ideal walking}\end{array}\right.\nonumber
\end{eqnarray}
From this, it can be seen that $\Lambda_w$ is very sensitive to the walking effect. The closer the system is to ideal walking, the bigger the value of $\Lambda_w$. This was further checked by calculating $\Lambda_w$ for several values of $\alpha_*/\alpha_c=1.04,1.06,1.08,1.1,1.12,1.14,1.16,1.18,1.2$. These points were then plotted as a curve in Fig.\ref{Lambdaw} to quantitatively show the sensitivity of $\Lambda_w$  to the degree of walking.
The small value of $\Lambda_w$ in our walking situation suggests that the walking effect in the present model is not large enough. In an ideal walking situation, $\Lambda_w$  is large and can be treated as $\Lambda_{\mathrm{ETC}}$.

\begin{figure}[t]
\caption{$S$ parameters for the running and walking cases.} \label{Sr}
\hspace*{-7cm}\vspace*{-0.7cm}\begin{minipage}[t]{8cm}
    \includegraphics[scale=0.8]{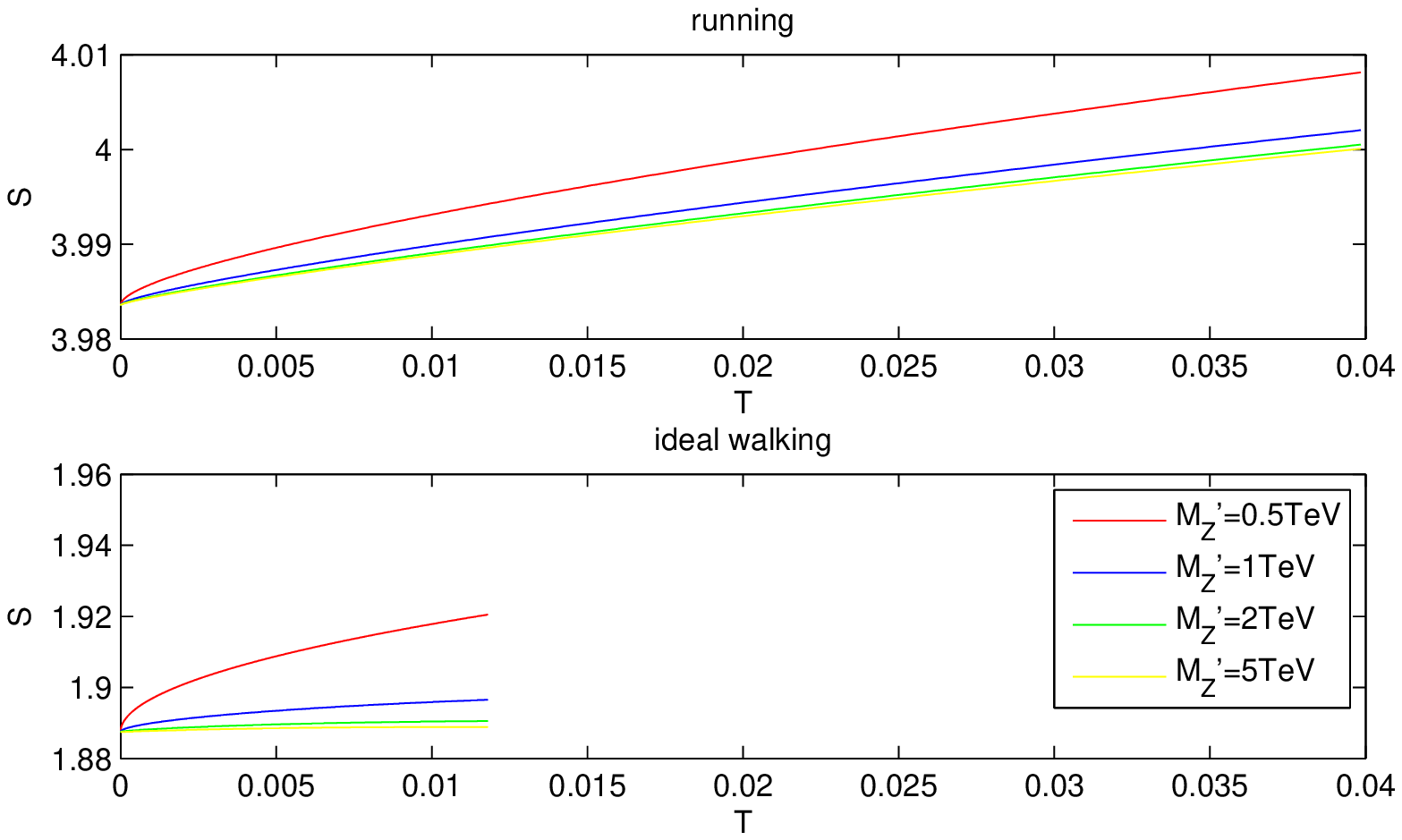}
\end{minipage}
\end{figure}

To show the effect of walking on the $S$ parameter, in Fig.\ref{Sr}, we show the value of $S$ for couplings corresponding to running and ideal walking. It can be seen that for ideal walking (the upper bound on $T$ is reduced to 0.012 in this case), $S$ is only slightly smaller than 2. Therefore, our prediction that $S$ is about 2 is not significantly altered, even as one approaches the walking region. However, Fig.\ref{Sr} shows that for running, $S$ is doubled by reaching a value of 4. This implies that because of the existence of the infrared fixed point, the walking only reduces the $S$ parameter by a factor of 2. Furthermore, comparing the values of the $S$ parameters at different couplings with their perturbative values $S_{\mathrm{pert}}=N_D*N/6\pi=9/\pi$, we found that the perturbative value of $S$ lies just between our realistic value and that of the running case.

For the effect of walking on the other EWCL LECs, our numerical calculation shows that for $\alpha_2,\alpha_3,\alpha_4$ walking reduces these LECs to roughly $65\%$ of their original values in the running case. $\alpha_5$, similar to the $S$ parameter, is reduced by the walking effect to half of its original value in the running case. $\alpha_6,\alpha_7,\alpha_9$ are reduced by one order of magnitude by the walking effect, but their signs are preserved. $\alpha_{10}$ is reduced by two orders of magnitude and changes in sign. Using the expression for $\alpha_{10}$ given by (\ref{alphaResult}), the numerical computation shows that some cancellations occur here. It is these cancellations that result in $\alpha_{10}$being the smallest among the EWCL LECs. Because of this cancellation, if the techniquark self-energy is changed, more sign changes may occur. This cancellation may reduce the reliability of our estimate of $\alpha_{10}$ and $\alpha_{10}$ may be seen as one of the limitations of the calculation for the approximations used. We found that not all LECs are sensitive to how close to ideal walking the theory is. The only major exception is $\alpha_{10}$.  Finally, we found that walking has almost no effect on the coloron mass. We interpret this to mean that the techniquark self-energy will change the value of the coloron mass significantly, but walking, which changes the form of the techniquark self-energy, does not have a large effect on the coloron mass. In fact, some quantities, such as $\Lambda_w$ are sensitive to this detailed form of the techniquark self-energy, but some other quantities, such as the coloron mass, are not.

\section{Summary}

In this paper, we discuss K. Lane's TC2 Model in the presence of nontrivial TC fermion condensation and walking. We focus on the walking effects in the model, which has not been discussed before. We also discuss the phase structure of the model in terms of the two-loop $\beta$ function of the TC coupling of the model. We found that to have both an infrared fixed point and spontaneous chiral symmetry breaking, the minimum $N$ for the TC group $SU(N)$ is $N=6$. This is the optimal choice because it is the value that is the most conformal that can be used in our model. Although this choice differs from the critical values, $N^c_{1,2}=5.42$ for the first and second sets of techniquarks and $N^c_3=4.93$ for the third set of techniquarks (Fig.\ref{fig-phase}), walking effects occur in the computed EWCL LECs. We can understand this explicit walking effect qualitatively through the relation, $N-N^c_i\ll N^c_i$ for $i=1,2,3$.  For $N=6$, using the technique used in our previous studies\cite{HongHao08,JunYi09,LangPLB},we derive the EWCL from Lane's model and calculate the EWCL LECs up to an order of $p^4$. We found that the primary contributions to the $p^4$ order coefficients arise from the three sets of techniquarks and $Z'$. There is no limit on the upper bound of the $Z'$ mass which differs from the TC2 models\cite{Hill95,Lane95,Sekhar} that we discussed previously. Moreover, all corrections from the $Z'$ particle are at least proportional to $\beta_1$ and vanish for a mixing of $\theta=0$.  It is especially important that the scale parameter, $\Lambda_w$, appears in the solution of the two-loop $\beta$. This signifies that the scale of walking cannot be assumed to be $\Lambda_{\mathrm{ETC}}$ in this model because, generally, $\Lambda_{\mathrm{TC}}\leq\Lambda_w\leq\Lambda_{\mathrm{ETC}}$. We found that $\Lambda_w=5.5$TeV. The value of $\Lambda_w$ is small because it is sensitive to the walking effect. However, our choice of $N$ differs from its critical value, and does not exhibit a sufficient walking effect. We verified that in a more ideal walking case, $\Lambda_w$ can be increased by at least two orders of magnitude. The ratio $(\Lambda_{\mathrm{ETC}}-\Lambda_w)/\Lambda_{\mathrm{ETC}}$ can be used as a measurement of the deviation of our theory from ideal walking. We also found that the coloron mass is roughly half of its expected value of 1 TeV and is independent of the walking effect.  The small coloron mass occurs as the result of including a correction from the coloron kinetic term for which the main contribution is from the techniquark self-energy. The $T$ and $U$ parameters are positive, and there is an upper bound for the $T$ parameter. For our choice of typical hyper-charges, the upper bound of the $T$ parameter is 0.035, which is well below the experimentally measured bound from PDG. The $S$ parameter is about 2 for our choice of typical hyper-charges, which already exceeds the experimentally verified constraint that it be half of the value from the running case, but similar to that of the ideal walking case. To reduce the value of the $S$ parameter, one can change hyper-charges. This can result in $S$ being negative for slightly larger values of $T$. This allows for a case in which both $S$ and $T$ are within the bound from PDG. The leftmost nine nonzero LECs, $\alpha_2,\alpha_3,\alpha_4,\alpha_5$ are on the order of $10^{-2}$ which matches the estimate obtained from naive dimensional analysis. $\alpha_6,\alpha_7,\alpha_9$ are on the order of $10^{-5}$ and $\alpha_{10}$ is on order of $10^{-10}$. This is because $\alpha_6,\alpha_7,\alpha_9$,and especially $\alpha_{10}$, are sensitive to walking effects. Comparing these results with the constraints imposed by the precision data\cite{precision}, we find that the results are consistent with the constraints from the precision data. However, $\alpha_3$ has the correct order of magnitude, but the wrong sign.

Previously, we investigated bosonic contribution to the EWCL LECs for most of the TC2 models. In the future, we will focus on calculating the EWCL LECs in four areas: The first will be to explore new physics models, including the non-TC2-type models. The second will be to investigate the part of the EWCL dealing with matter. In particular, we will focus on the top quark. The third will be to deepen our understanding of the structure of the model we are currently discussing in areas such as phase diagrams and the infrared behavior of the gauge coupling constant. The fourth will be to improve the precision of the computation and reduce the number of approximations necessary. With an increasing number of models in our EWCL platform, it will be effective for future investigations of the electroweak symmetry breaking mechanisms.

\section*{Acknowledgments}

This work was  supported by the National  Science Foundation of China
(NSFC) under Grants No. 10875065 and 11075085.

\appendix

\section{Process of integrating out the techniquarks}\label{InteOutTCq}

To integrate out the techniquarks, which we have done in previous studies\cite{HongHao08,JunYi09,LangPLB}, we assume only four fermion interactions in (\ref{action-TC1}), because a naive dimensional analysis indicates that the contributions from higher dimensional operators are usually suppressed in the low energy region. Also, this approximation leads to the conventional ladder approximation, which is often used in discussions of the SDE. This yields:
\begin{eqnarray}
&&\hspace*{-0.5cm}iS_{\mathrm{TC}}[\bar{T}_\xi,T_\xi,\bar{\psi},\psi]\approx\int d^4x_1d^4x_2\frac{(-ig_{\mathrm{TC}})^2}{2}G_{\mu_1\mu_2}^{\alpha_1\alpha_2}(x_1,x_2)
J_{\alpha_1}^{\mu_1}(x_1)J_{\alpha_2}^{\mu_2}(x_2)\nonumber\\
&&\hspace{-0.5cm}=-\frac{g^2_{\mathrm{TC}}}{2}\int d^4x_1d^4x_2G_{\mu_1\mu_2}^{\alpha_1\alpha_2}(x_1,x_2)\bigg[
\bar{\psi}(x_1)\tilde{t}^{\alpha_1}\gamma^{\mu_1}\psi(x_1)\bar{\psi}(x_2)\tilde{t}^{\alpha_2}\gamma^{\mu_2}\psi(x_2)\nonumber\\
&&+{\displaystyle\sum_{i,j=1,2,l,t,b}}\bar{T}^i_{\xi}(x_1)t^{\alpha_1}\gamma^{\mu_1}T^i_\xi(x_1)\bar{T}^j_{\xi}(x_2)
t^{\alpha_2}\gamma^{\mu_2}T^j_\xi(x_2)
+2{\displaystyle\sum_{i=1,2,l,t,b}}\bar{\psi}(x_1)\tilde{t}^{\alpha_1}\gamma^{\mu_1}\psi(x_1)\bar{T}^i_{\xi}(x_2)
t^{\alpha_2}\gamma^{\mu_2}T^i_\xi(x_2)\bigg]\nonumber\\
&&\hspace{-0.5cm}\approx\int d^4x_1d^4x_2\bigg[\bar{\psi}^\sigma(x_1)\tilde{\Pi}_{\sigma\rho}(x_1,x_2)\psi^\rho(x_2)
+{\displaystyle\sum_{i,j=1,2,l,t,b}}\bar{T}^{i\sigma}_{\xi}(x_1)\Pi^{ij}_{\sigma\rho}(x_1,x_2)\bar{T}^{j\rho}_{\xi}(x_2)\bigg]\;,\label{PiApprox}
\end{eqnarray}
where we have used (\ref{Jdef}) and (\ref{tildeJdef}). And
\begin{eqnarray}
\tilde{\Pi}_{\sigma\rho}(x_1,x_2)&\equiv&-g^2_{\mathrm{TC}}G_{\mu_1\mu_2}^{\alpha_1\alpha_2}(x_1,x_2)
\tilde{t}^{\alpha_1}\gamma^{\mu_1}_{\sigma\sigma_1}\langle\psi^{\sigma_1}(x_1)\bar{\psi}^{\rho_2}(x_2)\rangle\tilde{t}^{\alpha_2}\gamma^{\mu_2}_{\rho_2\rho}
\label{tildePidef}\\
\Pi^{ij}_{\sigma\rho}(x_1,x_2)&\equiv&-g^2_{\mathrm{TC}}G_{\mu_1\mu_2}^{\alpha_1\alpha_2}(x_1,x_2)
t^{\alpha_1}\gamma^{\mu_1}_{\sigma\sigma_1}\langle T^{i\sigma_1}(x_1)\bar{T}^{j\rho_2}(x_2){\rangle}t^{\alpha_2}\gamma^{\mu_2}_{\rho_2\rho}\;.
\label{Pidef}
\end{eqnarray}
To obtain (\ref{PiApprox}), we have used the average field approximation and approximated the four-fermion interactions using their vacuum expectation values (VEVs). Furthermore, we used the result: $\langle\bar{\psi}(x)\gamma^\mu\psi(x)\rangle=\langle\bar{T}^i(x)\gamma^{\mu}T^j(x)\rangle=0$,  which can be obtained from
the Lorentz invariance; $\langle\bar{\psi}(x)T^i(x)\rangle=\langle\bar{T}^i(x)\psi(x)\rangle=0$, which was assumed in Lane's original paper \cite{Lane96} and can be verified as a solution to the SDE. In fact, one can confirm that the VEVs between the different sets of techniquarks vanish and VEVs among the different techniquarks of the second set also vanish. For (\ref{PiApprox}), this yields:
\begin{eqnarray}
iS_{\mathrm{TC}}[\bar{T}_\xi,T_\xi,\bar{\psi},\psi]
&\approx&\int d^4x_1d^4x_2\bigg[\bar{\psi}^\sigma(x_1)\tilde{\Pi}_{\sigma\rho}(x_1,x_2)\psi^\rho(x_2)
+{\displaystyle\sum_{i,j=1,2}}\bar{T}^{i\sigma}_{\xi}(x_1)\bar{\Pi}^{ij}_{\sigma\rho}(x_1,x_2)T^{j\rho}_{\xi}(x_2)\nonumber\\
&&+{\displaystyle\sum_{i=l,t,b}}\bar{T}^{i\sigma}_{\xi}(x_1)\hat{\Pi}_{\sigma\rho}(x_1,x_2)T^{i\rho}_{\xi}(x_2)\bigg]
\label{PiApprox1}
\end{eqnarray}
with
\begin{eqnarray}
\Pi^{ij}_{\sigma\rho}(x_1,x_2)=\left\{\begin{array}{lll}\bar{\Pi}^{ij}_{\sigma\rho}(x_1,x_2)&~~~&i,j=1,2\\&&\\
\hat{\Pi}_{\sigma\rho}(x_1,x_2)&&i,j=l,t,b\end{array}\right.\;.\label{Pibarhat}
\end{eqnarray}
Therefore $\bar{\Pi}$, $\hat{\Pi}$ and $\tilde{\Pi}$ represent the fermion self-energies for the first, second, and third sets of techniquarks, respectively. Following the treatment in our previous studies\cite{HongHao08,JunYi09,LangPLB}, these techniquark self-energies can be approximated as:
\begin{eqnarray}
&&\hspace*{-0.5cm}\hat{\Pi}^{ij}_{\sigma\rho}(x,y)\approx-\delta_{\sigma\rho}[\hat{\Sigma}(\overline{\nabla}_x^2)\delta^4(x\!-\!y)]_{ij}
\hspace*{0.5cm}
\tilde{\Pi}_{\sigma\rho}(x,y)\approx-\delta_{\sigma\rho}\tilde{\Sigma}(\partial_x^2)\delta^4(x\!-\!y)\hspace*{0.5cm}
\overline{\nabla}^\mu\!=\!\partial^\mu\!-\!iV_{2\xi}^\mu~~~\label{PiApprox2}\\
&&\hspace*{-0.5cm}\bar{\Pi}^{ij}_{\sigma\rho}(x,y)\approx-[\delta_{\sigma\rho}\bar{\Sigma}(\hat{\nabla}_x^2)+i\gamma^5_{\sigma\rho}
\tau^2\bar{\Sigma}_5(\hat{\nabla}_x^2)]_{ij}\delta^4(x-y)
\hspace*{2.5cm}
\hat{\nabla}^\mu=\partial^\mu-iV_{1\xi}^\mu\bigg|_{v_1=0}\;,~~~~~\label{PiApprox3}
\end{eqnarray}
where $V_{2\xi}^\mu$, $V_{1\xi}^\mu$ and $v_1^\mu$ will be discussed later in the appendices. The above approximation is the lowest order of a dynamical perturbation originally proposed by Pagels and Stokar in Ref.\cite{DPT}. In this perturbation, all source dependent parts are expressed in terms of the techniquark self-energy and the detailed dependence is determined by including the minimal contribution that is covariant with the local chiral symmetry. An important result of this dynamical perturbation is that the lowest order, which includes the fermion loop terms, yields spontaneous chiral symmetry breaking and is dominated by the fermion self-energy. In our previous studies\cite{HongHao08,JunYi09,LangPLB}, the $\Pi$ functions are diagonal in the spinor space, but in this model, $\bar{\Pi}_{\sigma\rho}(x,y)$ in (\ref{PiApprox3}) differs from the conventional expression. In this case, there is an extra term ($\bar{\Sigma}_5$)  that is proportional to $\gamma^5$ and $\tau^2$ (in isospin space) because of the special model arrangement that generates nontrivial twisted TC fermion condensation. This condensation will stimulate topcolor symmetry breaking: $SU(3)_1\otimes SU(3)_2\rightarrow SU(3)_c$ and generate the coloron mass. Later,
we will discuss the appearance of this term and determine the functions corresponding to $\hat{\Sigma}$, $\tilde{\Sigma}$, $\bar{\Sigma}$ and $\bar{\Sigma}_5$ .

With the results from (\ref{PiApprox1})-(\ref{PiApprox3}), the techniquark interactions in (\ref{IntegralAfterRotation1}) become bilinear, and we can complete the integration over the techniquarks and obtain (\ref{SEW0}), which is given in the text. Where:
\begin{eqnarray}
  &&\hspace*{-0.5cm}V_{1\xi}=\begin{pmatrix}
    v_1\!+v_2-g_3\frac{\lambda^A}{2}B^A\cot\theta'& 0\\
    0 &v_1\!+v_2+g_3\frac{\lambda^A}{2}B^A\tan\theta'
  \end{pmatrix}
   \hspace*{0.7cm}A_{1\xi}=\begin{pmatrix}
    a_1\!+a_2&\\
    &a_1\!-a_2
  \end{pmatrix}~~~~~\label{V1A1}\\
  &&\hspace*{-0.5cm}V_{2\xi}=\begin{pmatrix}
    v_l&0&0\\
    0&v_t&0\\
    0&0&v_b
  \end{pmatrix}
  \hspace*{7.7cm}
  A_{2\xi}=\begin{pmatrix}
    a_l&0&0\\
    0&a_t&0\\
    0&0&a_b
  \end{pmatrix}\;.
\end{eqnarray}
The prime in $\mathrm{Tr'}$ denotes the trace of the extra $2\times 2$ space for the first two sets of techniquarks, and the double prime in $\mathrm{Tr"}$ denotes the trace of the extra $3\times 3$ space for the third set of techniquarks with:
\begin{eqnarray}
  &&\hspace*{-0.5cm}v_1=-\frac{1}{2}g_2\frac{\tau^a}{2}W_\xi^a
  -\frac{1}{2}g_1\frac{\tau^3}{2}(B_\xi-Z'\tan\theta)\nonumber\\
  &&\hspace*{-0.5cm}
  v_2=-\frac{1}{2}g_1(u_2+v_2)(B_\xi-Z'\tan\theta)-\frac{1}{2}g_1(u_1+v_1)(B_\xi+Z'\cot\theta)\\
  &&\hspace{-0.5cm}a_1=\frac{1}{2}g_2\frac{\tau^a}{2}W_\xi^a
  -\frac{1}{2}g_1\frac{\tau^3}{2}(B_\xi-Z'\tan\theta)\hspace*{2cm}
  a_2=\frac{1}{2}g_1 (u_1-v_1)(\cot\theta+\tan\theta)Z'~~~~
  \end{eqnarray}
  \begin{eqnarray}
  &&\hspace{-0.5cm}v_i=-\frac{1}{2}g_2\frac{\tau^a}{2}W_\xi^a
  -\frac{g_1}{2}\frac{\tau^3}{2}(B_\xi\!-\!Z'\tan\theta)
  -\frac{g_1}{2}(x^i_2\!+x^{i\prime}_2)(B_\xi\!-\!Z'\tan\theta)-\frac{g_1}{2}(x^{i}_1\!+x^{i\prime}_1)(B_\xi\!+\!Z'\cot\theta)\nonumber\\
  &&\hspace*{-0.5cm}a_i=\frac{1}{2}g_2\frac{\tau^a}{2}W_\xi^a
  -\frac{1}{2}g_1\frac{\tau^3}{2}(B_\xi-Z'\tan\theta)+\frac{1}{2}g_1 (x^i_1-x^{i\prime}_1)(\cot\theta+\tan\theta)Z'\hspace*{1.5cm}i=l,t,b\;.\label{altb}
\end{eqnarray}
We have used the relation
\begin{eqnarray}
&&\hspace*{-0.5cm}iq_1\xi{B}_{1\xi,\mu}P_L-iq_2\xi{B}_{2\xi,\mu}P_L+iq_1\xi_1'{B}_{1\xi,\mu}P_R-iq_2\xi'{B}_{2\xi,\mu}P_R
=-ig_1(\cot\theta+\tan\theta)\xi Z'_\mu\gamma^5\\
&&\hspace*{-0.5cm}-\!h_1\frac{\lambda^{A}}{2}\slashed{A}_1^A\!\!-\!g_2\frac{\tau^{a}}{2}\slashed{W}_\xi^{a}P_L\!\!
-\!q_1u_1\slashed{B}_{1\xi}P_L\!\!-\!q_2u_2\slashed{B}_{2\xi}P_L\!\!-\!q_1v_1\slashed{B}_{1\xi}P_R\!\!-\!q_2(v_2\!\!+\!\frac{\tau^3}{2})
\slashed{B}_{2\xi}P_{R}\nonumber\\
&&\hspace*{-0.5cm}=\slashed{v}_1\!+\slashed{v}_2-g_3\frac{\lambda^A}{2}\slashed{B}^A\cot\theta'+(\slashed{a}_1+\slashed{a}_2)\gamma^5\\
&&\hspace*{-0.5cm}-\!h_2\frac{\lambda^{A}}{2}\slashed{A}_2^A\!\!-\!g_2\frac{\tau^a}{2}\slashed{W}_\xi^aP_L\!\!
-\!q_1v_1\slashed{B}_{1\xi}P_L\!\!-\!q_2v_2\slashed{B}_{2\xi}P_L\!\!-\!q_1u_1\slashed{B}_{1\xi}P_R\!\!
-\!q_2(u_2\!\!+\!\frac{\tau^3}{2})\slashed{B}_{2\xi}P_R\nonumber\\
&&\hspace*{-0.5cm}=\slashed{v}_1\!+\slashed{v}_2+g_3\frac{\lambda^A}{2}\slashed{B}^A\tan\theta'+(\slashed{a}_1-\slashed{a}_2)\gamma^5\\
&&\hspace*{-0.5cm}-g_2\frac{\tau^a}{2}\slashed{W}_\xi^aP_L\!-q_1x_1\slashed{B}_{1\xi}P_L\!-q_2x_2\slashed{B}_{2\xi}P_L\!
-q_1x_1^{\prime}\slashed{B}_{1\xi}P_R\!-q_2(x_2^{\prime}\!+\!\frac{\tau^3}{2})\slashed{B}_{2\xi}P_R=\slashed{v}_l+\slashed{a}_l\gamma^5\\
&&\hspace*{-0.5cm}-g_2\frac{\tau^a}{2}\slashed{W}_\xi^aP_L\!-q_1y_1\slashed{B}_{1\xi}P_L\!-q_2y_2\slashed{B}_{2\xi}P_L\!
-q_1y_1^{\prime}\slashed{B}_{1\xi}P_R\!-q_2(y_2^{\prime}\!+\!\frac{\tau^3}{2})\slashed{B}_{2\xi}P_R=\slashed{v}_t+\slashed{a}_t\gamma^5\\
&&\hspace*{-0.5cm}-g_2\frac{\tau^a}{2}\slashed{W}_\xi^aP_L\!-q_1z_1\slashed{B}_{1\xi}P_L\!-q_2z_2\slashed{B}_{2\xi}P_L\!
-q_1z_1^{\prime}\slashed{B}_{1\xi}P_R\!-q_2(z_2^{\prime}\!+\!\frac{\tau^3}{2})\slashed{B}_{2\xi}P_R=\slashed{v}_t+\slashed{a}_t\gamma^5\;.
\end{eqnarray}

\section{Derivation of the Schwinger-Dyson equations for the techniquark self-energies}\label{DerSDE}

In this appendix, we derive the SDE for the techniquark self-energies. We start from the path integral given in (\ref{IntegralAfterRotation1}), and fix the functional integration over the $U$, $B^A_\mu$ and $Z'_\mu$ fields. The total functional derivative of the integrand with respect to $\bar{\psi}$
 and $\bar{T}_\xi^i$ is zero, which yields:
\begin{eqnarray}
0&=&\int\mathcal{D}\mu(\psi,T)~\frac{\delta}{\delta \bar{\psi}^\sigma(x)}e^{iS_{\mathrm{TC}}+iS_{\mathrm{TC1}}+iS_{\mathrm{source}}}\bigg|_{A^A_\mu=0}
\label{SDEpsi0}\\
0&=&\int\mathcal{D}\mu(\psi,T)
\frac{\delta}{\delta \bar{T}^{i,\sigma}_\xi(x)}e^{iS_{\mathrm{TC}}+iS_{\mathrm{TC1}}+iS_{\mathrm{source}}}\bigg|_{A^A_\mu=0}
\label{SDET0}\\
\mathcal{D}\mu(\psi,T)&\equiv&\mathcal{D}\bar{\psi}\mathcal{D}\psi\mathcal{D}\bar{T}^1_\xi\mathcal{D}T^1_\xi\mathcal{D}\bar{T}^2_\xi\mathcal{D}T^2_\xi
\mathcal{D}\bar{T}^l_\xi\mathcal{D}T^l_\xi\mathcal{D}\bar{T}^t_\xi\mathcal{D}T^t_\xi\mathcal{D}\bar{T}^b_\xi\mathcal{D}T^b_\xi\;,
\end{eqnarray}
In this case, we have introduced source terms with external sources $\bar{I}$ and $\bar{J}$ to help to derive the SDEs:
\begin{eqnarray}
 iS_{\mathrm{source}}=\int d^4x\bigg[\bar{\psi}(x)I(x)+{\displaystyle\sum_{i=1,2,l,t,b}}\bar{T}^i(x)J^i(x)\bigg]\;.
 \end{eqnarray}
 We derive $I^\rho(y)$ for both sides of (\ref{SDEpsi0}) and remove all external sources. We obtain:
\begin{eqnarray}
&&\hspace*{-0.5cm}0=S_{\psi\sigma\rho}^{-1}(x,y)+i[i\slashed{\partial}_x\!+g_1(\cot\theta\!+\!\tan\theta)\xi\slashed{Z}'\gamma^5]_{\sigma\rho}
\delta(x\!-\!y)
-g_{\mathrm{TC}}^2G_{\mu_1\mu_2}^{\alpha_1\alpha_2}(x,y)[\tilde{t}^{\alpha_1}\gamma^{\mu_1}S(x,y)\tilde{t}^{\alpha_2}\gamma^{\mu_2}]_{\sigma\rho}\nonumber\\
&&\label{SDEpsi1}\\
&&\hspace*{-0.5cm}S_{\psi\sigma\rho}(x,y)\equiv \langle
\psi^\sigma(x)\bar{\psi}^\rho(y)\rangle=
\frac{\int\mathcal{D}\mu(\psi,T)~\psi^\sigma(x)\bar{\psi}^\rho(y)~e^{iS_{\mathrm{TC}}+iS_{\mathrm{TC1}}}}
{\int\mathcal{D}\mu(\psi,T)~e^{iS_{\mathrm{TC}}+iS_{\mathrm{TC1}}}}\bigg|_{A^A_\mu=0}\;.
\end{eqnarray}
(\ref{SDEpsi1}) is the SDE in coordinate space for the third set of techniquarks. Combining (\ref{tildePidef}) and (\ref{SDEpsi1}), we find that $S_{\psi\sigma\rho}(x,y)$, which is determined by the SDE, relates to $\tilde{\Pi}_{\sigma\rho}(x,y)$, introduced in (\ref{tildePidef}), through:
\begin{eqnarray}
&&\hspace*{-0.5cm}0=S_{\psi\sigma\rho}^{-1}(x,y)+i[i\slashed{\partial}_x\!+g_1(\cot\theta\!+\!\tan\theta)\xi\slashed{Z}'\gamma^5]_{\sigma\rho}
\delta(x\!-\!y)+\tilde{\Pi}_{\sigma\rho}(x,y)=0\;.\label{SDEpsi2}
\end{eqnarray}
Similarly we derive $J^{j\rho}(y)$ for both sides of (\ref{SDET0}), and remove all external sources, We obtain:
\begin{eqnarray}
&&\hspace*{-0.5cm}0=S_{T\sigma\rho}^{ij,-1}(x,y)+i[i\slashed{\partial}_x\!+\!\slashed{V}_{1\xi}\!+\!\slashed{A}_{1\xi}\gamma^5]^{ij}_{\sigma\rho}
\delta(x\!-\!y)
-g_{\mathrm{TC}}^2G_{\mu_1\mu_2}^{\alpha_1\alpha_2}(x,y)[t^{\alpha_1}\gamma^{\mu_1}S(x,y)t^{\alpha_2}\gamma^{\mu_2}]^{ij}_{\sigma\rho}\nonumber\\
&&\hspace{11cm}i,j=1,2\label{SDET1}\\
&&\hspace*{-0.5cm}0=S_{T\sigma\rho}^{ij,-1}(x,y)+i[i\slashed{\partial}_x\!+\!\slashed{V}_{2\xi}\!+\!\slashed{A}_{2\xi}\gamma^5]^{ij}_{\sigma\rho}
\delta(x\!-\!y)
-g_{\mathrm{TC}}^2G_{\mu_1\mu_2}^{\alpha_1\alpha_2}(x,y)[t^{\alpha_1}\gamma^{\mu_1}S(x,y)t^{\alpha_2}\gamma^{\mu_2}]^{ij}_{\sigma\rho}\nonumber\\
&&\hspace*{11cm}i,j=l,t,b\;,\label{SDET2}
\end{eqnarray}
where
\begin{eqnarray}
S^{ij}_{T\sigma\rho}(x,y)\equiv \langle
T^{i\sigma}(x)\bar{T}^{j\rho}(y)\rangle=
\frac{\int\mathcal{D}\mu(\psi,T)~T^{i\sigma}(x)\bar{T}^{j\rho}(y)~e^{iS_{\mathrm{TC}}+iS_{\mathrm{TC1}}}}
{\int\mathcal{D}\mu(\psi,T)~e^{iS_{\mathrm{TC}}+iS_{\mathrm{TC1}}}}\bigg|_{A^A_\mu=0}\;.
\end{eqnarray}
(\ref{SDET1}) and (\ref{SDET2}) are the SDEs in the coordinate space of the first and second sets of techniquarks. Combining (\ref{Pidef}),  (\ref{Pibarhat}), (\ref{SDET1}) and (\ref{SDET2}), we find that $S^{ij}_{T\sigma\rho}(x,y)$ which is determined by the SDE, relates to $\bar{\Pi}^{ij}_{\sigma\rho}(x,y)$ and $\hat{\Pi}^{ij}_{\sigma\rho}(x,y)$, introduced in (\ref{Pidef}) and (\ref{Pibarhat}), through:
\begin{eqnarray}
&&\hspace*{-0.5cm}0=S_{T\sigma\rho}^{ij,-1}(x,y)+i[i\slashed{\partial}_x\!+\!\slashed{V}_{1\xi}\!+\!\slashed{A}_{1\xi}\gamma^5]^{ij}_{\sigma\rho}
\delta(x\!-\!y)+\bar{\Pi}^{ij}_{\sigma\rho}(x,y)
\hspace{1cm}i,j=1,2\label{SDET3}\\
&&\hspace*{-0.5cm}0=S_{T\sigma\rho}^{ij,-1}(x,y)+i[i\slashed{\partial}_x\!+\!\slashed{V}_{2\xi}\!+\!\slashed{A}_{2\xi}\gamma^5]^{ij}_{\sigma\rho}
\delta(x\!-\!y)+\hat{\Pi}^{ij}_{\sigma\rho}(x,y)
\hspace*{1cm}i,j=l,t,b\;.\label{SDET4}
\end{eqnarray}
Following the treatment in our previous works \cite{HongHao08,JunYi09,LangPLB}, the techniquark self-energies $\hat{\Sigma}$ and $\tilde{\Sigma}$ in (\ref{PiApprox2}) and $\bar{\Sigma}$, $\bar{\Sigma}_5$ in (\ref{PiApprox3}) are determined by removing the gauge fields in the SDEs. Using this approximation, we find the three sets of techniquarks:
\begin{eqnarray}
&&\hspace*{-0.5cm}S_{\psi\sigma\rho}(x,y)=\!\int\!\frac{d^4p}{(2\pi)^4}e^{-ip(x-y)}
\bigg[\frac{i}{\slashed{p}\!-\!\tilde{\Sigma}(-p^2)}\bigg]_{\sigma\rho}
\hspace*{1cm}S_{T\sigma\rho}^{ij}(x,y)=\!\int\!\frac{d^4p}{(2\pi)^4}e^{-ip(x-y)}
\bigg[\frac{i\delta_{ij}}{\slashed{p}\!-\!\hat{\Sigma}(-p^2)}\bigg]_{\sigma\rho}
\nonumber\\
&&\hspace*{9.7cm}i,j=l,t,b\\
&&\hspace*{-0.5cm}S_{T\sigma\rho}^{ij}(x,y)=\int\frac{d^4p}{(2\pi)^4}e^{-ip(x-y)}
\bigg[\frac{i}{\slashed{p}-\bar{\Sigma}(-p^2)-i\gamma_5\tau^2\bar{\Sigma}_5(-p^2)}\bigg]^{ij}_{\sigma\rho}\hspace*{2cm}i,j=1,2\;,
\end{eqnarray}
In Euclidean space, we obtain(\ref{SDEtildeSigma}), (\ref{SDEhatSigma}), (\ref{SDEbarSigma}) and (\ref{SDEbarSigma5})in the main text.

In terms of $\hat{\Sigma}$, comparing(\ref{SDEhatSigma}) with (\ref{SDEbarSigma}) and (\ref{SDEbarSigma5}),we can construct $\bar{\Sigma}$ and $\bar{\Sigma}_5$ as follows:
\begin{eqnarray}
\bar{\Sigma}(p_E^2)=\hat{\Sigma}(p_E^2)\cos\Theta\hspace*{2cm}\bar{\Sigma}_5(p_E^2)=\hat{\Sigma}(p_E^2)\sin\Theta\;.\label{ThetaDef}
\end{eqnarray}
$\Theta$  at the present stage in the computation is an arbitrary constant, and we have verified that the vacuum energy generated by $\bar{\Sigma}$ and $\bar{\Sigma}_5$ only depends on $\bar{\Sigma}^2+\bar{\Sigma}^2_5=\hat{\Sigma}^2$, which is independent of $\Theta$. Later we show that the coloron mass is dependent on $\Theta$ and the present model gives a relatively small coloron mass (several hundred GeV). In practice, we use the value of $\Theta$ which offers the largest coloron mass. Once nonzero techniquark self-energies are present, we will have nonzero techniquark condensates:
\begin{eqnarray}
&&\hspace*{-0.5cm}\langle\bar{T}^i_L(x)T^j_R(x)\rangle=-2N\!\int\!\frac{d^4p_E}{(2\pi)^4}\bigg[\frac{\delta_{ij}\bar{\Sigma}(p_E^2)}{p_E^2\!
+\!\bar{\Sigma}^2(p_E^2)\!+\!\bar{\Sigma}_5^2(p_E^2)}-\frac{i\tau^2_{ij}\bar{\Sigma}_5(p_E^2)}{p_E^2\!
+\!\bar{\Sigma}^2(p_E^2)\!+\!\bar{\Sigma}_5^2(p_E^2)}\frac{p_E^2\!-\!\bar{\Sigma}^2(p_E^2)}{p_E^2\!+\!\bar{\Sigma}^2(p_E^2)}\bigg]\nonumber\\
&&\hspace*{12.7cm}i,j=1,2,\\
&&\hspace{-0.5cm}\langle\bar{T}^i_L(x)T^j_R(x)\rangle=-2N\delta_{ij}\int\frac{d^4p_E}{(2\pi)^4}\frac{\hat{\Sigma}(p_E^2)}{p_E^2
+\hat{\Sigma}^2(p_E^2)}\hspace*{5cm}i,j=l,t,b,\\
&&\hspace*{-0.5cm}\langle\bar{\psi}_{L}(x)\psi_R(x)\rangle=-N(N-1)\int\frac{d^4p_E}{(2\pi)^4}\frac{\tilde{\Sigma}(p_E^2)}{p_E^2
+\tilde{\Sigma}^2(p_E^2)}\;.
\end{eqnarray}
Note that the first techniquark set has a nontrivial twisted condensation: $\langle\bar{T}^1_L(x)T^2_R(x)\rangle=-\langle\bar{T}^2_L(x)T^1_R(x)\rangle\neq 0$ resulting from the nonzero self-energies.

\section{ Integrating out the colorons and the low energy expansion}\label{LEexp}

The coefficients in (\ref{coloronAction}) are,
\begin{eqnarray}
&&\hspace*{-0.5cm}C=\int d^4\tilde{k}[-2\tau+\tau^2k_E^2+16\tau^2\bar{\Sigma}^2_5]\\
&&\hspace*{-0.5cm}\mathcal{K}=-\frac{1}{48\pi^2}[\ln\frac{\kappa^2}{\Lambda^2}+\gamma]\hspace*{2cm}\kappa,\Lambda~\mbox{: infrared and ultraviolet cutoffs}\label{kappDef}\\
&&\hspace*{-0.5cm}\hat{E}=\int d^4\tilde{k}[\tau^2+16\tau^2\bar{\Sigma}_5\bar{\Sigma}_5^\prime
+4\tau^2k_E^2\bar{\Sigma}^{\prime2}+8\tau^2k_E^2\bar{\Sigma}_5\bar{\Sigma}_5^{\prime\prime}
+4\tau^2k_E^2\bar{\Sigma}_5^{\prime2}-\frac{1}{3}\tau^3k_E^2-\frac{16}{3}\tau^3\bar{\Sigma}_5^2
\notag\\
&&\hspace*{1cm}-\frac{2}{3}\tau^3k_E^2\bar{\Sigma}\bar{\Sigma}^\prime-6\tau^3k_E^2\bar{\Sigma}_5\bar{\Sigma}_5^\prime
-\frac{32}{3}\tau^3\bar{\Sigma}\bar{\Sigma}^\prime\bar{\Sigma}_5^2
-\frac{32}{3}\tau^3\bar{\Sigma}_5^3\bar{\Sigma}_5^\prime
-\frac{2}{9}\tau^3k_E^4\bar{\Sigma}\bar{\Sigma}^{\prime\prime}
-\frac{2}{9}\tau^3k_E^4\bar{\Sigma}^{\prime2}\notag\\
&&\hspace*{1cm}-\frac{2}{9}\tau^3k_E^4\bar{\Sigma}_5\bar{\Sigma}_5^{\prime\prime}
-\frac{2}{9}\tau^3k_E^4\bar{\Sigma}_5^{\prime2}
-\frac{32}{3}\tau^3k_E^2\bar{\Sigma}\bar{\Sigma}^\prime\bar{\Sigma}_5\bar{\Sigma}_5^\prime
-\frac{16}{3}\tau^3k_E^2\bar{\Sigma}\bar{\Sigma}^{\prime\prime}\bar{\Sigma}_5^2
-\frac{16}{3}\tau^3k_E^2\bar{\Sigma}^{\prime2}\bar{\Sigma}_5^2\notag\\
&&\hspace*{1cm}-16\tau^3k_E^2\bar{\Sigma}_5^2\bar{\Sigma}_5^{\prime2}
+\frac{1}{18}\tau^4k_E^4+\frac{4}{3}\tau^4k_E^2\bar{\Sigma}_5^2
+\frac{2}{9}\tau^4k_E^4\bar{\Sigma}\bar{\Sigma}^\prime
+\frac{2}{9}\tau^4k_E^4\bar{\Sigma}_5\bar{\Sigma}_5^\prime
+\frac{16}{3}\tau^4k_E^2\bar{\Sigma}\bar{\Sigma}^\prime\bar{\Sigma}_5^2\notag\\
&&\hspace*{1cm}+\frac{16}{3}\tau^4k_E^2\bar{\Sigma}_5^3\bar{\Sigma}^\prime
+\frac{2}{9}\tau^4k_E^4\bar{\Sigma}^2\bar{\Sigma}^{\prime2}
+\tau^4k_E^4\bar{\Sigma}\bar{\Sigma}^\prime\bar{\Sigma}_5\bar{\Sigma}_5^\prime
+\tau^4k_E^4\bar{\Sigma}_5^2\bar{\Sigma}_5^{\prime2}\notag\\
&&\hspace*{1cm}+\frac{16}{3}\tau^4k_E^2\bar{\Sigma}^2\bar{\Sigma}^{\prime2}\bar{\Sigma}_5^2
+\frac{32}{3}\tau^4k_E^2\bar{\Sigma}\bar{\Sigma}^\prime\bar{\Sigma}_5^3\bar{\Sigma}_5^\prime
+\frac{16}{3}\tau^4k_E^2\bar{\Sigma}_5^4\bar{\Sigma}_5^{\prime2}]\\
&&\hspace*{-0.5cm}\int d^4\tilde{k}=N\int^{\infty}_{\frac{1}{\Lambda^2}}\frac{d\tau}{\tau} \int\frac{d^4k_E}{(2\pi)^4}e^{-\tau[k_E^2+\bar{\Sigma}^2(k_E^2)]},\hspace*{1cm}\bar{\Sigma}=\bar{\Sigma}(k_E^2),
\hspace*{1cm}\bar{\Sigma}_5=\bar{\Sigma}_5(k_E^2)\;,
\end{eqnarray}
Where $\hat{\mathcal{K}}_{13}^{\Sigma\neq0}$ are the coefficients that are introduced later in (\ref{Trlnpexp}), $\Lambda$ is a cutoff that is not sensitive to changes for values between 10 TeV and 100 TeV for our walking theory. In our practical calculation, we set it to 40 TeV. Combining the standard coloron kinetic term in (\ref{SEW0}) and the techniquark quantum loop correction given by (\ref{coloronAction}), we obtain the formula for the coloron mass (\ref{coloronmass}) given in the text.
With the coloron mass from (40), we can discuss coloron field integration in (\ref{coloronmass}), we then discuss coloron field integration in (\ref{SEW0}). This can be achieved using the standard loop expansion:
\begin{eqnarray}
&&\hspace*{-0.5cm}\int\mathcal{D}B_\mu^A~\exp\bigg[i\int d^4x[-\frac{1}{4}(A_{1\mu\nu}^{A}A^{A,1\mu\nu}
+A_{2\mu\nu}^{A}A^{A,2\mu\nu}+W_{\mu\nu}^aW^{a,\mu\nu}+B_{1,\mu\nu}B^{1,\mu\nu}+B_{2,\mu\nu}B^{2,\mu\nu})]\nonumber\\
&&\hspace*{-0.5cm}+\mathrm{Trln}[i\slashed{\partial}+g_1(\cot\theta\!+\tan\theta)\xi\slashed{Z}^\prime\gamma^5-\tilde{\Sigma}(\partial^2)]
+\mathrm{Tr"ln}[i\slashed{\partial}+\slashed{V}_{2\xi}\!+\slashed{A}_{2\xi}\gamma^5\!-\hat{\Sigma}(\overline{\nabla}^2)]\nonumber\\
&&\hspace*{-0.5cm}+\mathrm{Tr'ln}[i\slashed{\partial}+\!\slashed{V}_{1\xi}\!+\slashed{A}_{1\xi}\gamma^5\!
-\bar{\Sigma}(\hat{\nabla}^2)\!-i\gamma_5\tau^2\bar{\Sigma}_5(\hat{\nabla}^2)]\bigg]_{A^A_\mu=0}\nonumber\\
&&\hspace*{-0.5cm}=\exp\bigg[i\int d^4x[-\frac{1}{4}(A_{1\mu\nu}^{A}A^{A,1\mu\nu}
+A_{2\mu\nu}^{A}A^{A,2\mu\nu}+W_{\mu\nu}^aW^{a,\mu\nu}+B_{1,\mu\nu}B^{1,\mu\nu}+B_{2,\mu\nu}B^{2,\mu\nu})]\nonumber\\
&&+\mathrm{Trln}[i\slashed{\partial}+g_1(\cot\theta\!+\tan\theta)\xi\slashed{Z}^\prime\gamma^5-\tilde{\Sigma}(\partial^2)]
+\mathrm{Tr"ln}[i\slashed{\partial}+\slashed{V}_{2\xi}\!+\slashed{A}_{2\xi}\gamma^5\!-\hat{\Sigma}(\overline{\nabla}^2)]\nonumber\\
&&+\mathrm{Tr'ln}[i\slashed{\partial}+\!\slashed{V}_{1\xi}\!+\slashed{A}_{1\xi}\gamma^5\!
-\bar{\Sigma}(\hat{\nabla}^2)\!-i\gamma_5\tau^2\bar{\Sigma}_5(\hat{\nabla}^2)]+\mbox{loop corrections}\bigg]_{A^A_\mu=0,B^A_\mu=B^A_{\mu,c}}\;.
\end{eqnarray}
And $B^A_{\mu,c}$ is determined by requiring that the result reach its extremum at $B^A_\mu=B^A_{\mu,c}$. One can show that $B^A_{\mu,c}=0$ is one solution. Consequently, (\ref{SEW0}) becomes:
 \begin{eqnarray}
e^{iS_{\mathrm{EW}}[W_\mu^a,B_\mu]}
&=&e^{i\int d^4x[-\frac{1}{4}W_{\mu\nu}^aW^{a,\mu\nu}-\frac{1}{4}B_{\mu\nu}B^{\mu\nu}]}\int \mathcal{D}_\mu(U)\mathcal{F}[O_\xi]\delta(O_\xi-O^\dag_\xi)\int\mathcal{D}Z_\mu^\prime\nonumber\\
&&\exp\bigg[i\int d^4x[-\frac{1}{4}Z'_{\mu\nu}Z^{\prime\mu\nu}]+\mathrm{Trln}[i\slashed{\partial}+g_1(\cot\theta\!
+\tan\theta)\xi\slashed{Z}^\prime\gamma^5-\tilde{\Sigma}(\partial^2)]\nonumber\\
&&+\mathrm{Tr'ln}[i\slashed{\partial}+\!\slashed{V}_{1\xi}\!+\slashed{A}_{1\xi}\gamma^5\!
-\bar{\Sigma}(\hat{\nabla}^2)\!-i\gamma_5\tau^2\bar{\Sigma}_5(\hat{\nabla}^2)]\nonumber\\
&&+\mathrm{Tr"ln}[i\slashed{\partial}+\slashed{V}_{2\xi}\!+\slashed{A}_{2\xi}\gamma^5\!-\hat{\Sigma}(\overline{\nabla}^2)]+
\mbox{loop corrections}\bigg]_{A^A_\mu=B^A_\mu=0}\;.\label{SEW1}
\end{eqnarray}
Note that we are interested in the bosonic part of the EWCL, those operators involve explicit top quark fields, which belong to the part of the EWCL dealing with matter,  are beyond the scope of this paper. The top quark loop term (especially the top quark condensate) is expected to essentially contribute only to the top quark mass and not to the W and Z masses in TC2 models. This suggests that the contribution from top quark condensation to the bosonic part of the EWCL may also be small (we will show this in the future in a separate paper). Consequently, colorons, which are important in the formation of top-quark condensates and contribute the majority of the top-quark mass, only play a passive role in our present calculations. From (\ref{A1A2-AB}),the requirement, $A^A_\mu=B^A_\mu=0$ in (\ref{SEW1})  is equivalent to the requirement, $A^A_{1\mu}=A^A_{2\mu}=0$.

Now, with the help of a technique used in our previous studies\cite{HongHao08,JunYi09,LangPLB}, we take low energy expansion for the three TrLn terms in (\ref{SEW1}):
\begin{eqnarray}
&&\hspace*{-0.5cm}\mathrm{Trln}[i\slashed{\partial}+g_1(\cot\theta\!
+\tan\theta)\xi\slashed{Z}^\prime\gamma^5-\tilde{\Sigma}(\partial^2)]\bigg|_{\mathrm{normal~part}}\label{Trlnexp}\\
&&\hspace*{-0.5cm}=i\int d^4x(\cot\theta\!+\!\tan\theta)^2\bigg[\tilde{F}_0^2g_1^2\xi^{2}Z^{\prime
2}-(\mathcal{K}+\tilde{\mathcal{K}}_2^{\Sigma\neq
0})g_1^{2}\xi^{2}{Z}^\prime_{\mu\nu}{Z}^{\prime\mu\nu}-\tilde{\mathcal{K}}_1^{\Sigma\neq
0}g_1^{2}\xi^{2}(\partial^{\mu}Z_{\mu}^{\prime})^2\nonumber\\
&&+(\tilde{\mathcal{K}}_3^{\Sigma\neq 0}+\tilde{\mathcal{K}}_4^{\Sigma\neq
0})g_1^{4}(\cot\theta +\tan\theta)^2\xi^{4}Z^{\prime4}\bigg]+O(p^6)\nonumber
\end{eqnarray}
\begin{eqnarray}
&&\hspace*{-0.5cm}\mathrm{Tr'ln}[i\slashed{\partial}+\!\slashed{V}_{1\xi}\!+\slashed{A}_{1\xi}\gamma^5\!
-\bar{\Sigma}(\hat{\nabla}^2)\!-i\gamma_5\tau^2\bar{\Sigma}_5(\hat{\nabla}^2)]\bigg|_{\mathrm{normal~part}}\label{Trlnpexp}\\
&&\hspace*{-0.5cm}=i\int d^4x\bigg\{\hat{F}_0^2A_{1\xi}^2-8F^{\prime2}_0g_1^2u^2(\cot\theta+\tan\theta)^2Z^{\prime 2}-\frac{1}{2}\mathcal{K}\bigg[
g_2^2W^a_{\mu\nu}W^{a\mu\nu}+g_1^2[1+4(u_1+u_2)^2\nonumber\\
&&+4(v_1+v_2)^2]B_{\mu\nu}B^{\mu\nu}+g_1^2[4(u_2\tan\theta-u_1\cot\theta)^2
+4(v_2\tan\theta-v_1\cot\theta)^2+\tan^2\theta\nonumber\\
&&+4\hat{D}_0u^2(\cot\theta+\tan\theta)^2]Z'_{\mu\nu}{Z'}^{\mu\nu}
-2g_1^2[4(u_1+u_2)(u_2\tan\theta-u_1\cot\theta)\nonumber\\
&&+4(v_1+v_2)(v_2\tan\theta-v_1\cot\theta)+\tan\theta]B_{\mu\nu}{Z'}^{\mu\nu}\bigg]
+\mathrm{tr}\bigg[-\hat{\mathcal{K}}_1^{\Sigma\neq0}(d_{\mu}A_{1\xi}^\mu)^2+\hat{\mathcal{K}}_3^{\Sigma\neq0}(A_{1\xi}^2)^2\nonumber\\
&&-\hat{\mathcal{K}}_2^{\Sigma\neq0}(d_{\mu}A_{1\xi\nu}-d_{\nu}A_{1\xi\mu})^2
+\hat{\mathcal{K}}_4^{\Sigma\neq0}(A_{1\xi\mu}A_{1\xi\nu})^2
-\hat{\mathcal{K}}_{13}^{\Sigma\neq0}V_{1\xi\mu\nu}V^{\mu\nu}_{1\xi}
+i\hat{\mathcal{K}}_{14}^{\Sigma\neq0}V_{1\xi\mu\nu}A_{1\xi}^{\mu}A_{1\xi}^\nu\bigg]\nonumber\\
&&-8[\hat{D}_1a_0^4+\hat{D}_2a_0^2a_3^2]Z^{\prime4}+\hat{D}_3a_0^2Z^{\prime2}\mathrm{tr}(X^\mu X_\mu)+2\hat{D}_4a_0^2Z'_\mu Z'_\nu\mathrm{tr}(X^\mu X^\nu)\nonumber\\
&&+4i\hat{D}_2a_0^2a_3Z^{\prime2}Z'_\mu\mathrm{tr}(X^\mu\tau^3)\bigg\}+O(p^6)\nonumber
\end{eqnarray}
\begin{eqnarray}
&&\hspace*{-0.5cm}\mathrm{Tr"ln}[i\slashed{\partial}+\slashed{V}_{2\xi}\!+\slashed{A}_{2\xi}\gamma^5\!-\hat{\Sigma}(\overline{\nabla}^2)]
\bigg|_{\mathrm{normal~part}}\label{Trlnppexp}\\
&&\hspace*{-0.5cm}=i\int d^4x\sum_{\eta=l,t,b}\mathrm{tr}_f\bigg[\hat{F}_0^2a^{\eta2}-\hat{\mathcal{K}}_1^{\Sigma\neq0}(d_\mu
a^{\eta\mu})^2-\hat{\mathcal{K}}_2^{\Sigma\neq0}(d_\mu a_{\nu}^{\eta}-d_\nu a_{\mu }^{\eta})^2
+\hat{\mathcal{K}}_3^{\Sigma\neq0}(a^{\eta2})^2+\hat{\mathcal{K}}_4^{\Sigma\neq0}(a_{\mu}^{\eta} a_{\nu }^{\eta})^2\nonumber\\
&&-\hat{\mathcal{K}}_{13}^{\Sigma\neq0}v_{\mu\nu }^{\eta}v^{\eta\mu\nu}+i\hat{\mathcal{K}}_{14}^{\Sigma\neq0}a_{\mu }^{\eta} a_{\nu}^{\eta} v^{\eta\mu\nu}\bigg]+O(p^6)\;,\nonumber
\end{eqnarray}
where
\begin{eqnarray}
&&\hspace*{-0.5cm}d_{\mu}A_{1\xi\nu}=\partial_\mu A_{1\xi\nu}-i[V_{1\xi\mu},A_{1\xi\nu}]\hspace*{2cm}
V_{1\xi\mu\nu}=\partial_\mu V_{1\xi\nu}-\partial_\nu V_{1\xi\mu}-i[V_{1\xi\mu},V_{1\xi\nu}]\\
&&\hspace*{-0.5cm}d_{\mu}a^\eta_{\nu}=\partial_\mu a^\eta_\nu-i[v^\eta_\mu,a^\eta_\nu]\hspace*{3.8cm}
v^\eta_{\mu\nu}=\partial_\mu v^\eta_\nu-\partial_\nu v^\eta_\mu-i[v^\eta_\mu,v^\eta_\nu]\\
&&\hspace*{-0.5cm}F^{\prime2}_0=\int d^4\tilde{k}~2\tau\bar{\Sigma}^2_5\\
&&\hspace{-0.5cm}\hat{D}_0=\int d^4\tilde{k}
~[2\tau^2\bar{\Sigma}_5 \bar{\Sigma}^\prime_5
+\tau^2k_E^2\bar{\Sigma}_5 \bar{\Sigma}^{\prime\prime}_5
-\frac{2}{3}\tau^3\bar{\Sigma}^2_5
-\frac{2}{3}\tau^3k_E^2\bar{\Sigma}_5 \bar{\Sigma}^\prime_5
-\frac{4}{3}\tau^3\bar{\Sigma}\bar{\Sigma}^\prime \bar{\Sigma}_5^2
-\frac{4}{3}\tau^3\bar{\Sigma}_5^3 \bar{\Sigma}_5^\prime
\notag\\
&&\hspace{0.5cm}-\frac{4}{3}\tau^3k_E^2\bar{\Sigma} \bar{\Sigma}^\prime \bar{\Sigma}_5 \bar{\Sigma}_5^\prime
-\frac{2}{3}\tau^3k_E^2\bar{\Sigma} \bar{\Sigma}^{\prime\prime} \bar{\Sigma}_5^2
-\frac{2}{3}\tau^3k_E^2\bar{\Sigma}^{\prime2} \bar{\Sigma}_5^2
-\frac{2}{3}\tau^3k_E^2\bar{\Sigma}_5^3 \bar{\Sigma}_5^{\prime\prime}-\frac{10}{3}\tau^3 k_E^2\bar{\Sigma}_5^2 \bar{\Sigma}_5^{\prime2}+\frac{1}{6}\tau^4k_E^2\bar{\Sigma}_5^2\notag\\
&&\hspace{0.5cm}
+\frac{2}{3}\tau^4k_E^2\bar{\Sigma} \bar{\Sigma}^\prime \bar{\Sigma}_5^2
+\frac{2}{3}\tau^4k_E^2\bar{\Sigma}_5^3 \bar{\Sigma}_5^\prime
+\frac{2}{3}\tau^4k_E^2\bar{\Sigma}^2 \bar{\Sigma}^{\prime2} \bar{\Sigma}_5^2+\frac{4}{3}\tau^4k_E^2\bar{\Sigma} \bar{\Sigma}^\prime
\bar{\Sigma}_5^3 \bar{\Sigma}_5^\prime
+\frac{2}{3}\tau^4k_E^2\bar{\Sigma}_5^4 \bar{\Sigma}_5^{\prime2}]\label{D0def}
\end{eqnarray}
\begin{eqnarray}
&&\hspace{-0.5cm}\hat{D}_1=\int d^4\tilde{k}
~[2 \tau^3 \bar{\Sigma}_5^2
 -\frac{1}{3} \tau^4 k_E^2 \bar{\Sigma}_5^2
 -\frac{4}{3} \tau^4 \bar{\Sigma}^2  \bar{\Sigma}_5^2
 -\frac{2}{3} \tau^4 \bar{\Sigma}_5^4 ]\\
&&\hspace{-0.5cm}\hat{D}_2=\int d^4\tilde{k}
~[2 \tau^3 \bar{\Sigma}_5^2
 +\frac{1}{3} \tau^4 k_E^2 \bar{\Sigma}_5^2
 -4 \tau^4 \bar{\Sigma}^2  \bar{\Sigma}_5^2]\\
&&\hspace{-0.5cm}\hat{D}_3=\int d^4\tilde{k}
~[ \frac{1}{3} \tau^4 k_E^2 \bar{\Sigma}_5^2
 -\frac{4}{3} \tau^4 \bar{\Sigma}^2  \bar{\Sigma}_5^2]\\
&&\hspace{-0.5cm}\hat{D}_4=\int d^4\tilde{k}
~[ \tau^3 \bar{\Sigma}_5^2
 - \tau^4 \bar{\Sigma}^2  \bar{\Sigma}_5^2
 -\frac{1}{3} \tau^4 \bar{\Sigma}_5^4 ]\;.
\end{eqnarray}
$\hat{F}_0^2$ and $\hat{\mathcal{K}}_i^{\Sigma\neq0}$ are functions of the techniquark self-energy $\hat{\Sigma}(p_E^2)$  which is determined by (\ref{SDEhatSigma}). Detailed expressions for these quantities are given in (\ref{F0}) and (\ref{Kresult}) of Appendix.\ref{Kexpression}.
Similarly, $\tilde{F}_0^2$ and $\tilde{\mathcal{K}}_i^{\Sigma\neq0}$  are functions of the techniquark self-energy $\tilde{\Sigma}(p_E^2)$ , which is determined by (\ref{SDEtildeSigma}). Detailed expressions for these quantities are given in (\ref{F0}) and (\ref{Kresult}) of Appendix.\ref{Kexpression}. In this case, the substitution, $\hat{\Sigma}\rightarrow\tilde{\Sigma}$ is used.

With expansions (\ref{Trlnexp}),(\ref{Trlnpexp}) and (\ref{Trlnppexp}) and (\ref{V1A1})-(\ref{altb}), and by ignoring loop corrections, we can express (\ref{SEW1}) as (\ref{SEW2}) in the text. In this case, $S_0$ and $S_{Z'}$ are $Z'$ independent and dependent parts of the actions:
\begin{eqnarray}
S_0&=&\int d^4x\bigg\{
-(\frac{5}{4}\mathcal{K}+\frac{1}{4}\hat{\mathcal{K}}_2^{\Sigma\neq0}+\frac{5}{8}\hat{\mathcal{K}}_2^{\Sigma\neq0}+\frac{3}{8}\hat{\mathcal{K}}_{13}^{\Sigma\neq0})g_2^2W_{\mu\nu}^a
W^{a,\mu\nu}
-[(\frac{5}{4}+2\hat{u}+2\hat{x})\mathcal{K}+\frac{5}{8}\hat{\mathcal{K}}_2^{\Sigma\neq0}\nonumber\\
&&\hspace*{-1cm}+(\frac{5}{8}+2\hat{u}+2\hat{x})\hat{\mathcal{K}}_{13}^{\Sigma\neq0}]g_1^2B_{\mu\nu}B^{\mu\nu}
+(\frac{5}{8}\hat{\mathcal{K}}_{1}^{\Sigma\neq0}+\frac{5}{32}\hat{\mathcal{K}}_{3}^{\Sigma\neq0}-
\frac{5}{32}\hat{\mathcal{K}}_{4}^{\Sigma\neq0}-\frac{5}{8}\hat{\mathcal{K}}_{13}^{\Sigma\neq0}+\frac{5}{16}\hat{\mathcal{K}}_{14}^{\Sigma\neq0}
)(\mathrm{tr}[X_\mu X^\mu])^2\nonumber\\
&&+(\frac{5}{16}\hat{\mathcal{K}}_{4}^{\Sigma\neq0}+\frac{5}{8}\hat{\mathcal{K}}_{13}^{\Sigma\neq0}-
\frac{5}{16}\hat{\mathcal{K}}_{14}^{\Sigma\neq0})\mathrm{tr}[X^\mu X_\nu]\mathrm{tr}[X_\mu X^\nu]
+(\frac{5}{4}\hat{\mathcal{K}}_{2}^{\Sigma\neq0}-\frac{5}{4}\hat{\mathcal{K}}_{13}^{\Sigma\neq0})g_1\mathrm{tr}[\overline{W}^{\mu\nu}\tau^3]B_{\mu\nu}\nonumber\\
&&+(-\frac{5}{2}\hat{\mathcal{K}}_{13}^{\Sigma\neq0}+\frac{5}{8}\hat{\mathcal{K}}_{14}^{\Sigma\neq0})i\mathrm{tr}[\overline{W}_{\mu\nu}X^\mu X^\nu]+(-\frac{5}{4}\hat{\mathcal{K}}_{13}^{\Sigma\neq0}+\frac{5}{16}\hat{\mathcal{K}}_{14}^{\Sigma\neq0})ig_1B_{\mu\nu}\mathrm{tr}[\tau^3 X^\mu X^\nu]\nonumber\\
&&+\frac{1}{2}\hat{\mathcal{K}}_1^{\Sigma\neq0} \mathrm{tr}[U^{\dag}(D^{\mu}D_{\mu}U)U^{\dag}(D^{\nu}D_{\nu}U)+2U^{\dag}(D^{\mu}D_{\mu}U)(D^{\nu}U^{\dag})(D_{\nu}U)]\nonumber\\
&&+\frac{3}{4}\hat{\mathcal{K}}_1^{\Sigma\neq0}\mathrm{tr}[U^{\dag}(D^{\mu}D_{\mu}U)U^{\dag}(D^{\nu}D_{\nu}U)+2U^{\dag}(D^{\mu}D_{\mu}U)(D^{\nu}U^{\dag})(D_{\nu}U)]
\bigg\}\;,\label{S0def}
\end{eqnarray}
where
\begin{eqnarray}
&&\hspace*{-0.5cm}U(x)=\xi_L^{\dag}(x)\xi_R(x)\hspace*{2.5cm}X_{\mu}=U^{\dag}(D_{\mu}U)\hspace*{2.5cm}\overline{W}_{\mu\nu}=U^{\dag}g_{2}\frac{\tau^{a}}{2}W^{a}_{\mu\nu}U
~~~~~\\
&&\hspace*{-0.5cm}D_{\mu}U=\partial_{\mu}U+ig_{2}\frac{\tau^a}{2}W_{\mu}^aU-ig_{1}U\frac{\tau^3}{2}B_{\mu}\hspace*{1cm}
D_{\mu}U^{\dag}=\partial_{\mu}U^{\dag}-ig_{2}U^{\dag}\frac{\tau^a}{2}W_{\mu}^a+ig_{1}\frac{\tau^3}{2}B_{\mu}U^{\dag}~~~~~~~\\
&&\hspace*{-0.5cm}\hat{x}=(x_1+x_2)^2+(y_1+y_2)^2+(z_1+z_2)^2\hspace*{3cm}\hat{u}=(u_1+u_2)^2+(v_1+v_2)^2\;.
\end{eqnarray}
While
\begin{eqnarray}
S_{Z'}&=&\int d^4x~\bigg[\frac{1}{2}Z'_{\mu}D_Z^{-1,\mu\nu}Z'_{\nu}+Z^{\prime,\mu}J_{Z,\mu}+Z^{\prime2}Z_{\mu}'J^{\mu}_{3Z}
+g_{4Z}Z^{\prime4}\bigg]\;,~~~~\label{SZ'}
\end{eqnarray}
with
\begin{eqnarray}
&&\hspace{-0.5cm}D_Z^{-1,\mu\nu}=g^{\mu\nu}(c_{Z'}^2\partial^2+\bar{M}^2_{Z'})
-(1+\lambda_Z)\partial^{\mu}\partial^{\nu}+\Delta^{\mu\nu}_Z(X)\label{DZdef}\\
&&\hspace{-0.5cm}J_Z^\mu=J_{Z0}^\mu+g_1\gamma\partial^{\nu}B_{\mu\nu}+\tilde{J}_Z^\mu\label{JZdef}
\end{eqnarray}
\begin{eqnarray}
&&\hspace{-0.5cm}g_{4Z}=[10a_3^4+12a_3^2(2a_0^2+\hat{a}_0^2)+4a_0^4+2\hat{a}_0^4](\hat{\mathcal{K}}_{3}^{\Sigma\neq0}
+\hat{\mathcal{K}}_{4}^{\Sigma\neq0})\nonumber\\
&&\hspace{0.6cm}+g_1^4(\tan\theta+\cot\theta)^4\xi^4(\tilde{\mathcal{K}}_{3}^{\Sigma\neq0}\!+\!\tilde{\mathcal{K}}_{4}^{\Sigma\neq0})
-8\hat{D}_1a_0^4-8\hat{D}_2a_0^2a_3^2~~~\\
&&\hspace{-0.5cm}J_{3Z}^\mu=-i[(10a_3^3+12a_0^2a_3^2+6\hat{a}_0^2a_3)(\hat{\mathcal{K}}_{3}^{\Sigma\neq0}+\hat{\mathcal{K}}_{4}^{\Sigma\neq0})
+4a_0^2a_3\hat{D}_2]\mathrm{tr}[X^\mu\tau^3]\;,
\end{eqnarray}
where
\begin{eqnarray}
&&\hspace{-0.5cm}\bar{M}^2_{Z'}=2\tilde{F}_0^2g_1^2(\cot\theta+\tan\theta)^2\xi^{2}+4\hat{F}_0^2(2a_0^2+\hat{a}_0^2+5a_3^2)
-8F^{\prime2}_0a_0^2\label{barMZ'}\\
&&\hspace{-0.5cm}c^2_{Z'}=1+[4(\cot\theta+\tan\theta)^2\xi^2+2\tan^2\theta+8\hat{v}+3\tan^2\theta+\hat{y}]\mathcal{K}g_1^2
+4(\cot\theta+\tan\theta)^2\xi^2\tilde{\mathcal{K}}_2^{\Sigma\neq0}g_1^2\nonumber\\
&&\hspace*{0.6cm}+8(2a_0^2+\hat{a}_0^2+5a_3^2)\hat{\mathcal{K}}_2^{\Sigma\neq0}+[40a_3^2+2(\hat{t}+\hat{s})g_1^2]\hat{\mathcal{K}}_{13}^{\Sigma\neq0}
-16\hat{D}_0a_0^2\label{cZ'}\\
&&\hspace{-0.5cm}\lambda_Z=-2g_1^2(\tan\theta+\cot\theta)^2\tilde{\mathcal{K}}_{1}^{\Sigma\neq0}
-4(2a_0^2+\hat{a}_0^2+5a_3^2)\hat{\mathcal{K}}_{1}^{\Sigma\neq0}
\end{eqnarray}
\begin{eqnarray}
&&\hspace{-0.5cm}\Delta^{\mu\nu}_Z(X)=[40a_3^2\hat{\mathcal{K}}_{1}^{\Sigma\neq0}-(4a_0^2+2\hat{a}_0^2)\hat{\mathcal{K}}_{3}^{\Sigma\neq0}
-(4a_0^2+2\hat{a}_0^2+10a_3^2)\hat{\mathcal{K}}_{4}^{\Sigma\neq0}-20a_3^2\hat{\mathcal{K}}_{13}^{\Sigma\neq0}
+10a_3^2\hat{\mathcal{K}}_{14}^{\Sigma\neq0}\nonumber\\
&&\hspace*{1.5cm}+2a_0^2\hat{D}_4]\mathrm{tr}[X^\mu X^\nu]-(20\hat{\mathcal{K}}_{1}^{\Sigma\neq0}
+5\hat{\mathcal{K}}_{3}^{\Sigma\neq0}-10\hat{\mathcal{K}}_{13}^{\Sigma\neq0}
+5\hat{\mathcal{K}}_{14}^{\Sigma\neq0})a_3^2\mathrm{tr}[X^\mu\tau^3]\mathrm{tr}[X^\nu\tau^3]\nonumber\\
&&\hspace*{1.5cm}+g^{\mu\nu}[(5a_3^2+2a_0^2+\hat{a}_0^2)\hat{\mathcal{K}}_{3}^{\Sigma\neq0}+(2a_0^2+2\hat{a}_0^2-5a_3^2)\hat{\mathcal{K}}_{4}^{\Sigma\neq0}
-20a_3^2\hat{\mathcal{K}}_{13}^{\Sigma\neq0}+10a_3^2\hat{\mathcal{K}}_{14}^{\Sigma\neq0}\nonumber\\
&&\hspace*{1.5cm}+a_0^2\hat{D}_3]\mathrm{tr}[X^{\lambda}X_\lambda]
-g^{\mu\nu}(5\hat{\mathcal{K}}_{4}^{\Sigma\neq0}+10\hat{\mathcal{K}}_{13}^{\Sigma\neq0}-5\hat{\mathcal{K}}_{14}^{\Sigma\neq0}
)a_3^2tr[X_\lambda\tau^3]\mathrm{tr}[X^\lambda\tau^3]\\
&&\hspace{-0.5cm}J_{Z0}^\mu=-5ia_3\hat{F}_0^2\mathrm{tr}[X^\mu\tau^3]\\
&&\hspace{-0.5cm}\gamma=2[5a_3\hat{\mathcal{K}}_2^{\Sigma\neq0}+(5a_3+4g_1\hat{w}+2g_1\hat{z})\hat{\mathcal{K}}_{13}^{\Sigma\neq0}
+(4\hat{w}+\frac{5}{2}\tan\theta+2\hat{z})g_1\mathcal{K}]\label{gammaDef}\\
&&\hspace{-0.5cm}\tilde{J}_{Z}^\mu=10(-\hat{\mathcal{K}}_{2}^{\Sigma\neq0}+\hat{\mathcal{K}}_{13}^{\Sigma\neq0})
a_3 \partial_\nu\mathrm{tr}[\overline{W}^{\mu\nu}\tau^3]
+10(\hat{\mathcal{K}}_{13}^{\Sigma\neq0}-\frac{1}{4}\hat{\mathcal{K}}_{14}^{\Sigma\neq0})ia_3\partial_\nu\mathrm{tr}[X^\mu X^\nu\tau^3]\nonumber\\
&&\hspace{0.5cm}+5(\frac{1}{4}\hat{\mathcal{K}}_{3}^{\Sigma\neq0}-\frac{1}{4}\hat{\mathcal{K}}_{4}^{\Sigma\neq0}
-\hat{\mathcal{K}}_{13}^{\Sigma\neq0}+\frac{1}{2}\hat{\mathcal{K}}_{14}^{\Sigma\neq0})ia_3\mathrm{tr}[X^\nu X_\nu]tr[X^\mu\tau^3]\nonumber\\
&&\hspace{0.5cm}+5(\frac{1}{2}\hat{\mathcal{K}}_{4}^{\Sigma\neq0}+\hat{\mathcal{K}}_{13}^{\Sigma\neq0}
-\frac{1}{2}\hat{\mathcal{K}}_{14}^{\Sigma\neq0})ia_3\mathrm{tr}[X^\mu X_\nu]\mathrm{tr}[X^\nu\tau^3]\nonumber\\
&&\hspace{0.5cm}+(-5\hat{\mathcal{K}}_{13}^{\Sigma\neq0}+\frac{5}{4}\hat{\mathcal{K}}_{14}^{\Sigma\neq0})
a_3\mathrm{tr}[\overline{W}^{\mu\nu}(X_\nu\tau^3-\tau^3 X_\nu)]\nonumber\\
&&\hspace{0.5cm}+5ia_3\hat{\mathcal{K}}_1^{\Sigma\neq0}\mathrm{tr}\bigg[U^{\dag}(D^{\nu}D_{\nu}U)U^{\dag}D^{\mu}U\tau^{3}
-U^{\dag}(D^{\nu}D_{\nu}U)\tau^{3}U^{\dag}D^{\mu}U-\partial^{\mu}[U^{\dag}(D^{\nu}D_{\nu}U)\tau^{3}]\bigg]\nonumber\\
&&\hspace{0.5cm}+i\hat{a}_0{\mathcal{K}}_1^{\Sigma\neq0}\partial^\mu\mathrm{tr}[X^\nu X_\nu-U^\dag(D^\nu D_\nu U)]
\end{eqnarray}
in which
\begin{eqnarray}
&&\hspace{-0.5cm}a_0=\frac{1}{2}g_1(u_1-v_1)(\cot\theta-\tan\theta)\hspace*{3cm}a_3=\frac{1}{4}g_1\tan\theta\label{a0a3}\\
&&\hspace{-0.5cm}\hat{a}^2_0=\frac{1}{4}g^2_1(\tan\theta+\cot\theta)^2[(x_1-x_1')^2+(y_1-y_1')^2+(z_1-z_1')^2]\\
&&\hspace{-0.5cm}\hat{a}^4_0=\frac{1}{16}g_1^4(\tan\theta+\cot\theta)^4[(x_1-x_1')^4+(y_1-y_1')^4+(z_1-z_1')^4]
\end{eqnarray}
\begin{eqnarray}
&&\hspace{-0.5cm}\hat{v}=(u_2\tan\theta-u_1\cot\theta)^2+(v_2\tan\theta-v_1\cot\theta)^2\nonumber\\
&&\hspace{-0.5cm}\hat{w}=(u_1+u_2)(u_2\tan\theta-u_1\cot\theta)+(v_1+v_2)(v_2\tan\theta-v_1\cot\theta)\\
&&\hspace{-0.5cm}\hat{t}=2[(u_2+v_2)\tan\theta-(u_1+v_1)\cot\theta]^2\\
&&\hspace{-0.5cm}\hat{y}=(x_2'\tan\theta-x_1'\cot\theta)^2+(x_2\tan\theta-x_1\cot\theta)^2+(y_2'\tan\theta-y_1'\cot\theta)^2\nonumber\\
&&\hspace*{0.2cm}+(y_2\tan\theta-y_1\cot\theta)^2+
(z_2'\tan\theta-z_1'\cot\theta)^2+(z_2\tan\theta-z_1\cot\theta)^2\\
&&\hspace{-0.5cm}\hat{z}=(x_1+x_2)[(x_2'+x_2)\tan\theta-(x_1'+x_1)\cot\theta]+
(y_1+y_2)[(y_2'+y_2)\tan\theta-(y_1'+y_1)\cot\theta]\nonumber\\
&&\hspace{0.2cm}+(z_1+z_2)[(z_2'+z_2)\tan\theta-(z_1'+z_1)\cot\theta]\\
&&\hspace{-0.5cm}\hat{s}=[(x_2'+x_2)\tan\theta-(x_1'+x_1)\cot\theta]^2+[(y_2'+y_2)\tan\theta-(y_1'+y_1)\cot\theta]^2\nonumber\\
&&\hspace*{0.2cm}+[(z_2'+z_2)\tan\theta-(z_1'+z_1)\cot\theta]^2
\end{eqnarray}
From (\ref{SZ'}) and (\ref{DZdef}), it can be seen that the $Z'$ mass squared, $M_{Z'}^2$, is determined by:
\begin{eqnarray}
M_{Z'}^2=\frac{\bar{M}_{Z'}^2}{c_{Z'}^2}\;.\label{Z'mass}
\end{eqnarray}
\section{Process of integrating out $Z'$}\label{IntOutZ'}

From (\ref{SZ'}), the solution of Eq.(\ref{Z'Eq}) is
\begin{eqnarray}
Z_c^{\prime\mu}(x)=-D^{\mu\nu}_ZJ_{Z,\nu}(x)+O(p^3)+\mbox{loop corrections}\;,
\end{eqnarray}
then
\begin{eqnarray}
\bar{S}_{Z'}&=&\int d^4x\bigg[-\frac{1}{2}J_{Z,\mu}D_Z^{\mu\nu}J_{Z,\nu}
-J_{3Z,\mu'}(D_Z^{\mu'\nu'}J_{Z,\nu'})(D_Z^{\mu\nu}J_{Z,\nu})^2+g_{4Z}(D_Z^{\mu\nu}J_{Z,\nu})^4\bigg]\nonumber\\
&&+\mbox{loop corrections}\;,~~~\label{SZ'out}
\end{eqnarray}
where
\begin{eqnarray}
D_Z^{-1,\mu\nu}D_{Z,\nu\lambda}=D_Z^{\mu\nu}D_{Z,\nu\lambda}^{-1}=g^\mu_\lambda\;,
\end{eqnarray}
It can be shown that if our accuracy is on the order of $p^4$, then $p^1$ order $Z_c'$ solution is sufficient because all contributions from $p^3$ order $Z'_c$ are at least on the order of $p^6$.

Combining (\ref{SZ'out}), (\ref{DZdef}) and (\ref{JZdef})and ignoring loop corrections, we obtain:
\begin{eqnarray}
\bar{S}_{Z'}=\!\int\!
d^4x\bigg[-\frac{1}{2}J_{Z0,\mu}D_Z^{\mu\nu}J_{Z0,\nu}
-\frac{1}{\bar{M}_{Z'}^2}J_{Z0,\mu}(\tilde{J}^\mu_Z\!+\!g_1\gamma\partial_{\nu}B^{\mu\nu})-\frac{1}{\bar{M}_{Z'}^6}J_{3Z,\mu}J_{Z0}^{\mu}J_{Z0}^2
+\frac{g_{4Z}}{\bar{M}_{Z'}^8}J_{Z0}^4\bigg]\;.~~~~\label{DeltaSZ'}
\end{eqnarray}

With the help of the following algebraic relations,
\begin{eqnarray}
&&\partial_\mu\mathrm{tr}[\tau^3X^\mu]=0\nonumber\\
 &&\mathrm{tr}[\tau^3(\partial_\mu X_\nu-\partial_\nu X_\mu)]=
-2\mathrm{tr}(\tau^3X_\mu X_\nu)
+i\mathrm{tr}(\tau^3\overline{W}_{\mu\nu})-ig_1B_{\mu\nu}\nonumber\\
&&{\rm tr}(\tau^3X_\mu X_\nu){\rm tr}(\tau^3X^\mu
X^\nu)\\
&&=[{\rm tr}(X_\mu X_\nu)]^2-[{\rm tr}(X_\mu X^\mu)]^2-{\rm
tr}(X_\mu X_\nu){\rm tr}(\tau^3X^\mu){\rm tr}(\tau^3X^\nu)+{\rm
tr}(X_\mu X^\mu)[{\rm tr}(\tau^3X_\nu)]^2\nonumber\\
&&\mathrm{tr}(TA)\mathrm{tr}(TBC)+\mathrm{tr}(TB)\mathrm{tr}(TCA)+\mathrm{tr}(TC)\mathrm{tr}(TAB)
=2\mathrm{tr}(ABC)\;,\nonumber
\end{eqnarray}
where $\mathrm{tr}A=\mathrm{tr}B=\mathrm{tr}C=0$ and $T^2=1$.  We can simplify (\ref{DeltaSZ'}) into the form of the EWCL.
\section{$\mathcal{K}$ coefficients}\label{Kexpression}

In Minkowski space,
\begin{eqnarray}
\hat{F}_0^2&=&2\int d\tilde{p}\bigg[(-2\Sigma^2_p-p^2\Sigma_p\Sigma'_p)X_p^2
+(2\Sigma^2_p+p^2\Sigma_p\Sigma'_p)\frac{X_p}{\Lambda^2}\bigg],
\label{F0}\\
{\cal K}_{1}&=&2\int d\tilde{p}\bigg[-2A_pX_p^3
+2A_p\frac{X_p^2}{\Lambda^2}-A_p\frac{X_p}{\Lambda^4}
+\frac{p^2}{2}\Sigma'^2_p\frac{X_p}{\Lambda^2}-\frac{p^2}{2}\Sigma'^2_pX_p^2,
\bigg],\nonumber\\
{\cal K}_{2}&=&\int d\tilde{p}\bigg[-2B_pX_p^3
 +2B_p\frac{X_p^2}{\Lambda^2}-B_p\frac{X_p}{\Lambda^4}
+\frac{p^2}{2}\Sigma^{\prime 2}_p\frac{X_p}{\Lambda^2},
-\frac{p^2}{2}\Sigma^{\prime 2}_pX_p^2\bigg],
\nonumber\\
{\cal K}_{3}&=&2\int d\tilde{p}\bigg[
(\frac{4\Sigma^4_p}{3}-\frac{2p^2\Sigma^2_p}{3}+\frac{p^4}{18})(
 6X_p^4-\frac{6X_p^3}{\Lambda^2}+\frac{3X_p^2}{\Lambda^4}
-\frac{X_p}{\Lambda^6}),
\nonumber\\
&&+(-4\Sigma^2_p+\frac{p^2}{2})(-2X_p^3
+\frac{2X_p^2}{\Lambda^2}-\frac{X_p}{\Lambda^4})
-\frac{X_p}{\Lambda^2}+X_p^2\bigg],
\nonumber\\
{\cal K}_{4}&=&\int d\tilde{p}\bigg[
(\frac{-4\Sigma^4_p}{3}+\frac{2p^2\Sigma^2_p}{3}+\frac{p^4}{18})(
6X_p^4-\frac{6X_p^3}{\Lambda^2}+\frac{3X_p^2}{\Lambda^4}
-\frac{X_p}{\Lambda^6})
+4\Sigma^2_p(-2X_p^3+\frac{2X_p^2}{\Lambda^2}
\nonumber\\
&&-\frac{X_p}{\Lambda^4})+\frac{X_p}{\Lambda^2}-X_p^2\bigg],
\nonumber\\
{\cal K}_5&=&{\cal K}_6=0,
\nonumber\\
{\cal K}_7&=&2\int d\tilde{p}\bigg[(3\Sigma^2_p+2p^2\Sigma_p\Sigma'_p)X_p^2
+[-2\Sigma^2_p-p^2(1+2\Sigma_p\Sigma'_p)]\frac{X_p}{\Lambda^2}\bigg],
\nonumber\\
{\cal K}_8&=&0,
\nonumber\\
{\cal K}_9&=&2\int d\tilde{p}\bigg[(\Sigma^2_p+2p^2\Sigma_p\Sigma'_p)X_p^2
-p^2(1+2\Sigma_p\Sigma'_p)\frac{X_p}{\Lambda^2}\bigg],
\nonumber\\
{\cal K}_{10}&=&0,
\nonumber\\
{\cal K}_{11}&=&4\int d\tilde{p}\bigg[(-4\Sigma^3_p+p^2\Sigma_p)X_p^3
+(4\Sigma^3_p-p^2\Sigma_p)\frac{X_p}{\Lambda^2}
-(2\Sigma^3_p-\frac{1}{2}p^2\Sigma_p)\frac{X_p}{\Lambda^4}
+3\Sigma_p\frac{X_p}{\Lambda^2}
\nonumber\\
&&-3\Sigma_p X_p^2\bigg],
\nonumber\\
{\cal K}_{12}&=&0,
\nonumber\\
{\cal K}_{13}&=&\int d\tilde{p}\bigg[
(\frac{1}{2}p^2\Sigma'_p\Sigma''_p+\frac{1}{6}p^2\Sigma_p\Sigma'''_p)X_p
+(C_p-D_p)\frac{X_p}{\Lambda^2}
-(C_p-D_p)X_p^2-2E_pX_p^3
\nonumber\\
&&+2E_p\frac{X_p^2}{\Lambda^2}
-E_p\frac{X_p^2}{\Lambda^4}\bigg],
\nonumber\\
{\cal K}_{14}&=&-4\int d\tilde{p}\bigg[
-2F_pX_p^3+2F_p\frac{X_p^2}{\Lambda^2}
-F_p\frac{X_p}{\Lambda^4}
+\frac{p^2}{2}\Sigma_p^{\prime 2}\frac{X_p}{\Lambda^2}
-\frac{p^2}{2}\Sigma^{\prime 2}_pX_p^2\bigg],
\nonumber\\
{\cal K}_{15}&=&-4\int d\tilde{p}\bigg[
-(\Sigma_p+\frac{1}{2}p^2\Sigma'_p)\frac{X_p}{\Lambda^2}
+(\Sigma_p+\frac{1}{2}p^2\Sigma'_p)X_p^2\bigg],\nonumber\\
{\cal K}^{\Sigma\neq 0}_i&=&{\cal K}_i-{\cal K}_i\bigg|_{\hat{\Sigma}=0}\hspace*{2cm}i=1,2,\ldots,15
\label{Kresult}
\end{eqnarray}
in which the short notations are
\begin{eqnarray}
&&\hspace*{-1.5cm}\int d\tilde{p}\equiv
iN\int\frac{d^4p}{(2\pi)^4}e^{\frac{p^2-\hat{\Sigma}^2(p^2)}{\Lambda^2}},
\label{measure}\\
\Sigma_p&\equiv&\hat{\Sigma}(p^2),
\nonumber\\
X_p&\equiv&\frac{1}{p^2-\hat{\Sigma}^2(p^2)},
\nonumber\\
A_p&=&-\frac{2}{3}p^2\Sigma_p\Sigma'_p(-1-2\Sigma_p\Sigma'_p)-\frac{1}{3}
\Sigma^2_p(-1-2\Sigma_p\Sigma'_p)
+\frac{1}{3}p^2\Sigma^2_p(-\Sigma^{\prime 2}_p-
\Sigma_p\Sigma''_p)
\nonumber\\
&&-\frac{1}{6}p^4(-\Sigma^{\prime
2}_p -\Sigma_p\Sigma''_p),
\nonumber\\
B_p&=&-\frac{2}{3}p^2\Sigma_p\Sigma'_p(-1-2\Sigma_p\Sigma'_p)-\frac{1}{3}
\Sigma^2_p(-1-2\Sigma_p\Sigma'_p)+\frac{1}{3}p^2\Sigma^2_p(-\Sigma^{\prime
2}_p -\Sigma_p\Sigma''_p)
\nonumber\\
&&-\frac{1}{18}p^4(-\Sigma^{\prime 2}_p-\Sigma_p\Sigma''_p)
-\frac{1}{6}p^2(-1-2\Sigma_p\Sigma'_p),
\nonumber\\
C_p&=&\frac{1}{3}-\frac{1}{3}\Sigma_p\Sigma'_p
-\frac{1}{2}p^2\Sigma^{\prime 2}_p,
\nonumber\\
D_p&=&\frac{1}{2}p^2\Sigma^{\prime
2}_p-\frac{1}{3}p^2\Sigma_p\Sigma''_p
(-1-2\Sigma_p\Sigma'_p)-\frac{2}{9}p^4\Sigma'_p\Sigma''_p
(-1-2\Sigma_p\Sigma'_p)] -\frac{2}{9}p^4\Sigma^{\prime
2}_(p-\Sigma^{\prime 2}_p-\Sigma_p\Sigma''_p)
\nonumber\\
&&-\frac{1}{3}p^2\Sigma_p\Sigma'_p(-\Sigma^{\prime
2}_p-\Sigma_p\Sigma^{\prime\prime}_p),
\nonumber\\
E_p&=&-\frac{1}{6}p^2\Sigma_p\Sigma'_p(-1-2\Sigma_p\Sigma'_p)^2
-\frac{1}{9}k
p^4\Sigma^{\prime
2}_p(-1-2\Sigma_p\Sigma'_p)^2,
\nonumber\\
F_p&=&-\frac{4}{3}p^2\Sigma_p\Sigma'_p+
\frac{4}{3}p^2(\Sigma_p\Sigma'_p)^2-\frac{2}{3}\Sigma^2_p
+\frac{2}{3}\Sigma_p^3\Sigma'_p
+\frac{1}{3}p^2\Sigma^2_p(-\Sigma^{\prime
2}_p-\Sigma_p\Sigma''_p)
\nonumber\\
&&-\frac{1}{9}p^4(-\Sigma^{\prime 2}_p-\Sigma_p\Sigma''_p)
-\frac{1}{3}p^2(-1-2\Sigma_p\Sigma'_p)-\frac{1}{2}p^2.
\end{eqnarray}




\begin{thebibliography}{1}\label{biblio}

\bibitem{Hill95}
C.T.Hill, Phys.Lett.B {\bf 345}, 483(1995).

\bibitem{CDT}
R. S. Chivukula, B. A. Dobrescu, and J. Terning, Phys. Lett. B
{\bf 353}, 289(1995).

\bibitem{Lane95}
K.Lane and E.Eichten, Phys.Lett. B {\bf 352}, 382(1995).

\bibitem{Kominis}
D. Kominis, Phys. Lett. B {\bf358}, 312(1995); G. Buchalla, G.
Burdman, C. T. Hill, and D. Kominis, Phys. Rev. D {\bf 53}, 5185(1996).

\bibitem{Lane96}
K.Lane, Phys. Rev. D {\bf 54}, 2204(1996).

\bibitem{EWCL}
 T.Appelquist and G-H. Wu, Phys. Rev. D {\bf 48},
3235(1993); D {\bf 51}, 240(1995).

\bibitem{1D}
E. Farhi and L. Susskind, Phys. Rep. {\bf 74}, 277 (1981) and
references therein.

\bibitem{HongHao08}
H.H.Zhang, S.Z.Jiang, J.Y.Lang and Q.Wang, Phys. Rev. D. {\bf 77},
055003(2008).

\bibitem{JunYi09}
J.Y.Lang, S.Z.Jiang and Q.Wang, Phys. Rev. D. {\bf 79}, 015002(2009).

\bibitem{Sekhar}
F.Braam, M.Flossdorf, R.S.Chivukula, S.DiChiara and E.H.Simmons,
Phys. Rev. D {\bf 77}, 055005(2008).

\bibitem{LangPLB}
J.Y.Lang, S.Z.Jiang and Q.Wang, Phys. Lett. B {\bf 673}, 63(2009).

\bibitem{EWCLfermion}
E.Bagan, D.Espriu, J.Manzano, Phys. Rev. D {\bf 60}, 114035(1999).

\bibitem{NewWalking}
F. Sannino, arXiv:0804.0182[hep-ph].

\bibitem{ETC}
T.Appelquist and F.Sannino, Phys.Rev. D {\bf 59}, 067702(1999).

\bibitem{ConformalWindow}
D. D. Dietrich and F. Sannino,  Phys. Rev. D. {\bf 75},
085018(2007).

\bibitem{BZ}
T. Banks and A. Zaks, Nucl. Phys. B {\bf 196}, 189(1982).

\bibitem{Yamawaki}
K.Yamawaki, Int. J. Mod. Phys. A {\bf 25}, 5128(2010).

\bibitem{Runalpha}
K.I.Aoki, M.Bando, T.Kugo, M.G.Mitchard and N.Nakatani, Prog. Theor. Phys. {\bf 84}, 683(1990).

\bibitem{Lattice}
T.Appelquist, G.T.Fleming and E.T.Neil, Phys. Rev. D
{\bf 79}, 076010(2009).

\bibitem{LambdaTC}
T.Appelquist, M.Piai, and R.Shrock, Phys. Rev. D {\bf 69}, 015002(2004).

\bibitem{DPT}
H.Pagels and S.Stokar, Phys. ReV. D {\bf  20}, 2947(1979).

\bibitem{precision}
S.Dutta, K.Hagiwara, Q.S.Yan, K.Yoshida, Nucl. Phys. B {\bf 790}, 111(2008).


\end{thebibliography}
\end{document}